\documentclass[preprintnumbers,aps,prd,floatfix,nofootinbib,onecolumn]{revtex4}
\usepackage{qcd}
\usepackage{epsfig}
\usepackage{graphicx}
\usepackage[pagebackref]{hyperref}
\usepackage{bm}
\usepackage{float}
\usepackage{orcidlink}

\newcommand{\tref}[1]{Table~\ref{t.#1}}

\newcommand{\xf}{x_{\text{F}}}
\newcommand{\yh}{y_{\text{h}}}

\DeclareMathOperator{\arcsinh}{arcsinh}

\newcommand{\mspec}{M_{\text{F}}}

\begin{document}

\title{Phenomenology of TMD parton distributions in Drell-Yan \\ and $Z^0$ boson production in a hadron structure oriented approach}

\preprint{JLAB-THY-24-3987}

\author{F.~Aslan \orcidlink{0000-0001-5781-4361}\,}
\email{fpaslan@jlab.org}
 \affiliation{Department of Physics, University of Connecticut, Storrs, CT 06269, U.S.A.}
 \affiliation{Jefferson Lab, 12000 Jefferson Avenue, Newport News, VA 23606, USA}

  \affiliation{Center for Nuclear Femtography, SURA,
 12000 Jefferson Avenue, Newport News, VA 23606, USA}

\author{M.~Boglione \orcidlink{0000-0002-3647-1731}\,}
\email{boglione@to.infn.it}
\affiliation{Dipartimento di Fisica, Universit\`a degli Studi di Torino, Via P. Giuria 1, I-10125, Torino, Italy}
\affiliation{INFN, Sezione di Torino, Via P. Giuria 1, Torino, I-10125, Italy}

\author{J.~O.~Gonzalez-Hernandez \orcidlink{0000-0003-4352-9564}\,}
\email{joseosvaldo.gonzalezhernandez@unito.it}
\affiliation{Dipartimento di Fisica, Universit\`a degli Studi di Torino, Via P. Giuria 1, I-10125, Torino, Italy}
\affiliation{INFN, Sezione di Torino, Via P. Giuria 1, Torino, I-10125, Italy}
\author{T.~Rainaldi \orcidlink{0000-0002-8342-6765}\,}
\email{train005@odu.edu}
\affiliation{Department of Physics, Old Dominion University, Norfolk, VA 23529, USA}
\author{T.~C.~Rogers \orcidlink{0000-0002-0762-0275}\,}
\email{trogers@odu.edu}
\affiliation{Department of Physics, Old Dominion University, Norfolk, VA 23529, USA}
\affiliation{Jefferson Lab, 12000 Jefferson Avenue, Newport News, VA 23606, USA}
\author{A.~Simonelli \orcidlink{0000-0003-2607-9004}\,}
\email{andsim@jlab.org}
\affiliation{Department of Physics, Old Dominion University, Norfolk, VA 23529, USA}
\affiliation{Jefferson Lab, 12000 Jefferson Avenue, Newport News, VA 23606, USA}

\begin{abstract}
We present a first practical implementation of a recently proposed hadron structure oriented (HSO) approach to TMD phenomenology applied to Drell-Yan like processes, including lepton pair production at moderate $Q^2$ and $Z^0$ boson production. 
We compare and contrast general features of our methodology with other common practices and emphasize the improvements derived from our approach that we view as essential for applications where extracting details of nonperturbative transverse hadron structure is a major goal. These include the HSO's preservation of a basic TMD parton-model-like framework even while accounting for full TMD factorization and evolution, explicit preservation of the integral relationship between TMD and collinear pdfs, and the ability to meaningfully  compare different theoretical models of nonperturbative TMD parton distributions.
In our examples, we show that there is significant sensitivity at moderate $Q^2$ to both the form of the nonperturbative transverse momentum dependence and the parametrization of collinear parton densities. However, we also find that evolving to $Q^2 = M_Z^2$, without fitting, results in a satisfactory postdiction of existing data for $Z^0$ production, nearly independently of the modeling of nonperturbative transverse momentum behavior.  We argue that this demonstrates that moderate $Q$ measurements should be given greater weight than high $Q$ measurements in extractions of nonperturbative transverse momentum dependence. 
We also obtain new extractions of the nonperturbative Collins-Soper kernel within the HSO approach. We discuss its features and compare with some earlier extractions.
\end{abstract}

\date{January 25, 2024}

\maketitle

%%%%%%%%%%%%%%%%%%%%%%%%%%%%%%%%%%%%%%%%%%%%%
\section{Introduction}
\label{s.intro}

The techniques of transverse momentum dependent (TMD) factorization and evolution have found applications both in very high energy phenomenology and in studies of nonperturbative hadron structure.  However, phenomenological treatments that merge the two types of applications in a coherent way that preserves the desired features of both a hadron structure viewpoint and a high energy evolution viewpoint have remained elusive. To address this, recent work by three of us reformulated the details of the steps for applying TMD factorization in a way that is simultaneously optimized for phenomenological studies of hadron structure at moderate-to-low $Q$ and for evolution to the very large scales relevant to high energy phenomenology \cite{Gonzalez-Hernandez:2022ifv,Gonzalez-Hernandez:2023iso}. Reference~\cite{Gonzalez-Hernandez:2022ifv} referred to this as a ``bottom up'' approach while in Ref.~\cite{Gonzalez-Hernandez:2023iso} it was called a ``hadron structure oriented'' (HSO) approach. The HSO strategy is to construct phenomenological parametrizations for the TMD correlation functions, the TMD parton density functions (pdfs) and the TMD fragmentation functions (ffs), while strictly adhering to the logic of the TMD factorization and evolution derivations at each stage in the process. The purpose of the present paper is to begin the processes of putting these steps into practice.

The basic expression of TMD factorization for a process like Drell-Yan scattering valid in the small transverse momentum limit is
%%%%%%%%%%%%%%%%%%%%%%
\begin{equation}
\label{e.basicTMDmodel}
\frac{\diff{\sigma}}{\diff{^4\T{q}{}} \diff{\Omega}} \sim 
\SumInt \diff{^2 \T{k}{a}}{} \diff{^2 \T{k}{b}}{} f_{j/h_a}(x_a,\T{k}{a};\mu_Q,Q^2) f_{\bar{\jmath}/h_b}(x_b,\T{k}{b};\mu_Q,Q^2) \delta^{(2)} (\T{q}{} - \T{k}{a} - \T{k}{b}) \ ,
\end{equation}
%%%%%%%%%%%%%%%%%%%%%%
which exactly matches a TMD parton model description (e.g., Refs.~\cite{Gardiner:1970wy,Tangerman:1994eh}), except with evolution scales $Q$ and $\mu_Q$ as explicit auxiliary arguments of the TMD pdfs. The partons of flavor $j$ and $\bar{\jmath}$ are carried inside hadrons $h_a$ and $h_b$ with collinear momentum fractions $x_a$($x_b$) and transverse momenta $\T{k}{a}$($\T{k}{b}$). 
The coordinate space solution to the evolution equations for each of the TMD correlation functions is rather simple and takes the form
%%%%%%%%%%%%%%%%%%%%%%
\begin{equation}
\tilde{f}_{j/h}(x,\T{b}{};\mu_{Q},Q^2) = \tilde{f}_{j/h}(x,\T{b}{};\mu_{Q_0},Q_0^2) E(\Tsc{b}{},Q/Q_0) \, , \label{e.evofactor}
\end{equation}
%%%%%%%%%%%%%%%%%%%%%%
where $Q_0$ is an input scale and $E(\Tsc{b}{},Q/Q_0)$ is a collection of well-known exponential factors that implement evolution and whose only $\Tsc{b}{}$-dependence resides in the Collins-Soper (CS) kernel. Therefore, once a parametrization of a TMD pdf has been established at an input scale $Q_0$ and for all $\Tsc{b}{}$, evolving it to a higher $Q$ and using \eref{basicTMDmodel} becomes in principle very simple.
By comparison, the role of nonperturbative input parametrizations is somewhat obscured in the more complicated ways that evolved Drell-Yan cross sections are typically expressed.
Maintaining the factorization formula in the straightforward form in \erefs{basicTMDmodel}{evofactor} allows one to deal directly with issues related to the input parametrization that are often overlooked. 

The HSO approach simultaneously addresses a number of long-standing issues including:
%%%%%%%%%%%%%%%%%%%%%%
\begin{enumerate}[label=\arabic*)]
\item The need to preserve the integral normalizations that connect TMD and collinear correlation functions 
%%%%%%%%%%%%
\begin{equation}
\label{e.int_rel_basic}
f_{i/h}(x) \approx \int \diff{^2 \T{k}{}}{} f_{i/h}(x,\T{k}{}) \, ,
\end{equation}
%%%%%%%%%%%%
which map to an approximate probability interpretation, even at moderate $Q$. More traditional TMD parametrizations either lack this constraint, or they express it in a naive parton model form that does not include evolution. 
\item The need to match to a fixed order perturbative tail when transverse momentum is comparable to the hard scale. The parametrizations of TMD pdfs and ffs should match the large transverse momentum asymptotic behavior that is dictated by their operator definitions. 
\item The need to deal with a backwards evolution problem in TMD factorization. Specifically, data from high scale processes tend to have weak sensitivity to the nonperturbative parts in TMD parametrizations in comparison to what one finds at lower $Q$.  As such, extractions of nonperturbative transverse momentum dependence obtained from very large $Q$ measurements have errors that are amplified, and eventually blow up, as one evolves downward in $Q$. (We do emphasize, however, that understanding the nonperturbative contributions is relevant to reaching desired levels of precision at quite large scales. See, for example, \cite{ATLAS:2023lhg,Guzzi:2013aja}.)
\item The need for direct control, in the parametrizations themselves, over the transition between perturbative and nonperturbative descriptions of transverse momentum dependence as one moves from small to large transverse momentum. 
This is important for efforts to map out the regions in transverse momentum where different physical mechanisms dominate. The transition is smooth in the HSO approach, and it eliminates the arbitrary ``$\bmax$'' (and ``$\bmin$'') that appears in many standard high energy applications. Specifically, the scale at which a $\Tsc{b}{} Q_0 \to 0$ renormalization group improvement approach is imposed has been separated from the physical description of the transition between perturbative and truly nonperturbative regions.
See \aref{translation} and \sref{fullinptpara} below for a discussion of how these descriptions are connected.
\item The need for a recipe that maps any given model (say, from lattice QCD or other nonperturbative techniques) of TMD functions to the nonperturbative input of TMD factorization and evolution, and allows the predictive power of different models to be compared. 
\end{enumerate}
%%%%%%%%%%%%%%%%%%%%%%%%%%
Items 1) and 2) are essentially matters of internal consistency in the treatment of QCD factorization, so they are quite essential. Items 3) and 4) are important for applications to the study of hadron structure. Item 1) also plays an important role in the existing framework for interpreting TMD and collinear pdfs in terms of a parton model picture of hadronic structure. This can be seen, for example, in the ``prism'' diagrams that are frequently found in hadron structure literature~\cite[figure 1]{Lorce:2011dv}, where different types of correlation functions are linked by various limits and integrals. These integral connections between collinear and TMD functions have frequently been used in the past, and continue to be used, in parton model level phenomenological implementations, such as in Ref.~\cite[equations (1) and (2)]{Cammarota:2020qcw}. Thus, part of our goal in imposing conditions like 1) is also to bridge these  parton model motivated treatments with full TMD factorization treatments that include evolution.

The need to preserve predictive power has also been addressed, following somewhat different strategies, in Ref.~\cite{Qiu:2000ga,Grewal:2020hoc}. 
For our present approach, assembling all the pieces of a parametrization requires a nontrivial number of steps.
For these details, we refer the reader to both Ref.~\cite{Gonzalez-Hernandez:2022ifv} and Ref.~\cite{Gonzalez-Hernandez:2023iso}. For making the main points clearer, these earlier articles used a rather extensive notation (see, for example, the notation glossary in appendix~A of \cite{Gonzalez-Hernandez:2022ifv}). For instance, it was necessary there to carefully distinguish between concept of a correlation as an abstract theoretical object and the approximate parametrization that is modeled or extracted from a fit. To streamline the discussions in this paper, we revert back to simpler language and notation. It will be assumed that the reader is familiar enough with the steps in Refs.~\cite{Gonzalez-Hernandez:2022ifv,Gonzalez-Hernandez:2023iso} to resolve any ambiguities in notation, see also Ref.~\cite{youtubereview} for a review of the HSO approach.

It is important to reemphasize that the approach we are describing here does not relate to the underlying formalism, which is just the usual TMD factorization, but rather is merely a strategy for constructing phenomenological TMD parametrizations in a way that optimizes the handling and interpretation of explicitly nonperturbative parts of transverse momentum dependence. In particular, it is designed to allow one to confront questions about how to separately identify behavior that is irreducibly nonperturbative from contributions that can in principle be described through the use of collinear pdfs and perturbation theory. The aim is to place on a more rigorous footing the type of discussions about competing perturbative versus nonperturbative mechanisms in direct observations of data, such as that appearing in the discussion of figure 17 of Ref.~\cite{COMPASS:2017mvk}, where two apparently different mechanisms are invoked to describe small and large transverse momentum shapes. Or, very schematically, in the construction of a parametrization of an individual TMD pdf near the input scale, where one expects a separation into contributions like what is shown in \fref{PvNP}. The purple shaded area indicates sensitivity to parameters for the nonperturbative transverse momentum dependence, while the yellow shaded area represents behavior that is describable as perturbative radiation. A more precise statement of \eref{int_rel_basic} is 
\begin{equation}
\label{e.int_rel_basic2}
f_{i/h}(x;\mu) = \pi \int_0^{\mu^2} \diff{\Tscsq{k}{}}{} f_{i/h}(x,\T{k}{};\mu,\mu^2) + \Delta(f(x),\alpha_s(\mu)) + \text{power suppressed} \, . 
\end{equation}
With $\mu \approx Q$, it states that the area under the curve in \fref{PvNP} is constrained in terms of known collinear pdfs. A yellow shaded region around the $\Tsc{k}{} \approx Q$ cutoff must exist because variations with respect to this cutoff are associated with DGLAP evolution. But the relative contributions of the purple and yellow shaded areas cannot be adjusted independently from one another without violating \eref{int_rel_basic2}.

\begin{figure}[h!]
\centering
\includegraphics[width=10cm]
{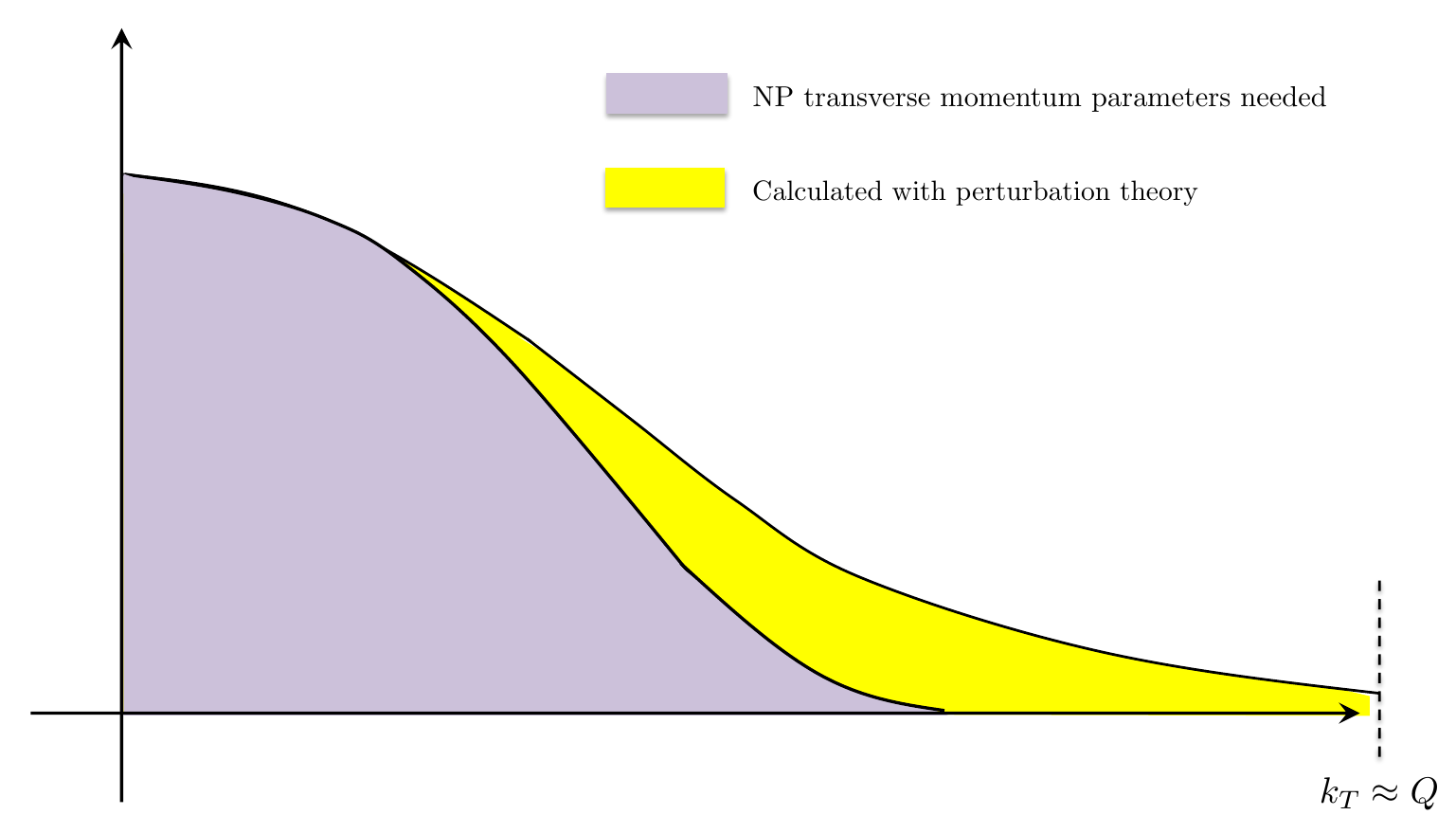}
\caption{Schematic separation of an input TMD pdf into perturbative and nonperturbative contributions. The total area under the curve is constrained by \eref{int_rel_basic2}. The purple and yellow shaded areas are not independently adjustable.}
\label{f.PvNP}
\end{figure}
The purpose of our approach is to build such interdependencies into the fitting parametrizations themselves, thus making it obvious, for example, how adjustments to nonperturbative transverse momentum parameters propagate to affect perturbative regions and vice-versa.

While we do present fits in this paper, the fits themselves are not the primary result that we wish to present, but rather they are used to demonstrate proof of principle for the methodology. 
The fits include far too small of 
a sample of data to adequately constrain nonperturbative TMD pdfs.
However, it is possible, for example, to demonstrate the general predictive power of TMD factorization applications in the HSO approach in fits to Drell-Yan scattering measurements.
We will focus on Drell-Yan-like processes (high mass lepton pair production, etc) because they have a number of attractive features for an initial phenomonenological application. First, they only involve TMD pdfs, not the admixture of pdfs and ffs that appear in semi-inclusive deep inelastic scattering (SIDIS). Second, data are available for a very wide range of $Q^2$, making studies of the effect of evolution feasible. 

We will explain the HSO fitting in a manner designed to highlight the points enumerated above:
\begin{itemize}
\item As a first step, we will fit the parameters of TMD pdfs to Drell-Yan data at moderate $Q$ in an HSO framework. (More specifically we will refer to the E288~\cite{Ito:1980ev} and E605~\cite{Moreno:1990sf}  experiments.) We will alternate between two different but classic model parametrizations of the nonperturbative transverse momentum ``core,'' namely a Gaussian parametrization and a spectator model, thus illustrating item 5) above.
We will find that both models give comparable and very satisfactory descriptions of the data.
\item
In an entirely separate step, we will next evolve the moderate $Q$ TMD extractions upward to make postdictions for $Z^0$ boson production, and we compare with existing CDF and D0 data without fitting. 
The aim of this exercise is to illustrate the predictive power of the framework by demonstrating that a reasonable description of the $Z^0$ boson cross section is obtained regardless of input model parametrizations. Furthermore, it highlights the importance of moderate scale measurements in constraining nonperturbative input. 
\end{itemize}

% The points enumerated above pertain to methodology. But there is a separate, related empirical observation that collinear and TMD factorization, while formally related, present some tension in phenomenological implemetations \cite{Gonzalez-Hernandez:2018ipj,Bacchetta:2019tcu,Boglione:2014oea}. For instance, some large-$Tsc{q}{}$ data cannot be described successfully  even with modern collinear pdf extractions, rendering the smooth matching between the TMD and collinear regions an unsolved phenomenological issue. First steps towards resolving this, were presented in~\cite{Gonzalez-Hernandez:2023iso}. We will not address this issue specifically in this paper because it is strongly tied to collinear pdf phenomenology. We remark, however, that our approach establishes the pathway for resolving the tension, by building models that manifestly relate TMDs to their collinear counterparts.  
The points enumerated above pertain to methodology. But there is a separate, related empirical observation that 
it is difficult to describe the large transverse momentum tails of transverse momentum distributions 
\cite{Boglione:2014oea}, and some observables at large $\Tsc{q}{}$ \cite{Gonzalez-Hernandez:2018ipj,Bacchetta:2019tcu},
entirely within collinear factorization. We will not address this issue specifically in this paper because it is primarily a matter of collinear pdf phenomenology. We remark, however, that our approach establishes one pathway for resolving the tension, since it 
allows for combining
% combines 
collinear and pdf fitting in one step. Indeed, between the TMD and collinear pdfs, it is the TMD pdfs that are the more fundamental objects. The collinear pdfs are derived from the TMD pdfs through the application of transverse momentum integrals. Viewed from that perspective, the collinear parametrizations should be tailored to be consistent with a TMD treatment, not vice-versa. Ultimately, it is not sensible to address collinear and TMD phenomenology in entirely separate and discrete steps.
The setup in this paper is a first step toward a combined TMD-collinear phenomenological framework. We will discuss this issue in more detail in the conclusion.  

We end this introduction with some philosophical remarks contrasting our approach with broader phenomenological trends in partonic structure. There has been important progress in accommodating large quantities of data in global fits of transverse momentum dependence.
%, and while we acknowledge the importance of this for helping to establish the outlines of what is possible in future experiments, our concern is that an excessive focus on the indiscriminate global fitting of maximally large amounts of data obscures the basic TMD factorization structures that are of primary interest and also leads to overfitting. Taken to an extreme, this works against the ultimate goal of a refined understanding of hadron structure and nonperturbative physics. Our strategy deliberately departs from what we view as a trend toward excessive curve fitting as an end to itself. Our approach prioritizes instead the preservation of predictive power coming from the nonperturbative parts of TMD correlation functions. Our ultimate goal is a phenomenological framework that makes falsifiable predictions which can legitimately be said to test the underlying theory assumptions that are of greatest interest to hadron structure theorists and quantum field theorists more broadly. 
We acknowledge the importance of this for establishing the outlines of what is possible in future experiments. However, we hope to avoid an excessive amount of global fitting in our work that could obscure the basic physics of the TMD factorization structures, %that 
which are of primary interest. %that are of primary interest. 
Our approach deliberately prioritizes the preservation of predictive power coming from the nonperturbative parts of TMD correlation functions. Our ultimate goal is a phenomenological framework that makes falsifiable predictions which can legitimately be said to test the underlying theory assumptions that are of greatest interest to hadron structure theorists and quantum field theorists more broadly.

The paper is organized as follows. 
In \sref{sidis} we review the set-up of Drell-Yan kinematics, and we establish the notation and summarize the fundamental steps that lead to the factorization of Drell-Yan scattering and to TMD evolution. 
In \sref{cutoff-def-coll} we present the cutoff definition of collinear pdfs and discuss the large momentum asymptote of their convolution, as it appears in the TMD factorization formula for Drell-Yan processes. 
In \sref{pdfs} we demonstrate how the TMD parton distribution functions are parametrized, at the input scale, within the HSO approach. 
\sref{fittingmodQ} and \sref{test-large-Q} are devoted to the data analysis, to the fit of Drell-Yan measurements at moderate Q and to their corresponding  evolution to larger Q. In \sref{gKfunction} the CS kernel is specifically addressed and compared to other, independent extractions. \sref{comparison} is dedicated to a more general comparison of the results obtained by applying the HSO approach with other analyses and TMD extractions. Finally in  \sref{conclusion} we draw our conclusions.

\newpage

%%%%%%%%%%%%%%%%%%%%%%%%
\section{Unpolarized Drell-Yan scattering}
\label{s.sidis}

\begin{figure}[h!]
\centering
\includegraphics[width=10cm]%{DY_Kin.png}
{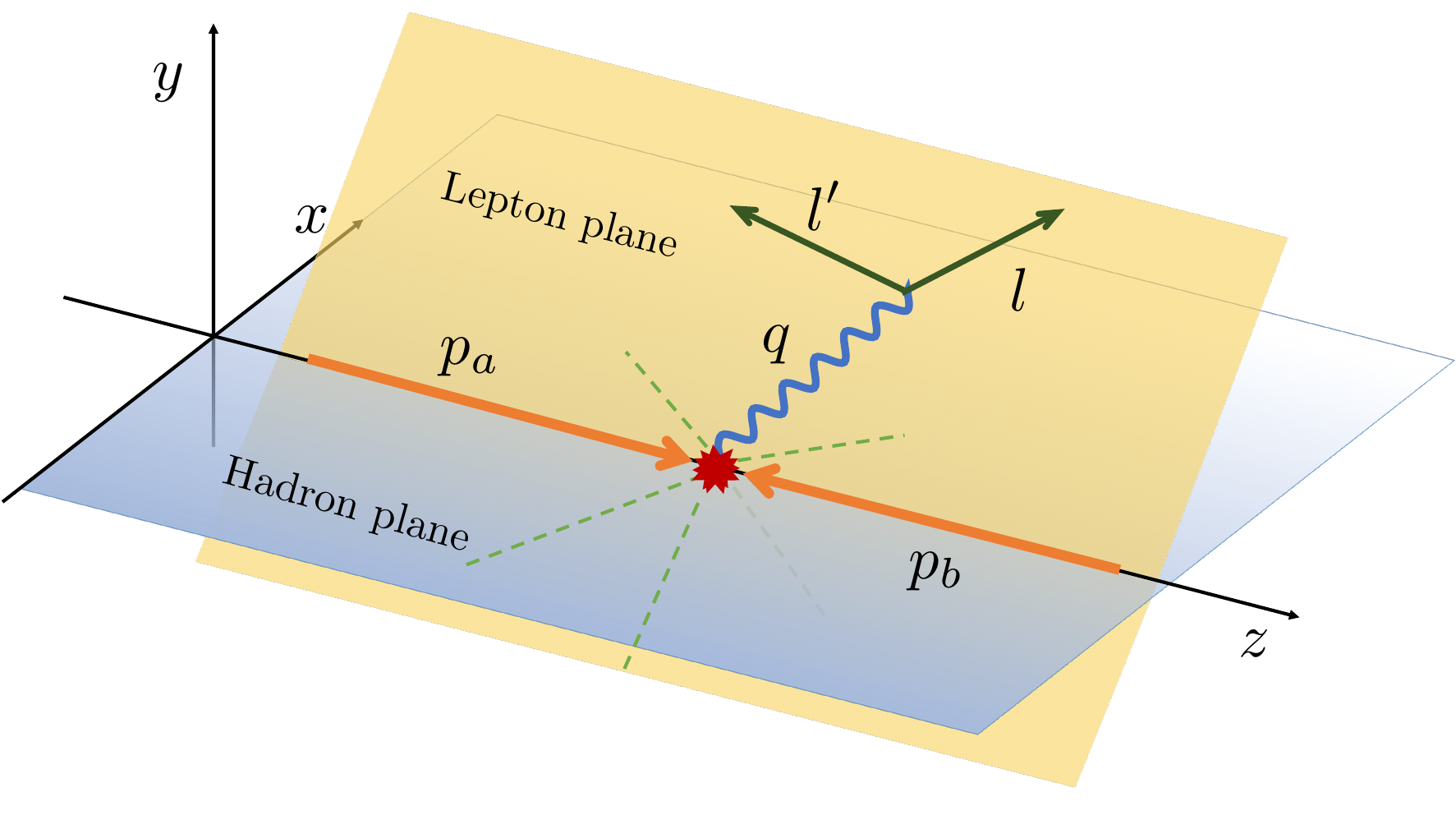}
\caption{Kinematic configuration of Drell-Yan process}
\label{fig:DY_kin}
\end{figure}

The process involves a collision between two hadrons with the inclusive production of a lepton-antilepton pair in the final state as shown in Fig.~\ref{fig:DY_kin}: 
%%%%%%%%%%%%%%%%
\begin{equation}
p_a + p_b \to l + l' + X \, .
\end{equation}
%%%%%%%%%%%%%%%%
We will label the momenta of the incoming hadrons (which could be nucleons, pions, nuclei, etc) by $p_a$ and $p_b$ respectively, $l$ and $l'$ are the final state lepton and antilepton momenta respectively, and $X$ is the unobserved integrated part of the final state. 
The four-momentum of the final state virtual photon is 
%%%%%%%%%%%%%%%%
\begin{equation}
q \equiv l + l' \;\; \text{with} \;\; q^2 = Q^2 = (l + l')^2 \, .
\end{equation}
%%%%%%%%%%%%%%%%

In setting up Drell-Yan and $W^{\pm}$/$Z^0$ production cross sections, we follow standard steps~\cite{Balazs:1997xd}.
The unpolarized Drell-Yan cross section is, following \cite{Arnold:2008kf}, decomposable into four structure functions that describe the angular dependence of the dilepton pair. It is first separated into leptonic and hadronic tensors,
%%%%%%%%%%%%%%%%
\begin{equation}
\label{e.xsection}
l_0 l_0^{\prime} \frac{\diff{\sigma}{}}{\diff{^3 \vect{l}} \, \diff{^3 \vect{l}^\prime}} = \frac{\alpha_\text{em}^2}{4 \sqrt{(p_a \cdot p_b)^2 - M_a^2 M_b^2} Q^4} L_{\mu \nu} W^{\mu \nu} \, 
\end{equation}
%%%%%%%%%%%%%%%%
where $M_a$ and $M_b$ are the masses of the colliding hadrons. 
The hadronic tensor is
%%%%%%%%%%%%%%%%
\begin{equation}
W^{\mu \nu}(p_a,p_b)  = \frac{1}{(2 \pi)^4} \int \diff{^4 z}{} e^{i q \cdot z} \langle p_a,p_b | j^\mu (0) j^\nu (z)   | p_a,p_b \rangle \, ,
\end{equation}
%%%%%%%%%%%%%%%%
and the lowest order leptonic tensor is
%%%%%%%%%%%%%%%%
\begin{equation}
\label{e.emtensor}
L_{\mu \nu} = 4 (l_\mu l^{\prime} _\nu+ l^{\prime}_ \mu l_\nu -  l \cdot l^\prime g_{\mu \nu}) \, .
\end{equation}
%%%%%%%%%%%%%%%%
 For simplicity we only deal with the electromagnetic case here. See below for the extension to electroweak bosons. We use the usual kinematical variables,
%%%%%%%%%%%%%%%%
\begin{equation}
x_a  \equiv \frac{Q^2}{2 p_a \cdot q}, \qquad x_b  \equiv \frac{Q^2}{2 p_b \cdot q}, \qquad s = (p_a + p_b)^2 \, .
\end{equation}
%%%%%%%%%%%%%%%%
Throughout the rest of this paper, we will use the approximation that the colliding hadron masses are negligible, $M_{a,b}^2 \ll Q^2$. 

\newpage

%%%-----------------------------
\subsection{The reference frames}

We will use the following two reference frames:

\subsubsection{The hadron frame}

The hadron frame is the specific center-of-mass frame in which the incoming hadrons both have exactly zero transverse momentum. In light-cone variables, 
%%%%%%%%%%%%%%
\begin{align}
p_{a,h} = \parz{p_a^+,0,\T{0}{}} = \parz{\sqrt{s/2},0,\T{0}{}} \, , \\
p_{b,h} = \parz{0,p_b^-,\T{0}{}} = \parz{0,\sqrt{s/2},\T{0}{}} \, ,
\end{align}
%%%%%%%%%%%%%
with $p_a^+ = p_b^-$. As usual, we work in the massless hadron approximation. 
The virtual photon has momentum 
%%%%%%%%%%%%%
\begin{equation}
q_{h} = \left(e^{y_h} \sqrt{\frac{Q^2+\Tscsq{q}{h}}{2}},e^{-y_h} \sqrt{\frac{Q^2+\Tscsq{q}{h}}{2}},\T{q}{h}\right) \, ,
\end{equation}
%%%%%%%%%%%%%
where $y_h$ is the hadron frame rapidity. The hadron-frame transverse momentum in terms of Lorentz invariants is
%%%%%%%%%%%%%
\begin{equation}
\Tscsq{q}{h} = \frac{2 p_a\cdot {q}\  p_b \cdot q}{p_a \cdot p_b} - Q^2 \, .
\end{equation}
%%%%%%%%%%%%%
Also,  
%%%%%%%%%%%%%
\begin{equation}
s = 2 p_a \cdot p_b = \frac{Q^2}{x_a x_b \parz{1 + \frac{\Tscsq{q}{h}}{Q^2}}} \, , \qquad y_h = \frac{1}{2} \ln \parz{\frac{x_a}{x_b}} \, ,
\end{equation}
and in the hadron frame $x_a/x_b = q_h^+/q_h^-$.
%%%%%%%%%%%%%
%%%%%%%%%%%%%%
Other commonly used variables are Feynman $x_F$ and $\tau$, 
%%%%%%%%%%%%%%
\begin{equation}
x_F \equiv x_a - x_b = \frac{2 q_h^z}{\sqrt{s} \parz{1 + \frac{\Tscsq{q}{h}}{Q^2}}}\, , \qquad \tau \equiv \frac{Q^2}{s} = x_a x_b \parz{1 + \frac{\Tscsq{q}{h}}{Q^2}} \, .
\end{equation}
%%%%%%%%%%%%%%
In the literature, the expressions for $s$, $x_F$ and $\tau$ are frequently used only in their $\Tscsq{q}{}/Q^2 \to 0$ limit. Since we ultimately wish to describe all $\Tsc{q}{}$ we have kept their $\Tsc{q}{}$-sensitive forms.

\subsubsection{The photon frame}
\label{s.photonframe}

A general photon frame is one where the virtual photon is at rest. Using Minkowski coordinates and (as usual) neglecting hadron masses, the hadron momenta in a photon frame are
%%%%%%%%%%%%%%%%%%%%
\begin{subequations}
\begin{align}
q_\gamma^\mu &= (Q,\vect{0}) \, , \\
p_{a,\gamma}^\mu &= \left|\vect{p}_{a,\gamma}\right|(1,\vect{n}_{a,\gamma}) \, ,  \\
p_{b,\gamma}^\mu &= \left|\vect{p}_{b,\gamma}\right|(1,\vect{n}_{b,\gamma}) \, . 
\end{align}
\end{subequations}
%%%%%%%%%%%%%%%%%%%%
It is convenient to define unit 3-vectors $\vect{n}_{a,\gamma}$ and $\vect{n}_{b,\gamma}$ that point along the incoming hadron momenta. 
Following steps similar to the $e^+ e^-$-annihilation case \cite{Collins:2011qcdbook}, we define unit four-vectors
%%%%%%%%%%%%%%%%%%%%
\begin{equation}
\label{e.unitvects}
Z_\gamma^\mu 
= \frac{(0,\vect{n}_{a,\gamma}-\vect{n}_{b,\gamma})}{|\vect{n}_{a,\gamma}-\vect{n}_{b,\gamma}|} \, , \ \ \ \ \ 
X_\gamma^\mu = \frac{(0,\vect{n}_{a,\gamma}+\vect{n}_{b,\gamma})}{|\vect{n}_{a,\gamma}+\vect{n}_{b,
\gamma}|} \, .
\end{equation}
%%%%%%%%%%%%%%%%%%%%
The Collins-Soper frame is a particular photon frame wherein the $z$-axis is fixed to align
along the spatial components of $Z_\gamma^\mu$ and the $x$-axis along the spatial components of $X_\gamma^\mu$. The $z$-axis then bisects the angle between
$\vect{p}_{a,\gamma}$ and $-\vect{p}_{b,\gamma}$. 
The lepton and antilepton momenta in the Collins-Soper frame are
%%%%%%%%%%%%%%%%
\begin{align}
l &{}= \frac{Q}{2} \parz{1,\sin \theta \cos \phi,\sin \theta \sin \phi ,\cos \theta} \, , \\
l' &{}= \frac{Q}{2} \parz{1,-\sin \theta \cos \phi,-\sin \theta \sin \phi ,-\cos \theta} \, .
\end{align}
%%%%%%%%%%%%%%%%

\subsection{Structure function decomposition}

In terms of $X^\mu$ and $Z^\mu$, the unpolarized parts of the hadronic tensor can be decomposed into the following conventional unpolarized structure functions,
%%%%%%%%%%%%%%%%
\begin{align}
W^{\mu \nu} &{}= \parz{-g^{\mu \nu} + \frac{q^\mu q^\nu}{Q^2} - Z^\mu Z^\nu} F^1_{UU} + Z^\mu Z^\nu F^2_{UU} \no 
&{}- \parz{X^\mu Z^\nu + Z^\mu X^\nu} F_{UU}^{\cos \phi} + \parz{-g^{\mu \nu} + \frac{q^\mu q^\nu}{Q^2} - 2 X^\mu X^\nu - Z^\mu Z^\nu }F_{UU}^{\cos 2 \phi} + \text{pol. dep.} \, ,
\end{align}
%%%%%%%%%%%%%%%%
The cross section in \eref{xsection} becomes
%%%%%%%%%%%%%%%%
\begin{equation}
l_0 l_0^{\prime} \frac{\diff{\sigma}{}}{\diff{^3 \vect{l}} \, \diff{^3 \vect{l}^\prime}} =  \frac{ \alpha_\text{em}^2}{s Q^2} \left\{ (1 + \cos^2 \theta) F^1_{UU} +  (1 - \cos^2 \theta) F^2_{UU} + \sin 2 \theta \cos \phi F_{UU}^{\cos \phi} + \sin^2 \theta \cos 2 \phi F_{UU}^{\cos 2 \phi}\right\} \, ,
\end{equation}
%%%%%%%%%%%%%%%%
where $\theta$ and $\phi$ are the polar and azimuthal angles respectively of lepton $l$ in the Collins-Soper frame. The structure functions are conveniently obtained from the hadronic tensor with the projection tensors,
%%%%%%%%%%%%%%%%
\begin{align}
P_1^{\mu \nu} &{}= -\frac{1}{2} \parz{g^{\mu \nu} + Z^\mu Z^\nu} \, , \label{e.tens1} \\
P_2^{\mu \nu} &{}= Z^\mu Z^\nu \, , \label{e.tens2} \\
P_{\phi}^{\mu \nu} &{}= -\frac{1}{2} \parz{X^\mu Z^\nu + Z^\mu X^\nu} \,  \label{e.tensphi} \\
P_{2\phi}^{\mu \nu} &{}= -\frac{1}{2} \parz{g^{\mu \nu} + Z^\mu Z^\nu} - X^\mu X^\nu \, . \label{e.tens2phi}
\end{align}
%%%%%%%%%%%%%%%%

Changing variables so that the cross section is differential in photon 4-momentum $q$, 
%%%%%%%%%%%%%%%%
\begin{equation}
\frac{\diff{\sigma}{}}{\diff{^4 q} \, \diff{\Omega}} =  \frac{ \alpha_\text{em}^2}{2 s Q^2} \left\{ (1 + \cos^2 \theta) F^1_{UU} +  (1 - \cos^2 \theta) F^2_{UU} + \sin 2 \theta \cos \phi F_{UU}^{\cos \phi} + \sin^2 \theta \cos 2 \phi F_{UU}^{\cos 2 \phi}\right\} \, ,
\end{equation}
%%%%%%%%%%%%%%%%
where ${\Omega}$ is the lepton solid angle. Integrating over all lepton angles gives the spin/polarization independent part of the cross section differential in $Q^2$, $\Tscsq{q}{h}$, and $y_h$ in the hadron frame, 
%%%%%%%%%%%%%%%%
\begin{equation}
\frac{\diff{\sigma}{}}{\diff{^2 \T{q}{h}} \diff{Q^2}{} \diff{y_h}{} } =  \frac{ 2 \pi \alpha_\text{em}^2}{3 s Q^2} \parz{2 F^1_{UU} + F^2_{UU}}  \, .
\end{equation}
%%%%%%%%%%%%%%%%
or,
%%%%%%%%%%%%%%%%
\begin{equation}
\frac{\diff{\sigma}{}}{\diff{\Tscsq{q}{h}} \diff{Q^2}{} \diff{y_h}{} } =  \frac{ 2 \pi^2 \alpha_\text{em}^2}{3 s Q^2} \parz{2 F^1_{UU} + F^2_{UU}}  \, . \label{e.finalcross}
\end{equation}
%%%%%%%%%%%%%%%%

\subsection{TMD factorization for Drell-Yan scattering}

The usual TMD-factorization expression for the hadronic tensor is
%%%%%%%%%%%%%%%%%%%%%%%%%%%%%%%
\begin{align}
&{} W^{\mu \nu}(x_a,x_b,Q,\T{q}{h}) \no
&{}= \sum_{j} H_{j\bar{\jmath}}^{\mu \nu} \int \diff{^2 \T{k}{a}}{} \diff{^2 \T{k}{b}}{} f_{j/ h_a}(x_a,\T{k}{a};\mu_Q,Q^2) f_{\bar{\jmath}/h_b}(x_b, \T{k}{b};\mu_Q,Q^2) \delta^{(2)} (\T{q}{h} - \T{k}{a} - \T{k}{b}) + \left(a\longleftrightarrow b\right) \no
&{}=\sum_{j} H_{j\bar{\jmath}}^{\mu \nu} \int \frac{\diff[2]{\T{b}{}}}{(2 \pi)^2}
    ~ e^{i\T{q}{h}\cdot \T{b}{} }
    ~ \tilde{f}_{j/h_a}(x_a,\T{b}{};\mu_Q,Q^2) 
    ~ \tilde{f}_{\bar{\jmath}/h_b}(x_b,\T{b}{};\mu_Q,Q^2)+ \left(a\longleftrightarrow b\right)  \no
&{}=\sum_{j} H_{j\bar{\jmath}}^{\mu \nu}\left\lbrace \left[ f_{j/h_a}, f_{\bar{\jmath}/h_b} \right] + \left[ f_{j/h_b}, f_{\bar{\jmath}/h_a} \right] \right\rbrace \, ,
\label{e.hadrotens}
\end{align}
%%%%%%%%%%%%%%%%%%%%%%%%%%%%%
where on the last line we have used the common bracket notation for abbreviating the transverse convolution integrals. The hard part in \eqref{e.hadrotens} reads, 
%%%%%%%%%%%%%%%%%%%%%%%%%%%%%
\begin{equation}
H_{j\bar{\jmath}}^{\mu \nu} = \frac{e_j^2}{2 Q^2 N_c} \trd{\sla{\hat{k}}_a \gamma^\mu \sla{\hat{k}}_b \gamma^\nu} \left| H_{j \bar{\jmath}}(\alpha_s(\mu),\mu/Q) \right|^2 \, .
\end{equation}
%%%%%%%%%%%%%%%%%%%%%%%%%%%%%
The basics of the TMD factorization theorem originate in the Collins-Soper-Sterman (CSS) formalism~\cite{Collins:1981uw,Collins:1981uk,Collins:1984kg} and its updated forms~\cite{Collins:2011qcdbook,Collins:2017oxh}. See Ref.~\cite{Rogers:2015sqa} for a review that includes a detailed list of references. 
The $\hat{k}_a$ and $\hat{k}_b$ partonic momenta are the hard approximate momenta in the Collins-Soper frame, 
%%%%%%%%%%%%%%%%
\begin{align}
\hat{k}_{a,\gamma} &{}= \parz{Q/\sqrt{2},0,\T{0}{}}  , \label{e.khata} \\
\hat{k}_{b,\gamma} &{}= \parz{0,Q/\sqrt{2},\T{0}{}}  \, . \label{e.khatb}
\end{align}
%%%%%%%%%%%%%%%%
We have already fixed the auxiliary renormalization group (RG) and rapidity scales, $\mu$ and $\zeta$, to $\mu_Q = C_2 Q$ and $\zeta = Q^2$ respectively in \eref{hadrotens} to optimize perturbation theory. We will use $C_2 = 1$ throughout this paper. 
The hard vertex factor for Drell-Yan scattering is \cite{Moch:2005id,Collins:2017oxh}
%%%%%%%%%%%%%%%%
\begin{equation}
   |H_{j \bar{\jmath}}(\alpha_s(\mu_Q),\mu_Q/Q)|^2 
  = 1 + \sum_{n=1}^{\infty} \parz{\frac{\alpha_s(\mu_Q)}{4 \pi}}^n \hat{H}_{j \bar{\jmath}}^{(n)} 
  \label{e.Hardfactor}
\end{equation}
%%%%%%%%%%%%%%%%
with the lowest order, 
%%%%%%%%%%%%%%%%
\begin{align}
  H_{j \bar{\jmath}}^{(1)}
  ={}&
    \delta_{j \bar{\jmath}} C_F \left( - 16 + \frac{7\pi^2}{3}  \right) \, . \label{eq:H.hat.1} 
\end{align}
%%%%%%%%%%%%%%%%
Here and throughout the rest of this paper we will keep $\order{\alpha_s}$ in perturbative expressions. However, most $\alpha_s^2$ contributions are available and we list them in \aref{higherorder} for future use.

The projection tensors in \erefs{tens1}{tens2phi} give the TMD factorization formula in terms of the unpolarized structure functions,
%%%%%%%%%%%%%%%%
\begin{align}
F_{UU}^1 &{}= 
\sum_j 
e_j^2
\frac{\left| H_{j \bar{\jmath}} \right|^2}{4 \pi^2 N_c}  
\int \diff{^2 \T{b}{}} e^{i \T{q}{h} \cdot \T{b}{}} \tilde{f}_{j/h_a}(x_a,\T{b}{};\mu_Q,Q^2) 
    ~ \tilde{f}_{\bar{\jmath}/h_b}(x_b,\T{b}{};\mu_Q,Q^2)+ \left(a\longleftrightarrow b\right) \label{e.Fuu} \\ \, 
&{}= 
\sum_j 
e_j^2
\frac{\left| H_{j \bar{\jmath}} \right|^2}{N_c} 
\int \diff{^2 \T{k}{a}}{} \diff{^2 \T{k}{b}}{} f_{j/ h_a}(x_a,\T{k}{a};\mu_Q,Q^2) f_{\bar{\jmath}/h_b}(x_b, \T{k}{b};\mu_Q,Q^2) \delta^{(2)} (\T{q}{h} - \T{k}{a} - \T{k}{b}) + \left(a\longleftrightarrow b\right) \, , \no
F_{UU}^2 &{}= 0 \, . \label{e.Fuu2}
\end{align}
%%%%%%%%%%%%%%%%

%%%-----------------------------
\subsection{$Z^0$ boson production}

For the calculation of cross sections at scales around the $Z^0$ boson mass, we will consider all contributions, namely, the photon, $Z^0$ and interference channels. 
% Note that, at these scales, the running of the QED coupling is relevant. We will use its value at the Z boson mass, $\alpha_{\text{em}}(M_Z)=1/127.951$~\cite{ParticleDataGroup:2022pth}.

To obtain expressions for the structure functions $F^{1,Z}_{UU}$ and $F^{2,Z}_{UU}$ in the $Z^0\to e^+e^-$ channel, one may replace in \erefs{Fuu}{Fuu2}
\begin{align}\label{e.replaceZ0}
e_j^2 \to 
\frac{Q^4}{\left(Q^2-M_Z^2\right)^2 + M_Z^2\,\Gamma_Z^2}
\left(\frac{1+\left[ 1 - 4  \sin^2 \theta_w \right]^2}{16 \cos^2 \theta_w \sin^2 \theta_w}\right)
\left(\frac{1+\left[ 1 - 4 | e_j | \sin^2 \theta_w \right]^2}{16 \cos^2 \theta_w \sin^2 \theta_w}\right)\,,
\end{align}
with $M_Z$, $\Gamma_Z$ the mass and the width of the $Z^0$ boson, respectively, 
and where $\theta_w$ is the Weinberg angle in the $\msbar$ renormalization scheme and at $\mu = M_z$.
The contributions from the interference of intermediate photon and $Z^0$ boson can be obtained similarly, by the replacement
\begin{align}\label{e.replaceinterf}
    e_j^2
\to&
-\frac{Q^2(Q^2-M_Z^2)}{(Q^2-M_Z^2)^2+M_Z^2 \Gamma_Z^2}\,
\frac{|e_j| \,(1-4\,\sin^2\theta_W)(1-4\,|e_j|\,\sin^2\theta_W)}{8\,\sin^2\theta_W\,\cos^2\theta_W}\,.
\end{align}
Note that, at these scales, the running of the QED coupling is relevant. We will use the value reported in Ref.~\cite{ParticleDataGroup:2022pth},  $\alpha_{\text{em}}(M_Z)=1/127.951$.
\subsection{Evolution}

The treatment of evolution for the TMD pdfs is entirely within the now standard approach. For a review with a list of references see, for instance, Ref.~\cite{Rogers:2015sqa}. 
A useful summary of the general logic of the evolution equations is also to be found in \cite{Collins:2012ss}.
The TMD pdfs exactly satisfy the following evolution equations in coordinate space, 
%%%%%%%%%%%%%%%%
\begin{align}
 \frac{\partial  \ln \tilde{f}_{j/p}(x,{\Tsc{b}{}}{};\mu,\zeta)}{\partial \ln \sqrt{\zeta} } ={}& \tilde{K}({\Tsc{b}{}}{};\mu)  \, , \label{e.CSeq}  \\
 \frac{\text{d}{ \tilde{K}({\Tsc{b}{}}{};\mu)}}{\text{d}{\, \ln \mu}{}} ={}&  - \gamma_K(\alpha_s(\mu))  \, , \label{e.CSrg} \\
\frac{\text{d}{\, \ln  \tilde{f}_{j/p}(x,{\Tsc{b}{}}{};\mu,\zeta)}}{\text{d}{\, \ln \mu}} ={}&  \gamma(\alpha_s(\mu);\zeta /\mu^2) = \gamma(\alpha_s(\mu);1) - \gamma_K(\alpha_s(\mu)) \frac{1}{2} \ln \left(\frac{\zeta}{\mu^2}\right) \, , \label{e.TMDrg}
\end{align}
%%%%%%%%%%%%%%%%
where $\tilde{K}$ is the Collins-Soper kernel, $\gamma_K$ is its anomalous dimension and $\gamma$ the TMD anomalous dimension.
After TMD evolution from an initially low input scale $Q_0$ to an arbitrary higher scale $Q$, \eref{Fuu} becomes
%%%%%%%%%%%%%%%%
\begin{align}
F_{UU}^1 &{}= 
\sum_j 
e_j^2
\frac{\left| H_{j \bar{\jmath}} \right|^2}{4 \pi^2 N_c}  
\int \diff{^2 \T{b}{}} e^{i \T{q}{h} \cdot \T{b}{}} \tilde{f}_{j/h_a}(x_a,\T{b}{};\mu_{Q_0},Q_0^2) 
    ~ \tilde{f}_{\bar{\jmath}/h_b}(x_b,\T{b}{};\mu_{Q_0},Q_0^2) \times \no 
    \qquad &{} \times \exp\left\{  
        \tilde{K}(\Tsc{b}{};\mu_{Q_0}) \ln \parz{\frac{ Q^2 }{ Q_0^2}}
           +\int_{\mu_{Q_0}}^{\mu_Q}  \frac{ \diff{\mu'} }{ \mu' }
           \biggl[ 2 \gamma(\alpha_s(\mu'); 1) 
                 - \ln\left(\frac{Q^2}{ {\mu'}^2 } \right)\gamma_K(\alpha_s(\mu'))
           \biggr]
  \right\} + \left(a\longleftrightarrow b\right) \, . \label{e.Fuuevolved} 
\end{align}
%%%%%%%%%%%%%%%%
We take $Q_0$ to be the lowest scale for which TMD factorization is to be considered trustworthy. All perturbatively calculable quantities will be kept through $\order{\alpha_s}$.

Within the HSO approach, the strategy is to construct parametrizations of $f_{j/h}(x,\T{k}{};\mu_{Q_0},Q_0^2)$ and $\tilde{K}(\Tsc{b}{};\mu_{Q_0})$ that simultaneously: 1.) are phenomenologically successful in the $Q_0$ regime, 2.) recover the perturbative expression for $\Tsc{k}{} \approx Q$, and 3.) obey the appropriate evolution equations when evolving to $Q \gg Q_0$. 

The implementation of \erefs{CSeq}{Fuuevolved} in this paper will make use of results for the anomalous dimensions and evolution kernels that were originally calculated in a range of different formalisms, some of whose connection to the basic TMD factorization in \eref{Fuuevolved} is not immediately obvious. Some translation is required, and for that we refer the reader to Ref.~\cite{Collins:2017oxh}. For example, expressions for $\tilde{K}{}$, $\gamma$, and $\gamma_K$ are from \cite{Davies:1984hs}, and extensions up to $\order{\alpha_s^3}$ can be obtained in, for example, Ref.~\cite{Moch:2005id,Li:2016ctv}.

%%%%%%%%%%%%%%%%%%%%%%%%%%%%%%%%%%%%%%%%%%%%%%%%%
\section{Cutoff collinear pdfs and the large transverse momentum asymptote}
\label{s.cutoff-def-coll}

As explained in \sref{intro}, the HSO approach  preserves the integral normalizations that relate TMD and collinear correlation functions, and ensures that TMD pdfs match the large transverse momentum asymptotic behavior dictated by the operator definitions. It is useful, therefore, to define a collinear pdf obtained by integrating the TMD pdf over $\Tsc{k}{}$,
%%%%%%%%%%%%%%%%%
\begin{align}
f^c_{i/p}(x;\mu_Q;\mu_c) &{}\equiv 2 \pi \int_0^{\mu_c} \diff{\Tsc{k}{}}{} \Tsc{k}{} f_{i/p}(x,\T{k}{};\mu_Q,Q^2) \, , \label{e.fc_deff} 
\end{align}
%%%%%%%%%%%%%%%%%
where $\mu_c = \mu_c(\mu_Q)$ is a cutoff on $\Tsc{k}{}$.  It coincides with the literal probability density interpretation that one has in the parton model, and it equals the $\msbar$ definition up to calculable $\order{\alpha_s}$ corrections and corrections suppressed by powers of $1/\mu$,
%%%%%%%%%%%%%%%%%
\begin{align}
f^c_{i/p}(x;\mu_Q;\mu_c) ={}& f^{\msbar}_{i/p}(x;\mu_Q)
+ \Delta_{i/p}(\alpha_s(\mu_Q),\mu_c/\mu_Q) + \order{\frac{m^2}{\mu_Q^2} } \, \label{e.cutpdf} 
\end{align}
%%%%%%%%%%%%%%%%%
where $\Delta$ is the correction, see  Sec.~III of \cite{Gonzalez-Hernandez:2022ifv} for the equivalent expression for fragmentation functions.
% \footnote{$m$ represents any small mass of the problem. The subleading errors in expressions like \eref{cutpdf} need not in general be exactly quadratic, but we will retain this notation for simplicity since the exact power is irrelevant for our purposes.}
\footnote{$m$ represents any mass scale that may be considered small relative to the hard scale, such as $\Lambda_\text{QCD}$, a light quark mass, or a small hadronic mass.
% , hadronic sized mass. 
The subleading errors in expressions like \eref{cutpdf} need not in general be exactly quadratic, but we will retain this notation for simplicity since the exact power is irrelevant for our purposes.}
For our applications, we will set $\mu_c = \mu_Q$ and drop the $\order{m/\mu_Q}$ errors in \eref{cutpdf} and 
express the cutoff definition (Eq.~(58) in Ref.~\cite{Gonzalez-Hernandez:2022ifv}) as 
%%%%%%%%%%%%%%%%%
\begin{align}
f^c_{i/p}(x;\mu_Q) = \lim_{\frac{m}{\mu_Q} \to 0} f^c_{i/p}(x;\mu_Q;\mu_Q) \, ,
\label{e.cutpdf2}
\end{align}
%%%%%%%%%%%%%%%%%
where we have dropped the underline on the right-hand side and the superscripts of Ref.~\cite{Gonzalez-Hernandez:2022ifv}.

When $\Tsc{k}{} \approx \mu \approx Q$, the perturbative tail approximation to a single TMD pdf (through $\mathcal{O}\left(\alpha_s\right)$) is 
%%%%%%%%%%%%%%%%%%%%%
\begin{align}
f^\text{pert}_{i/p}(\xbj,\Tsc{k}{};\mu_{Q},Q) &{}= 
\frac{1}{2 \pi} \frac{1}{\Tscsq{k}{}} \left[ A_{i/p}(\xbj;\mu_{Q}) + B_{i/p}(\xbj;\mu_{Q}) \ln \left(\frac{Q^2}{\Tscsq{k}{}}\right) + A_{i/p}^{g}(\xbj;\mu_{Q}) \right] \, . 
\label{e.pertf}
\end{align}
%%%%%%%%%%%%%%%%%%%%%
The convolution product that appears in the TMD factorization formula is 
%%%%%%%%%%%%%%%%%%%%%
\begin{equation}
\label{e.rewrite}
\left[ f_a, f_b \right] = \int \diff{^2 \T{k}{}}{} f_a(\xbj,-\T{k}{}+\T{q}{}/2;\mu_Q;Q^2) f_b(\xbj,\T{k}{}+\T{q}{}/2;\mu_Q;Q^2) \, . 
\end{equation}
%%%%%%%%%%%%%%%%%%%%%%%
% so that
In the limit  $\Tsc{q}{}\sim Q$, $Q\to\infty$, the bracket in \eref{rewrite} gives the so called ``asymptotic term''. This can be calculated entirely in collinear factorization, in terms of the perturbative tail of \eref{pertf}, up to power suppressed terms

\begin{align}\label{e.asy}
\left[ f_a, f_b \right]
&{}= f_a^\text{pert}(\xbj,\T{q}{};\mu_Q;Q^2) f_b^c(\xbj;\mu_Q) + f_b^\text{pert}(\xbj,\T{q}{};\mu_Q;Q^2) 
f_a^c(\xbj;\mu_Q)\no 
&{}+  \int \diff{^2 \T{k}{}}{} \left\{  f^\text{pert}_a(\xbj,-\T{k}{}+\T{q}{}/2;\mu_Q;Q^2) f^\text{pert}_b(\xbj,\T{k}{}+\T{q}{}/2;\mu_Q;Q^2) 
 \right. \no
&{} \left. \qquad \qquad -  f^\text{pert}_a(\xbj,\T{q}{};\mu_Q;Q^2) f^\text{pert}_b(\xbj,\T{k}{}+\T{q}{}/2;\mu_Q;Q^2) \Theta(\mu_Q - |\T{k}{}+\T{q}{}/2|) \right. \no &{} \left. \qquad \qquad -  f^\text{pert}_a(\xbj,-\T{k}{}+\T{q}{}/2;\mu_Q;Q^2) f^\text{pert}_b(\xbj,\T{q}{};\mu_Q;Q^2)  \Theta(\mu_Q - |-\T{k}{} + \T{q}{}/2|) \right\} + \order{\frac{m^2}{\Tscsq{q}{}}} \, \no
&{}= \left[ f_a, f_b \right]_{\text{ASY}} + \order{\frac{m^2}{\Tscsq{q}{}}} \, .
\end{align}
%%%%%%%%%%%%%%%%%%%%%%

The term in \eref{asy} is useful to implement large-$\Tsc{q}{}$ corrections to the TMD approximation. We refer the reader to sections V and VI of Ref.~\cite{Gonzalez-Hernandez:2023iso} for more details.

%%%%%%%%%%%%%%%%%%%%%%%%%%%%%%%%%%%%%%%%%%%%%%%%%%%%%%%%

\section{TMD parton distribution functions: parametrizations}
\label{s.pdfs}

In this section, we summarize the steps for setting up the parametrizations that we will use in later sections. 

%%%-----------------------------
\subsection{The input scale TMD pdfs}

\subsubsection{Two component setup}

We will use the same additive two component setup from \cite{Gonzalez-Hernandez:2023iso} as the basic TMD parametrization.
At an input scale $Q_0$ it is
%%%%%%%%%%%%%%%%%%%%%
\begin{align}
\inptp{f}{i/p}(\xbj,\Tsc{k}{};\mu_{Q_0},Q_0^2) &{}= 
\frac{1}{2 \pi} \frac{1}{\Tscsq{k}{} + m_{i,p,A}^2} A_{i/p}(\xbj;\mu_{Q_0}) + \frac{1}{2 \pi} \frac{1}{\Tscsq{k}{} + m_{i,p,B}^2} B_{i/p}(\xbj;\mu_{Q_0}) \ln \left(\frac{Q_0^2 }{\Tscsq{k}{}+m_{i,p,L}^2}\right)  \no
&{}+  \frac{1}{2 \pi} \frac{1}{\Tscsq{k}{} + m_{g,p}^2} A_{i/p}^{g}(\xbj;\mu_{Q_0}) \no
&{} 
+
C_{i/p} \,\np{f}{i/p}(\xbj,\Tsc{k}{};Q_0^2) \, .
\label{e.candidateqpdf}
\end{align}
%%%%%%%%%%%%%%%%%%%%%
The mass parameters in the different terms above can in principle all be different, though from here forward we will take them to be equal for simplicity, i.e. $m_{i,p}\equiv m_{i,p,A}=m_{i,p,B}=m_{i,p,L}$. 
Then, the other factors in \eref{candidateqpdf} are 
%%%%%%%%%%%%%%%%%%%%%
\begin{align}
&{} A_{i/p}(x;\mu_{Q_0}) \equiv \sum_{i'} \delta_{i'i} \frac{\alpha_s(\mu_{Q_0})}{\pi}
\left\{ \left[(P_{ii'}\otimes{f}_{i'/p})(x;\mu_{Q_0})\right] \vphantom{\frac{3 C_F}{2} d(z;\mu_{Q_0})} -  \frac{3 C_F}{2} f_{i'/p}(x;\mu_{Q_0})  \right\} \, , \label{e.A_def_pdf}  \\
&{} B_{i/p}(x;\mu_{Q_0}) \equiv \sum_{i'} \delta_{i'i} \frac{\alpha_s(\mu_{Q_0}) C_F}{\pi}f_{i'/p}(x;\mu_{Q_0})  \, , \label{e.B_def_pdf} \\
&{}A^{g}_{i/p}(x;\mu_{Q_0}) \equiv \frac{\alpha_s(\mu_{Q_0})}{\pi} \left[(P_{ig}\otimes{f}_{g/p})(x;\mu_{Q_0})\right] \, , \\
&{}C_{i/p} \equiv
\frac{1}{N_{i/p}}
\Bigg[
f_{i/p}(x;\mu_{Q_0}) - A_{i/p}(x;\mu_{Q_0}) \ln \parz{\frac{\mu_{Q_0}}{m_{i,p}}} -  B_{i/p}(x;\mu_{Q_0}) \ln \parz{\frac{\mu_{Q_0}}{m_{i,p}}} \ln \parz{\frac{Q_0^2}{\mu_{Q_0} m_{i,p}} } \,   \no
&{} - A^{g}_{i/p}(x;\mu_{Q_0}) \ln \parz{\frac{\mu_{Q_0}}{m_{g,p}}} +\frac{\alpha_s(\mu_{Q_0})}{2 \pi} 
\left\{ \sum_{i'} \delta_{i'i} [ \mathcal{C}_{\Delta}^{i/i'} \otimes f_{i'/p} ](x;\mu_{Q_0}) + [ \mathcal{C}_{\Delta}^{i/g} \otimes f_{g/p} ](x;\mu_{Q_0}) \right\}
\Bigg]
\, , \label{e.C_def_pdf} 
\end{align}
%%%
and
%%%%%%%%%%%%%%%%%%%%
\begin{align}
P_{qq}(x)&{}=P_{\bar{q}\bar{q}}(z)=C_F\left[\frac{1+x^2}{\parz{1-x}_+}+\frac{3}{2}\delta\parz{1-x}\right] \, , \label{e.pdists} \\
P_{ig}(x)&{}= T_F \left[x^2 + (1-x)^2 \right] \, , \\
\mathcal{C}_{\Delta}^{i/i}(x) &{}=   C_F (1 - x) -C_F\frac{\pi^2}{12}\delta(1-x) \, ,
\label{e.Delta_trans1pdf} \\
\mathcal{C}_{\Delta}^{i/g}(x) &{}=  2 T_F x (1-x)  \, ,
\label{e.Delta_trans2pdf} \\
N_{i/p}\equiv{}&
\,2\pi\,\int_{0}^{\infty}\diff{\Tsc{k}{}}{}\Tsc{k}{}\,\np{f}{i/p}(x, \Tsc{k}{};Q_0^2) \, .
\label{e.Fnorm}
\end{align}
%%%%%%%%%%%%%%%%%%%%%
The $\np{f}{i/p}(x, \Tsc{k}{};Q_0^2)$ functions parametrize a ``core'' or peak of the TMD pdf while the remaining terms interpolate to the $\order{\alpha_s}$ perturbative tail at large $\Tsc{k}{}$. The value of $C_{i/p}$ is fixed by requiring the TMD pdfs to match the corresponding collinear pdfs after they are integrated up to a cutoff $\mu_c$, up to correction terms to convert between different schemes as in \erefs{fc_deff}{cutpdf}. More explicitly, the correction term in Eq.~\eqref{e.cutpdf} with the choice $\mu_c=\mu_Q$ reads
\begin{equation}
\Delta_{i/p}(\alpha_s(\mu_Q),1) = {}\frac{\alpha_s(\mu_{Q})}{2 \pi} 
\left\{ \sum_{i'} \delta_{i'i} [ \mathcal{C}_{\Delta}^{i/i'} \otimes f_{i'/p} ](x;\mu_{Q}) + [ \mathcal{C}_{\Delta}^{i/g} \otimes f_{g/p} ](x;\mu_{Q}) \right\} + \mathcal{O}\left(\alpha^2_s(\mu_Q)\right) .
\end{equation}
The ``$\otimes$'' convolution symbol has the usual definition,
%%%%%%%%%%%%%%%%%%%%%%
\begin{equation}
(f\otimes{g})(x;\mu) \equiv \int_{x}^1\frac{\diff\xi}{\xi}f(x/\xi)g(\xi;\mu) \, .
\end{equation}
%%%%%%%%%%%%%%%%%%%%%%
The guiding principles motivating \eref{candidateqpdf} 
are devised to deliver a parametrization that
i) smoothly and gradually interpolates between a purely nonperturbative behavior at $\Tsc{k}{} \approx 0$ and a fixed-scale, fixed order perturbative tail at $\Tsc{k}{} \approx Q_0$ and, ii) that preserves the integral relation in \eref{int_rel_basic2}. Note that its Fourier-Bessel transform reproduces the operator product expansion at small $\Tsc{b}{}$. Thus, it is built following the recipe in Sec.~VI of \cite{Gonzalez-Hernandez:2022ifv}. The additive structure of the interpolation is not strictly necessary, but it makes performing integrals and Fourier-Bessel transforms simple. Indeed, we strongly emphasize that the HSO approach is not specific to the parametrization we choose here, or to any particular parametrization. It can accommodate any parametrization, which may be formulated either in transverse momentum or coordinate space. Our specific choice for this paper of a simple additive structure is motivated primarily by convenience, but it is likely to be updated or modified in future iterations.  

\subsubsection{Nonperturbative models of small transverse momentum}
\label{s.nonpertcore}

The details about the nonperturbative behavior are contained within the last line of \eref{candidateqpdf} in the $\np{f}{i/p}(x, \Tsc{k}{};Q_0^2)$ function. We will compare two basic forms for $\np{f}{i/p}(x, \Tsc{k}{};Q_0^2)$. One popular parametrization is a Gaussian shape,
%%%%%%%%%%%%%
\begin{align}
\label{e.npmodels}
\np{f}{i/p}^{\text{Gauss}}(\xbj,\T{k}{};Q_0^2)={}&
\frac{e^{-\Tscsq{k}{}/M_\text{F}^2}}{\pi  M_\text{F}^2}\, ,
\end{align}
%%%%%%%%%%%%%
where $M_{\text{F}}$ is a model parameter. 
The second core parametrization is the spectator model in Eq.~(44) of Ref.~\cite{Bacchetta:2008af},
\begin{align}
\label{e.spectatorpdf}
\np{f}{i/p}^{\text{Spect}}(\xbj,\T{k}{};Q_0^2)={}&
\frac{1}{\pi}
\frac{6 \, L^6}{L^2+2 (m_q+\xbj\,M_p)^2}
\frac{ \Tscsq{k}{}+(m_q+\xbj\,M_p)^2}{ \left(\Tscsq{k}{}+L^2\right)^4 }
\,,\quad 
L^2= (1-\xbj) \Lambda^2 +\xbj M_X^2
-\xbj(1-\xbj) M_p^2\,, 
\end{align}
where the quantities $m_q, M_X, \Lambda$ are model parameters and $M_p$ is the proton mass. 
The overall factors in \erefs{npmodels}{spectatorpdf} are chosen so that the core functions are normalized to unity, i.e.  $N_{i/p} = 1$. 

In the future, more sophisticated modeling may replace 
\erefs{npmodels}{spectatorpdf}. For example, the core models might be guided by work in Refs.~\cite{Bacchetta:2008af,Schweitzer:2012dd,Bacchetta:2017vzh,Guerrero:2020hom}. Developments in lattice QCD~\cite{LPC:2022zci} may also soon provide guidance.

\subsubsection{Coordinate space representation}

Since TMD evolution is usually performed in coordinate space, it will be convenient to write the coordinate space versions of the above parametrizations. They are, 
% %%%%%%%%%%%%%
% \begin{align}
% \label{e.coordspace_tmdpdfinput}
% \inptp{\tilde{f}}{j/p}(\xbj,\T{b}{};\mu_{Q_0},Q_0^2) &{}= \int \diff[2]{\T{k}{}}{} 
% e^{-i \T{k}{} \cdot \T{b}{} } \inptp{f}{j/p}(\xbj,\T{k}{};\mu_{Q_0},Q_0^2) \no
% &{}= K_0\parz{m_{i,p,A} \Tsc{b}{} } A_{i/p}(\xbj;\mu_{Q_0}) + K_0\parz{m_{i,p,B} \Tsc{b}{}} \ln \parz{\frac{Q_0^2 \Tsc{b}{}}{2 m_{i,p,B} e^{-\gamma_E}}} B_{i/p}(\xbj;\mu_{Q_0}) \no
% &{} \; + K_0\parz{m_{g,p} \Tsc{b}{}} A_{i/p}^{g}(\xbj;\mu_{Q_0}) + C_{i/p} \,\np{\tilde{f}}{i/p}(\xbj,\Tsc{b}{};Q_0^2) \, ,
% \end{align}
% %%%%%%%%%%%%
%%%%%%%%%%%%%
\begin{align}
\label{e.coordspace_tmdpdfinput}
\inptp{\tilde{f}}{j/p}(\xbj,\T{b}{};\mu_{Q_0},Q_0^2) &{}= \int \diff[2]{\T{k}{}}{} 
e^{-i \T{k}{} \cdot \T{b}{} } \inptp{f}{j/p}(\xbj,\T{k}{};\mu_{Q_0},Q_0^2) \no
&{}= K_0\parz{m_{i,p} \Tsc{b}{} } A_{i/p}(\xbj;\mu_{Q_0}) + K_0\parz{m_{i,p} \Tsc{b}{}} \ln \parz{\frac{Q_0^2 \Tsc{b}{}}{2 m_{i,p} e^{-\gamma_E}}} B_{i/p}(\xbj;\mu_{Q_0}) \no
&{} \; + K_0\parz{m_{g,p} \Tsc{b}{}} A_{i/p}^{g}(\xbj;\mu_{Q_0}) + C_{i/p} \,\np{\tilde{f}}{i/p}(\xbj,\Tsc{b}{};Q_0^2) \, ,
\end{align}
%%%%%%%%%%%%
with 
%%%%%%%%%%%%%
\begin{align}
\np{\tilde{f}}{i/p}^{\text{Gauss}}(\xbj,\Tsc{b}{};Q_0^2)={}&
e^{-\Tscsq{b}{} M_\text{F}^2/4} \, , \no
\np{\tilde{f}}{i/p}^{\text{Spectator}}(\xbj,\Tsc{b}{};Q_0^2)={}& \frac{1}{4}\parz{\frac{ (m_q+x M_p)^2 - L^2}{2 (m_q+x M_p)^2 + L^2}} \parz{L\Tsc{b}{}}^3 K_3(L \Tsc{b}{}) + \frac{3}{2}\parz{\frac{ L^2}{2 (m_q+x M_p)^2 + L^2}} \parz{L\Tsc{b}{}}^2 K_2(L \Tsc{b}{}) \, ,
\end{align}
%%%%%%%%%%%%
where  $K_0$, $K_2$ and $K_3$ are modified Bessel functions of the second kind.
% To keep the expressions simple, in \eref{coordspace_tmdpdfinput} we have set $m_{i,p,L} = m_{i,p,B}$. In phenomenological applications below, we will make this parameter assumption. 

As anticipated, \eref{coordspace_tmdpdfinput} matches the $\order{\alpha_s}$ operator product expansion (OPE) for small transverse sizes $\Tsc{b}{} \to 0$, up to errors suppressed by powers of $\Tsc{b}{}$. Since $K_0(m b_T) = - \ln{(m b_T/2e^{-\gamma_E})} + \mathcal{O}(b_T^2 m^2)$ and $\np{\tilde{f}}{i/p} = 1 + \order{b_T^a}$ for $a > 0$ (independently of which core parametrization is considered), the small-$b_T$ behavior of \eref{coordspace_tmdpdfinput} is
%%%%%%%%%%%%
\begin{align}
&\tilde{f}_{j/p}(\xbj,\T{b}{};\mu_{Q_0},Q_0^2) \to f_{i/p}(x;\mu_{Q_0}) - \parz{A_{i/p}(\xbj;\mu_{Q_0}) + A_{i/p}^{g}(\xbj;\mu_{Q_0})} \ln \parz{\frac{\Tsc{b}{} \mu_{Q_0}}{2 e^{-\gamma_E}}} \no 
&{} - B_{i/p}(\xbj;\mu_{Q_0}) \left[ \ln^2 \parz{\frac{\Tsc{b}{} \mu_{Q_0}}{2 e^{-\gamma_E}}} + \ln \parz{\frac{\Tsc{b}{} \mu_{Q_0}}{2e^{-\gamma_E}}} \ln \parz{\frac{Q_0^2}{\mu_{Q_0}^2}} \right] \no
&{} + \left\{ \sum_{i'} \delta_{i'i} [ \mathcal{C}_{\Delta}^{i/i'} \otimes f_{i'/p} ](x;\mu_{Q_0}) + [ \mathcal{C}_{\Delta}^{i/g} \otimes f_{g/p} ](x;\mu_{Q_0}) \right\} + \order{\Tscsq{b}{} m^2} \, ,
\end{align}
%%%%%%%%%%%%
where, crucially, all the dependence on the masses $m_{i,p}$ has been cancelled by the logarithms appearing in the expression for the coefficient $C_{i/p}$, defined in \eref{C_def_pdf}.

%%%-----------------------------
\subsection{The CS kernel: input scale parametrization}
\label{s.CSinput}
Next, we discuss the parametrization that we will use for the CS kernel through $\order{\alpha_s(\mu)}$. The analogous $\order{\alpha_s(\mu)^2}$ expressions are also straightforward to write down, but we will not use them for applications in this paper, so we include them in \aref{higherorder} for use in future work.
The perturbative CS kernel in coordinate space is 
%%%%%%%%%%%%%
\begin{align}
\label{e.pert_K}
\tilde{K}(\Tsc{b}{}; \mu)   
={}& -\frac{2 C_F \alpha_s(\mu)}{\pi} \ln \left( \frac{\Tsc{b}{} \mu}{2 e^{-\gamma_E}} \right)   + O(\alpha_s(\mu)^2) \, ,
\end{align}
%%%%%%%%%%%%%
with anomalous dimension 
%%%%%%%%%%%%%
\begin{align}
  \gamma_K(\alpha_s(\mu)) ={}&   
    \frac{2 C_F \alpha_s}{\pi}
    + \order{\alpha_s(\mu)^2} \, .
\end{align}
%%%%%%%%%%%%%
The HSO approach requires the renormalization group equation to be exactly satisfied through the working order of $\alpha_s(\mu)$ over the full range of $0 < \Tsc{b}{}< \infty$ (See Sec.~IV, Eq.(41) of \cite{Gonzalez-Hernandez:2022ifv}).  In our case, we work at $\order{\alpha_s(\mu)}$ so we need
%%%%%%%%%%%%%
\begin{align}
&{}\frac{\diff{}{}}{\diff{\ln \mu}{}} \tilde{K}(\Tsc{b}{};\mu) = - \gamma_K(\alpha_s(\mu)) = -\frac{2 C_F \alpha_s(\mu)}{\pi} + \order{\alpha_s(\mu)^2} \label{e.K_RG}\, .
\end{align}
%%%%%%%%%%%%%
The input transverse momentum space CS kernel is 
%%%%%%%%%%%%%
\begin{align}
\inpt{K}(\Tsc{k}{};\mu_{Q_0}) = 
 A_K^{(1)}(\mu_{Q_0}) \frac{1}{\Tscsq{k}{} + m_K^2} 
 + K_\text{core}(\Tsc{k}{}) + D_K(\mu_{Q_0}) \delta^{(2)}\parz{\T{k}{}} \, , \label{e.final_mom_K}
\end{align}
%%%%%%%%%%%%%
where 
%%%%%%%%%%%%%
\begin{align}
A_K^{(1)}(\mu_{Q_0}) = \frac{\alpha_s(\mu_{Q_0}) C_F}{\pi^2} \, .
\end{align}
%%%%%%%%%%%%%
The function $K_\text{core}(\Tsc{k}{})$ is analogous to 
$\np{f}{i/p}(\xbj,\Tsc{k}{};Q_0^2)$. It is used to describe the very large $\Tsc{b}{}$ behavior, and it is required to vanished like a power at small $\Tsc{b}{}$. It will generally introduce at least one extra parameter beyond $m_K$. In coordinate space, we will demand that $\tilde{K}(\Tsc{b}{};\mu_{Q_0})$ approach a negative constant $b_K$ (up to perturbative corrections) as $\Tsc{b}{} \to \infty$ (see the discussion in Sec.~VII of Ref.~\cite{Collins:2014jpa}). We will use a Gaussian for the core, 
%%%%%%%%%%%%%
\begin{equation}
\label{e.CSmodel}
K_\text{core}(\Tsc{k}{}) = \frac{b_K}{4 \pi m_K^2} e^{-\frac{\Tscsq{k}{}}{4 m_K^2}} \, .
\end{equation}
%%%%%%%%%%%%%
The last term in \eref{final_mom_K} has a $D_K(\mu_{Q_0})$, which is
%%%%%%%%%%%%%
\begin{align}
D_K(\mu_{Q_0}) &{}= - b_K + \frac{2 \alpha_s(\mu_{Q_0}) C_F}{\pi} \ln \parz{\frac{m_K}{\mu_{Q_0}}} \, .
\end{align}
%%%%%%%%%%%%%
Transforming \eref{final_mom_K} into coordinate space gives
%%%%%%%%%%%%%
\begin{align}
\label{e.final_coord_K2}
\inpt{\tilde{K}}(\Tsc{b}{};\mu_{Q_0}) = 2 \pi A_K^{(1)}(\mu_{Q_0}) K_0\parz{m_K \Tsc{b}{}} + b_K e^{-m_K^2 \Tscsq{b}{}} + D_K(\mu_{Q_0}) \, .
\end{align}
%%%%%%%%%%%%%
It is straightforward to verify that \eref{final_coord_K2} 
equals \eref{pert_K} when $m_k \Tsc{b}{} \to 0$. Using
%%%%%%%%%%%%%
\begin{equation}
\label{e.A_K_RG}
\frac{\diff{A_K^{(1)}(\mu_{Q_0})}{}}{\diff{\ln \mu_{Q_0}}{}} = \order{\alpha_s(\mu_{Q_0})^2} \, 
\end{equation}
%%%%%%%%%%%%%
also confirms that it is consistent with \eref{K_RG} for all $\Tsc{b}{}$.

The large-$\Tsc{b}{}$ limit of the CS kernel in \eref{final_coord_K2} is 
%%%%%%%%%%%%%
\begin{equation}
\label{e.CS_limit}
\lim_{\Tsc{b}{} \to \infty} \inpt{\tilde{K}}(\Tsc{b}{};\mu_{Q_0}) = D_{K}(\mu_{Q_0}) \, . 
\end{equation}
%%%%%%%%%%%%%
The equations above that relate objects like $\inpt{\tilde{K}}(\Tsc{b}{};\mu_{Q_0})$ and $\gamma_K(\alpha_s(\mu))$ become exact if the $\order{\alpha_s(\mu_{Q_0})^2}$ corrections are systematically dropped everywhere. It is noteworthy that the general behavior of this parametrization is consistent with trends in recent lattice QCD calculations -- see, for example, Fig.~7 of \cite{Shanahan:2021tst}.

%%%%%%%%%%%%%%%%%%%%%%%%%%%%%%%%%%%%%%%
\subsection{The full TMD pdf parametrizations at the input scale}
\label{s.fullinptpara}

In calculations at an input Drell-Yan scale of $Q = Q_0$, the appropriate auxiliary scales are $\mu = \sqrt{\zeta} = Q_0$, 
and the above parametrizations are perturbatively well-behaved 
at $\Tsc{b}{} \approx 1/Q_0$, but they give ultraviolet divergent logarithms when $\Tsc{b}{} \to 0$.  That is outside the region that is  physically probed when $Q \approx Q_0$, but it means the parametrizations are inadequate for evolving to $Q \gg Q_0$ where sensitivity to $\Tsc{b}{} \ll 1/Q_0$ grows. To fix this, we evolve $\inptp{f}{i/p}(\xbj,\Tsc{k}{};\mu_{Q_0},Q_0^2)$ and $\inpt{\tilde{K}}(\Tsc{b}{};\mu_{Q_0})$ again to a scale that smoothly transforms from $Q_0$ to $\sim 1/\Tsc{b}{}$. The functional form we will use for the transformation is 
%%%%%%%%%%%%%
\begin{align}
\label{e.qbar_param_a}
&\overline{Q}_0(\Tsc{b}{},a) = Q_0 \left[ 1 - \parz{1 - \frac{C_1}{Q_0 \Tsc{b}{}}}
e^{-\Tscsq{b}{} a^2} \right] \, ,\quad C_1 = 2e^{-\gamma_E}\approx 1.123,
\end{align}
%%%%%%%%%%%%%
see Appendix C of~\cite{Gonzalez-Hernandez:2022ifv},  
where $a$ is a parameter that controls exactly where in $\Tsc{b}{}$ the switch from $Q_0$ to $2 e^{-\gamma_E}/\Tsc{b}{}$ takes place. The result is simply a scheme transformation in a perturbatively controlled region, so any functional form is equally valid, with only $\order{\alpha_s(\mu_{Q_0})^2}$ sensitivity to the exact choice. In a general $\order{\alpha_s(\mu)^n}$ treatment, sensitivity to parameters like $a$ only arise at $\order{\alpha_s(\mu_{Q_0})^{n+1}}$.

On the surface, it is tempting to view the use of a $\overline{Q}_0(\Tsc{b}{},a)$ as nearly identical to the $\bstarsc$ prescription, and $Q_0$ as exactly a version of $1/\bmax$. However, this is not quite the case. To understand this, it is important to note that there are two separate transition scales involved in $\inptp{\tilde{f}}{i/p}(\xbj,\Tsc{b}{};\mu_{Q_0},Q_0^2)$. First, there is a transition between perturbative $1/\Tsc{b}{}$ scales comparable to $Q_0$ or larger, where one expects at least reasonable agreement with low order perturbation theory calculations at a fixed renormalization scale of $Q_0$. Second, there is a transition to the region of very small $\Tsc{b}{}$ where the additional renormalization group improved $\mu \sim 1/\Tsc{b}{}$ treatment that is needed to account for the $\Tsc{b}{} Q_0 \to 0$ limit.  In the usual way of implementing the $\bstarsc$ procedure, one is forced to treat both of these as exactly the same transition, and both are controlled by the numerical value of a $\bmax$, or other parameters in $\bstar(\Tsc{b}{})$. In other words, if $1/\Tsc{b}{}$ is small enough that it ceases to be a useful renormalization group scale, then it is taken to be the case that one has entered a purely nonperturbative regime. The procedure excludes the option of simply switching to a fixed scale above some $\Tsc{b}{}$ while continue to use perturbation theory. If $\bmax$ is taken to be large, then perturbation theory with a running renormalization scale $\mu \sim b_0/\Tsc{b}{}$ must be assumed to be valid even down to very low values. If $\bmax$ is taken to be small, then one is in danger of grouping perturbatively calculable behavior with the ``nonperturbative'' functions.

The HSO approach allows one to separate the treatment of these two transitions. First, the transition between purely perturbative and nonperturbative regions is controlled by the nonperturbative parameters in a model, like the $m_q$, $M_F$, $m_K$, etc above. Separately, the transition between a fixed $\mu = Q_0$ and the small-$Q_0 \Tsc{b}{}$ RG-improved scale is controlled by $\overline{Q}_0(\Tsc{b}{},a)$. The latter transition is just an RG scheme change, so its effect is a higher order contribution that can then be minimized in the HSO approach by including higher order calculations. The former transition, however, involves the actual physically meaningful description of the transition between a generally perturbative tail region and a purely nonperturbative region, so it requires nonperturbative models or calculations. Thus, the HSO approach allows one to model or parametrize the first transition while still exploiting RG scheme independence to deal with the latter transition. In the standard $\bstarsc$ approach, the roles of model parameters like $m_q$, $M_F$, $m_K$, etc and the roles of scheme change functions like $Q_0 \Tsc{b}{}$ are all collapsed into the parameters of $\bstar(\T{b}{})$.

 In our calculations, we will fix the value of $a$ to the input scale $Q_0$, but we have verified that the sensitivity to $a$ is negligible in that it has no impact either on the final fit results or the physical cross section in the TMD region, even at higher energies. See section VII A of Ref.~\cite{Gonzalez-Hernandez:2022ifv} for a related discussion. 

With this step complete, the final TMD pdf parametrizations, which we will substitute into 
\eref{Fuuevolved}, are 
%%%%%%%%%%%%%
\begin{widetext}
\begin{align}
&\tilde{f}_{i/p}(\xbj,\Tsc{b}{};\mu_{Q_0},Q_0^2) \no
&= \inptp{\tilde{f}}{i/p}(\xbj,\Tsc{b}{};\mu_{\overline{Q}_0},\overline{Q}_0^2)
\exp \left\{
\int_{\mu_{\overline{Q}_0}}^{\mu_{Q_0}} \frac{d \mu^\prime}{\mu^\prime} \left[\gamma(\alpha_s(\mu^\prime);1) 
- \ln \left(\frac{Q_0}{\mu^\prime} \right)\gamma_K(\alpha_s(\mu^\prime))
  \right] +\ln \left(\frac{Q_0}{\overline{Q}_0}\right) \tilde{K}(\Tsc{b}{};\mu_{\overline{Q}_0}) \right\} \, ,
  \label{e.evolvedd3p}
\end{align} 
\end{widetext}
%%%%%%%%%%%%%
with
%%%%%%%%%%%%%
\begin{equation}
\inptp{\tilde{f}}{i/p}(\xbj,\Tsc{b}{};\mu_{Q_0},Q_0^2)=\int d^2{\bf k}_T e^{-i{\bf k}_T\cdot {\bf b}_T} {f}_{\text{inpt},i/p}(\xbj,\Tsc{k}{};\mu_{Q_0},Q_0^2) \, ,
\end{equation}
%%%%%%%%%%%%%
and
%%%%%%%%%%%%%
\begin{equation}
\label{e.evol_paramb}
\tilde{K}(\Tsc{b}{};\mu_{Q_0}) = \inpt{\tilde{K}}(\Tsc{b}{};\mu_{\overline{Q}_0}) -
\int_{\mu_{\overline{Q}_0}}^{\mu_{Q_0}} \frac{\diff{\mu'}}{\mu'}  \gamma_K(\alpha_s(\mu')) \, ,
\end{equation}
%%%%%%%%%%%%%
and with the 
$\order{\alpha_s}$ evolution kernels. 
If the evolution kernels are expressed through order $\order{\alpha_s(\mu)}$ and $\overline{Q}_0$ is held fixed, then the $\mu_{Q_0}$ and $Q_0$ evolution in \eref{evol_paramb} is exact -- there is no $\order{\alpha_s(\mu_{Q_0})^2}$ error term. A similar statement applies to the evolution in \eref{evolvedd3p}. 

The full $\Tsc{k}{}$-space parametrization follows from Fourier-Bessel transforming \eref{evol_paramb} into $\Tsc{k}{}$ space. Evolving requires the kernel $\gamma(\alpha_s(\mu);1)$ in $\msbar$ renormalization which is 
%%%%%%%%%%%%%
\begin{align}
\gamma(\alpha_s(\mu);1)    
&{}= \frac{3 C_F \alpha_s(\mu)}{2 \pi} + \order{\alpha_s(\mu)^2} \, . \label{e.gammaexp}
\end{align}
%%%%%%%%%%%%%
Notice that, having fixed the parametrizations \eref{evolvedd3p} and \eref{evol_paramb} at the input scale, it becomes almost trivial to evolve the structure functions like \eref{Fuuevolved} to higher $Q$. This is the starting point for our implementation of phenomenological applications. Although this presentation adheres very closely to a TMD parton model picture, it has a somewhat different surface appearance relative to many typical CSS and related treatments as they are applied in high energy applications. However, the translation between the two ways of organizing TMD factorization is quite straightforward -- see \aref{translation} for a review of the steps and for a set of formulas that translate the above organization of expressions into the familiar ``$g$-functions'' of past implementations. 
%%%%%%%%%%%%%%%%%%%%%%%%%%%%%%%%%%%%%%%%%%%%%%%%%%%%
%%%%%%%%%%%%%%%%%%%%%%%%%%%%%%%%%%%%%%%%%%%%%%%%%%%%
\section{Fitting moderate $Q$ Drell-Yan Measurements}
\label{s.fittingmodQ}

In our first confrontation with data, we start by considering Drell-Yan scattering measurements at moderate hard scales.
We focus on the E288~\cite{Ito:1980ev} and E605~\cite{Moreno:1990sf} experiments and perform independent fits. In each case, we use {\tt Minuit2}~\cite{Hatlo:2005cj} to minimize the quantity
%%%%%%%%%%%%%%%%%%%%%%
\begin{align}
    \chi^2 =&  
    \frac{(1-N)^2}{\delta_N^2}
    +
    \sum_i \frac{(d_i- t_i/N)^2}{\sigma_i^2} \,,
\end{align}
%%%%%%%%%%%%%%%%%%%%%%
where $d_i, \sigma_i$ are the data points and their (uncorrelated) uncertainties, $t_i$ is the corresponding theory calculation, and $N$ is an overall normalization, a nuisance parameter, common to all the points in a given experiment.  The first term is the usual penalty related to $N$(see for instance \cite{DAgostini:1993arp,Barlow:2017xlo,Borsa:2023zxk}), which depends on the normalization uncertainty $\delta_N$ reported in each experiment. For all of our fits, we will calculate cross sections in the TMD approximation, neglecting for now contributions of the so-called $Y$ term. Thus, we will consistently impose the kinematical cut
%%%%%%%%%%%%%%%%%%%%%%
\begin{align}
\label{e.kincuts}
    \Tsc{q}{} \leq 0.2\,Q\, ,
\end{align}
%%%%%%%%%%%%%%%%%%%%%%
which is typical of TMD analyses.
We focus our attention to the region  away from the  $\Upsilon$ resonances, so we exclude the data bins with $9\,\text{GeV}<Q<11\,\text{GeV}$ for E288, and $9.5\,\text{GeV}<Q<10.5\,\text{GeV}$ for E605.

Tables reporting the minimal values of our model parameters will be presented in the following sections, as well as the reduced $\chi^2$
%%%%%%%%%%%%%%%%%%%%%%
\begin{align}
\label{e.chi2dof}
\chi^2_{\text{dof}}=&\chi^2/(n-p-1)\,,
\end{align}
%%%%%%%%%%%%%%%%%%%%%%
with $n$ the number of fitted data points and $p$ the number of free model parameters. The -1 in the denominator of~\eref{chi2dof} accounts for the estimation of the nuisance parameter $N$.  We will refrain from displaying correlation matrices for the parameters, as they do not play a central role in our discussions. Instead, uncertainty bands in the Hessian approximation will be shown in comparisons to data (see, for instance, \cite{Pumplin:2001ct}). For this, we determine the $p$ independent ``directions" in parameter space that diagonalize the Hessian matrix to obtain $p$ eigensets, and compute asymmetric errors as in Eqs.~(10-12) of Ref.~\cite{Buckley:2014ana}. In all fits, we will have $p=3$, corresponding to 3 eigensets. Furthermore, we choose $\Delta \chi^2=3.53$ which corresponds to a $1\sigma$ confidence region for varying 3 parameters, under the usual regularity conditions of Wilks' theorem~\cite{Wilks:1938dza}. For the \msbar~collinear pdfs, we use the NLO extraction {\tt MMHT2014} of Ref.~\cite{Harland-Lang:2014zoa}, accessible through {\tt LHAPDF6}~\cite{Buckley:2014ana}.    

%%%%%%%%%%%%%%%%%%%%%%%%%%%%%%%%%%%%%%%%%%%%%%%%% 
 \subsection{Fixed-target data sets}
\label{e.fixedtarData}
 
The E288 experiment~\cite{Ito:1980ev} measures final-state muon pairs for the scattering of protons off a fixed heavy target. The relevant observable is the $\Tsc{q}{}$-dependent cross section in \eref{finalcross}, integrated over intervals of $Q$. Data were taken for three different values of the beam energy, $E_{\text{beam}}=200, 300, 400\, \text{GeV}$. The experimental collaboration provides the following information regarding kinematic variables:
%%%%%%%%%%%%%%%%%%%%
\begin{align}
    q_{\text T}:&\qquad q_{\text T}^{\text {min}},\quad q_{\text T}^{\text {max}},\quad\langle q_{\text T}\rangle=\frac{1}{2}\left(q_{\text T}^{\text {min}}+q_{\text T}^{\text {max}}\right),\quad
    \Delta q_{\text{T}}=0.2\,\text{GeV}
    \nonumber\\
    \label{e.E288binning}
    y_{\text{h}}:&\qquad
    \langle y_{\text{h}}\rangle\quad\text{(for each $E_{\text{beam}}$)}\\
    Q:&\qquad 
    Q_{\text {min}},\quad Q_{\text {max}}\,.
    \nonumber
\end{align}
%%%%%%%%%%%%%%%%%%%%
For this observable, we evaluate \eref{finalcross} at the experimental average values of the dimuon rapidity  $\langle\yh\rangle$ 
and $\langle\Tsc{q}{}\rangle$,
and compute
only the integral over $Q^2$
%%%%%%%%%%%%%%%%%%%%
\begin{align}
\label{e.E288Obs}    
{\cal O}_{\text{E288}}=&
    \frac{1}{\pi}
\int dQ^2\frac{d^3\sigma}{d q_T^2 \,d\yh\,dQ^2}\Bigg|_{
\substack{
\yh=\langle\yh\rangle\\
\Tsc{q}{}=\langle\Tsc{q}{}\rangle
}
}\,,
\end{align}
%%%%%%%%%%%%%%%%%%%%
where the factor of $1/\pi$ comes from averaging over the azimuthal angle of the dimuon's momentum.  
The E605 \cite{Moreno:1990sf} experiment performs the same measurement with only a few differences: i) $E_{\text{beam}}=800\,\text{GeV}$, ii) large-$Q$ bins are wider with improved statistics, iii) instead of $\langle\yh\rangle$, 
 there is one bin in $\xf$
%%%%%%%%%%%%%%%%%%%%
\begin{align}
    \xf:&\qquad 
    {\xf}_{\text {min}}=-0.1\,,\quad {\xf}_{\text {max}}=0.2\,,\quad
    \langle\xf\rangle=0.1 \, ,
    \nonumber
\end{align}
%%%%%%%%%%%%%%%%%%%%
but otherwise, the same information as in \eref{E288binning} is provided.
For the E605 experiment we compute
%%%%%%%%%%%%%%%%%%%%
\begin{align}
\label{e.E605Obs}
{\cal O}_{\text{E605}}=&
\frac{1}{\pi}
\int dQ^2 \frac{d^3\sigma}{dq_{\text T}^2 \,d y_{\text{h}}\,dQ^2}\Bigg|_{
\substack{
\yh=\overline{\yh}\\
\Tsc{q}{}=\langle\Tsc{q}{}\rangle
}}\,,
\qquad
\text{with}
\quad
\overline{\yh}
=
\arcsinh
\left(
\frac{\sqrt{s}\sqrt{Q^2+\langle\Tsc{q}{}\rangle^2}}{2 Q^2}\,\langle\xf\rangle
\right)\,.
\end{align}
%%%%%%%%%%%%%%%%%%%%
For our purposes, it is sufficient to work within the approximations of \eref{E288Obs} and \eref{E605Obs}. Future refinements will include the explicit calculation of over-the-bin averages by integrating numerically over each bin of $\yh(\xf)$ and $\Tsc{q}{}$. 

Both experiments provide the cross section per nucleon, so we have to consider this in our calculations. We use a simplified model of the relationship between nuclear and proton TMD pdfs for a target with atomic number $Z$ and total nucleon number $A$
%%%%%%%%%%%%%%%%%%%%
\begin{align}
\label{e.tmdtarget}
f_{i/t}=&
\frac{Z}{A}\,f_{i/p}
+
\frac{A-Z}{A}\,f_{i/n}\,,
\end{align}
%%%%%%%%%%%%%%%%%%%%
where the neutron TMD $f_{i/n}$ is related to $f_{i/p}$ by isospin symmetry,
as it is usually done (see for instance \cite{Bacchetta:2017gcc,Moos:2023yfa}). 
The simple treatment of \eref{tmdtarget} is a useful point of departure for future refinements.

For both data sets we 
will start by using $Z=29, A=63$ for a copper target. Note, however, that the E288 experiment also uses a platinum target, but the proportion of different nuclei, or its effect on the observables, is not clear\footnote{One might consider different scenarios with either copper, platinum or both targets and perform tests as we propose here. However, for this article we assume a copper target.}.

%%%%%%%%%%%%%%%%%%%%%%%%%%%%%%%%%%%%%%%%
\subsection{Gaussian fits} 
\label{s.gauss}
For the nonperturbative description of very small transverse momentum, we start with 
the Gaussian models of \eref{npmodels} 
and set the nonperturbative masses to 
%%%%%%%%%%%%%%%%%%%%
\begin{align}
\label{e.masseschoices1}
\mspec \to M_0 + M_1\log(1/\xbj)\,,
\qquad
m_{i,p,A}=m_{i,p,B}=m_{i,p,L}=m_{g,p}=0.3\,\text{GeV}\,,
\end{align}
%%%%%%%%%%%%%%%%%%%%
where $M_0,M_1$ are two free parameters of the fit. Our choice of logarithmic dependence on $\xbj$ is typical of some early phenomenological analyses within the CSS formalism. (see, for instance, Ref.~\cite{Landry:1999an}.) 
For the nonperturbative behavior of the CS kernel we use~\eref{CSmodel}. There, we set the mass parameter to $m_K=0.3\,\text{GeV}$  and fit only $b_K$.  In total we have 3 free parameters and one additional nuisance normalization for each fit.
In keeping with the recipe from Sect.~VI of 
\cite{Gonzalez-Hernandez:2022ifv}, we compute  the $Q_0=4\,\text{GeV}$ input scale cross section using the functions from \eref{candidateqpdf} and \eref{final_mom_K}, with perturbative parts at $O(\alpha_s)$,
and refrain from implementing the RG improvements of \sref{fullinptpara} until a later stage when we evolve to larger $Q$.
Best-fit values of the parameters and $\chi^2$ are reported in \tref{paragauss}. Comparisons to fitted data are presented in \fref{E288_p-cu_1norm} and  \fref{E605_p-cu}, where both central lines and uncertainty bands are shown. 
% We have confirmed that switching from input to RG improved TMD pdfs is phenomenologically insignificant close to the input scale, by computing the corresponding $\chi^2$ with RG improvements and the same parameter values of \tref{paragauss}. Variation of the $\chi^2$ is about 0.15\,\%.
We have confirmed that switching from input to RG improved TMD pdfs is phenomenologically insignificant close to the input scale, by 
% computing
refitting
the $\chi^2$ including RG improvements. Differences in the minimal  $\chi^2$ is about 0.15\,\% and parameter values are unaffected.
We also checked that refitting with RG improvements, but increasing $a$ in the scale transformation of \eref{qbar_param_a} by a factor of 2, the effect on the minimal $\chi^2$ appears only in the fourth digit. See also Fig. 8 of \cite{Gonzalez-Hernandez:2022ifv}.

%%%%%%%%%%%%%%%%%%%%%
\begin{table}[h!]
\centering
\begin{tabular}{ c c c }
\multicolumn{3}{c}{Gaussian fits}\\
\hline
        &  E288 (130 pts.)     & E605 (52 pts.)  \\
\hline
$\chi^2_{\text{dof}}$ & 1.04 & 1.68\\
$M_0$ (GeV)  &    0.0576 & 0.404 \\
$M_1$ (GeV)  &    0.403  & 0.290 \\
$b_K$   &    2.12   & 0.744 \\
$N$(nuisance)   &    1.29  & 1.28 
\end{tabular}
%%%%%%%%%%%%%%%%%%%%%
\caption{Minimal parameters obtained by fitting E288 and E605 data independently, using the models of \eref{npmodels} and \eref{CSmodel}. Parameters are correlated, but we do not show correlation matrices. Uncertainties are  calculated by varying parameters along the ``plus'' and ``minus'' directions of the 3 eigensets in each case.}
\label{t.paragauss}
\end{table}
%%%%%%%%%%%%%%%%%%%%%%%%%%%%%%%%%%%%%%%%
\begin{figure}
    \centering
    \includegraphics[scale=0.51]{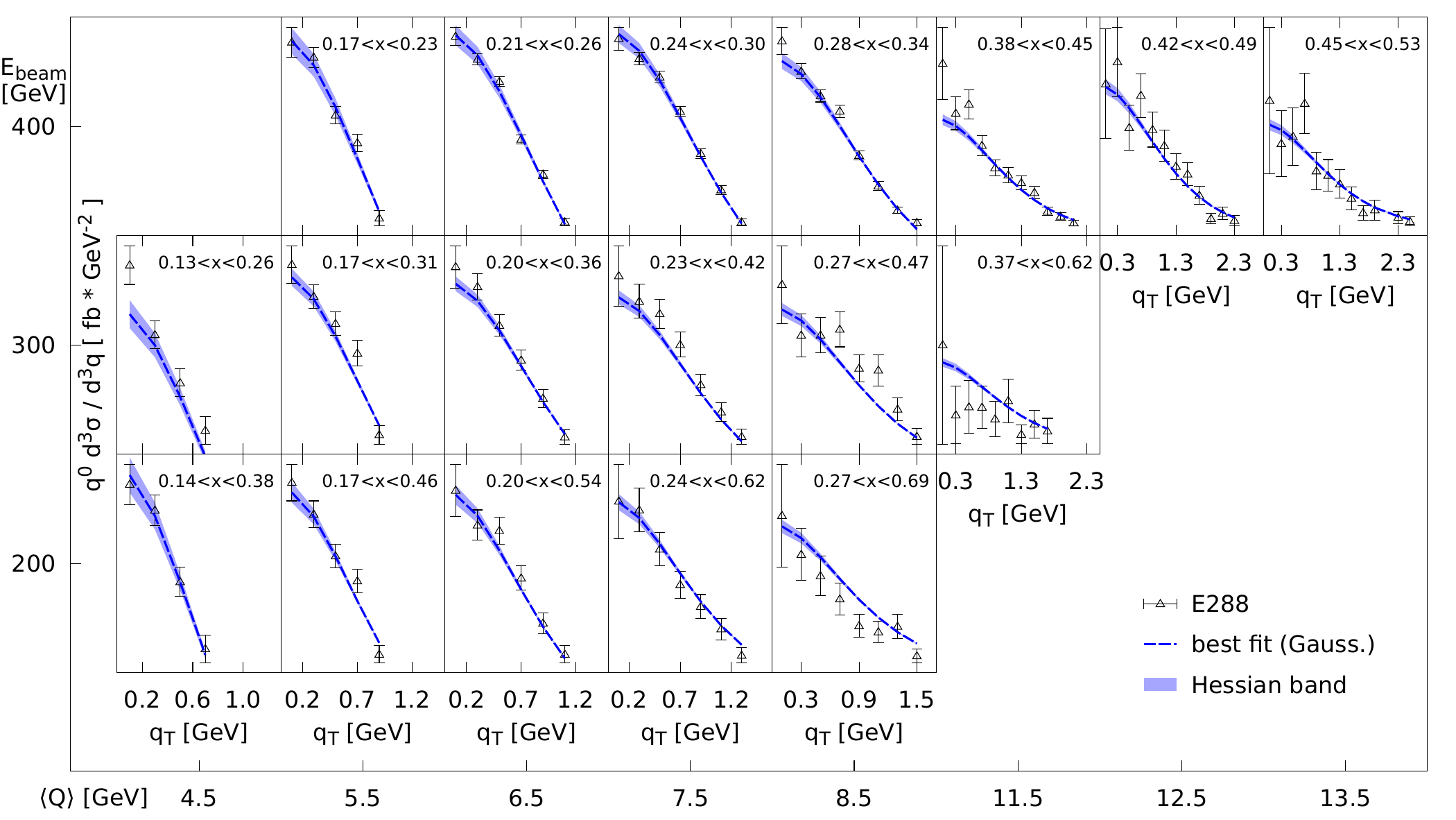}
    \caption{Comparison of data from the E288 Drell-Yan experiment \cite{Ito:1980ev} to the best-fit central lines using the HSO approach with the Gaussian core model of \eref{npmodels}. 
    Vertical ranges are different for each panel, so tick marks are not shown. The vertical scale is adjusted in each plot for better visibility.
    The nonperturbative 
    $\xbj$-dependence is parametrized by setting $M_F\to M_0+M_1 \log(1/\xbj)$ where both $M_0, M_1$ are free fit parameters. We have fixed the small masses in \eref{candidateqpdf} to  $m_{i,p,A}=m_{i,p,B}=m_{i,p,L}=m_{g,p}=0.3\,\text{GeV}$. The nonperturbative model for the large-$\Tsc{b}{}$ CS-kernel is that of \eref{CSmodel}, for which we fix $m_K=0.3\,\text{GeV}$ and leave $b_K$ free. The perturbative coefficients of Eqs.~(\ref{e.Hardfactor}), (\ref{e.candidateqpdf}) and (\ref{e.final_mom_K}) are calculated through $O(\alpha_s)$. Uncertainty bands are calculated with 3 eigensets and with $\Delta\chi^2=3.53$, as explained in the text.
    Both theory central lines and bands are multiplied by the corresponding minimal value for the nuisance parameter. For the central line this is $N=1.29$.
    }
    \label{f.E288_p-cu_1norm}
\end{figure}
%%%%%%%%%%%%%%%%%%%%%%%%%%%%%%%%%%%%%%%%
\begin{figure}[ht]
    \centering
    \includegraphics[scale=0.49]{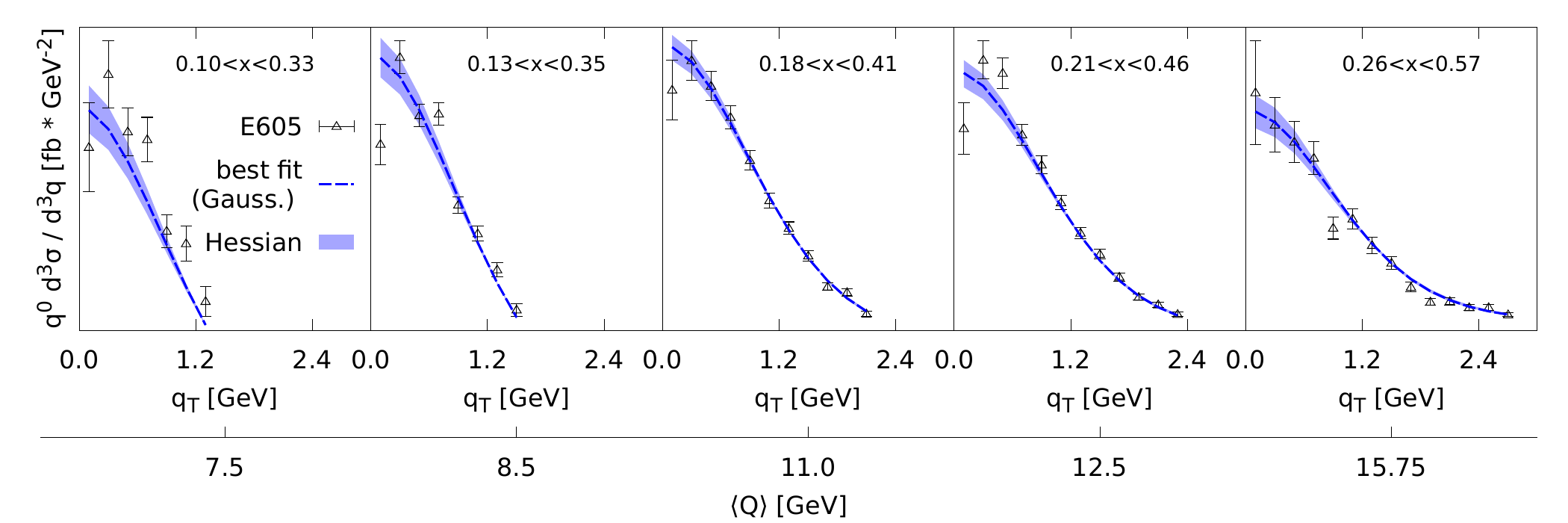} \caption{Comparison of data from the E605 experiment \cite{Moreno:1990sf} to best-fit central lines and hessian bands using the HSO approach with a Gaussian core model. Vertical ranges are different for each panel, so tick marks are not shown. The vertical scale is adjusted in each plot for better visibility. The model assumptions and calculation of bands are the same as in \fref{E288_p-cu_1norm}. 
    Both theory central lines and bands are multiplied by the corresponding minimal value for the nuisance parameter. For the central line this is $N=1.28$.
    }
    \label{f.E605_p-cu}
\end{figure}
\subsection{Spectator model fit} 
%%%%%%%%%%%%%%%%%%%%%%%%%%%%%%%%%%%%%%%%
We now turn to the model of \eref{spectatorpdf} for the TMD pdf core function. In contrast to the Gaussian case, this model implies its own explicit $\xbj$-dependence. In principle, \eref{spectatorpdf} depends on 3 mass parameters,  $m_q,\Lambda$ and $M_X$. But to make a more direct comparison to the Gaussian case, i.e. by keeping the same number of parameters, we set the ``quark'' mass to $m_q=0$, and leave $\Lambda$ and $M_X$ free in our fit. For the spectator model case, we present the fit for the E288 set only, since we find that the E605 data alone are not sufficient to constrain both the CS kernel and the TMD pdf. Apart from the use of the spectator model, all of our choices are the same as in the Gaussian case, namely, we use \eref{CSmodel} for the CS kernel with fixed $m_K=0.3\,\text{GeV}$ and with all other nonperturbative masses in \eref{candidateqpdf} also set to $m=0.3\,\text{GeV}$. Results are shown in \tref{paraspect}. We note that the minimal $\chi^2_{\text{dof}}$ is the same as in the Gaussian case to three significant figures. Although we do not show comparison to the fitted data, results are essentially identical as in the Gaussian  case, \fref{E288_p-cu_1norm}. Finally, using the parameter values of \tref{paraspect}, we have checked that RG improvements are phenomenologically irrelevant, as for the Gaussian case. 
% \tr{The effect on $\chi^2$ of varying the parameter $a$ by a factor of 2 appears only in the fifth digit. See also Fig. 8 of Ref.~\cite{Gonzalez-Hernandez:2022ifv}. }
This time, the variation of the minimal $\chi^2$ is about 0.26\,\%.
%%%%%%%%%%%%%%%%%%%%%
\begin{table}[h!]
\centering
\begin{tabular}{ c c }
\multicolumn{2}{c}{Spectator model fit}\\
\hline
    &  E288 (130 pts.)  \\
\hline
$\chi^2_{\text{dof}}$ & 1.04 \\
$\Lambda$ (GeV)  &    0.801 \\
$M_X$  (GeV) &    0.438   \\
$b_K$   &    1.90    \\
$N$(nuisance)  &  1.23  
\end{tabular}
\caption{Minimal parameters obtained by fitting E288 data with the models of \eref{spectatorpdf} and \eref{CSmodel}. Parameters are correlated, but we do not show correlation matrices. Uncertainties are  calculated by varying parameters along the ``plus" and ``minus" directions of the 3 eigensets.}
\label{t.paraspect}
\end{table}
%%%%%%%%%%%%%%%%%%%%%

%%%%%%%%%%%%%%%%%%%%%%%%%%%%%%%%%%%%%%%%%%%
\subsection{Results for TMD pdfs}

The behavior of the TMD pdfs determined by our fit to E288 experimental data are shown in~\fref{tmdbands}. Here we only show results from the Gaussian model \eref{npmodels} and postpone comparisons to the spectator model until~\sref{comparison}. 
The use of the HSO approach has guaranteed that the TMD pdf of \eref{candidateqpdf} (without RG improvements) asymptotes to the perturbative tail in~\eref{pertf} at the input scale.
This feature is preserved after implementing the RG improvements of \eref{evolvedd3p}, as seen in the different panels of~\fref{tmdbands} (blue lines). Upon evolution to larger scales, such agreement is improved for smaller values of  $x$ (top panels), as evidenced by the general trend of the TMD lines when compared to the perturbative tail (dot-dashed lines) and, in particular, the relative position of their nodes. Recall that the perturbative tails are determined entirely within collinear factorization for $\Tscsq{k}{} \approx Q^2$, while the full TMD pdfs involve evolution from the input scale, and as such the effects of the CS kernel play a role in their profile. Therefore, the observed agreement after evolution is not trivial. 
Note that for $Q=91\,\text{GeV}$, in the top panels of~\fref{tmdbands}, the solid lines closely trace the behavior of the tail. At larger values of $x$ (bottom panels), differences between the TMD pdfs and the perturbative tail are more visible, although still in reasonable agreement. Improvements to the parametrization are certainly possible, e.g. by carefully tuning the parameter $a$ in the scale transformation of \eref{qbar_param_a} or by including higher orders in $\alpha_s$, but we leave this for upcoming work. 
We stress that keeping track of how closely the extracted TMD merges with the large-$\Tsc{k}{}$ region is an important step in phenomenology.  
For instance, it can assist in preventing the parametrizations from becoming excessively flexible.
%%%%%%%%%%%%tmds gaussian
\begin{figure}[!h]
    \centering
    \includegraphics[scale=0.1667]{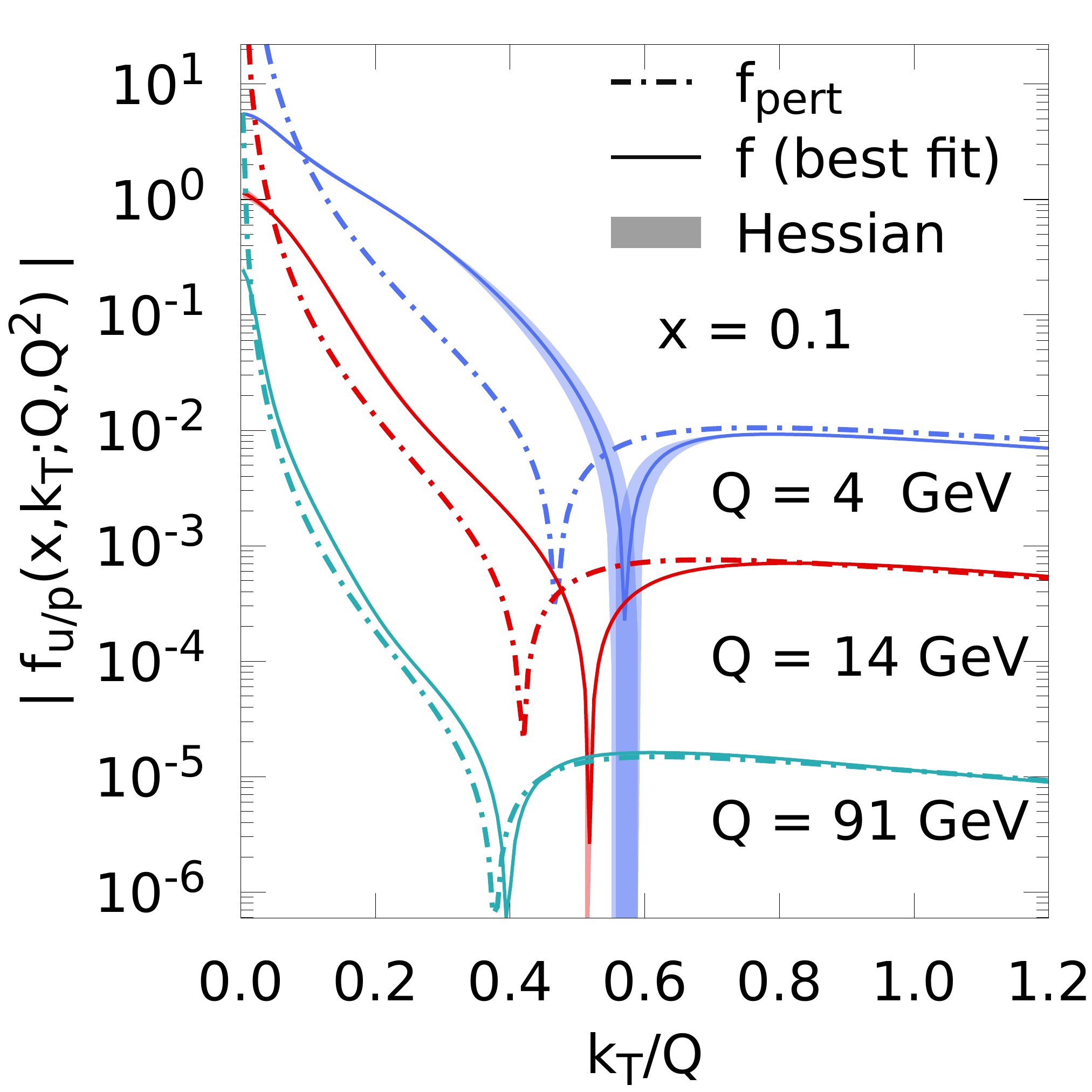}
    \includegraphics[scale=0.1667]{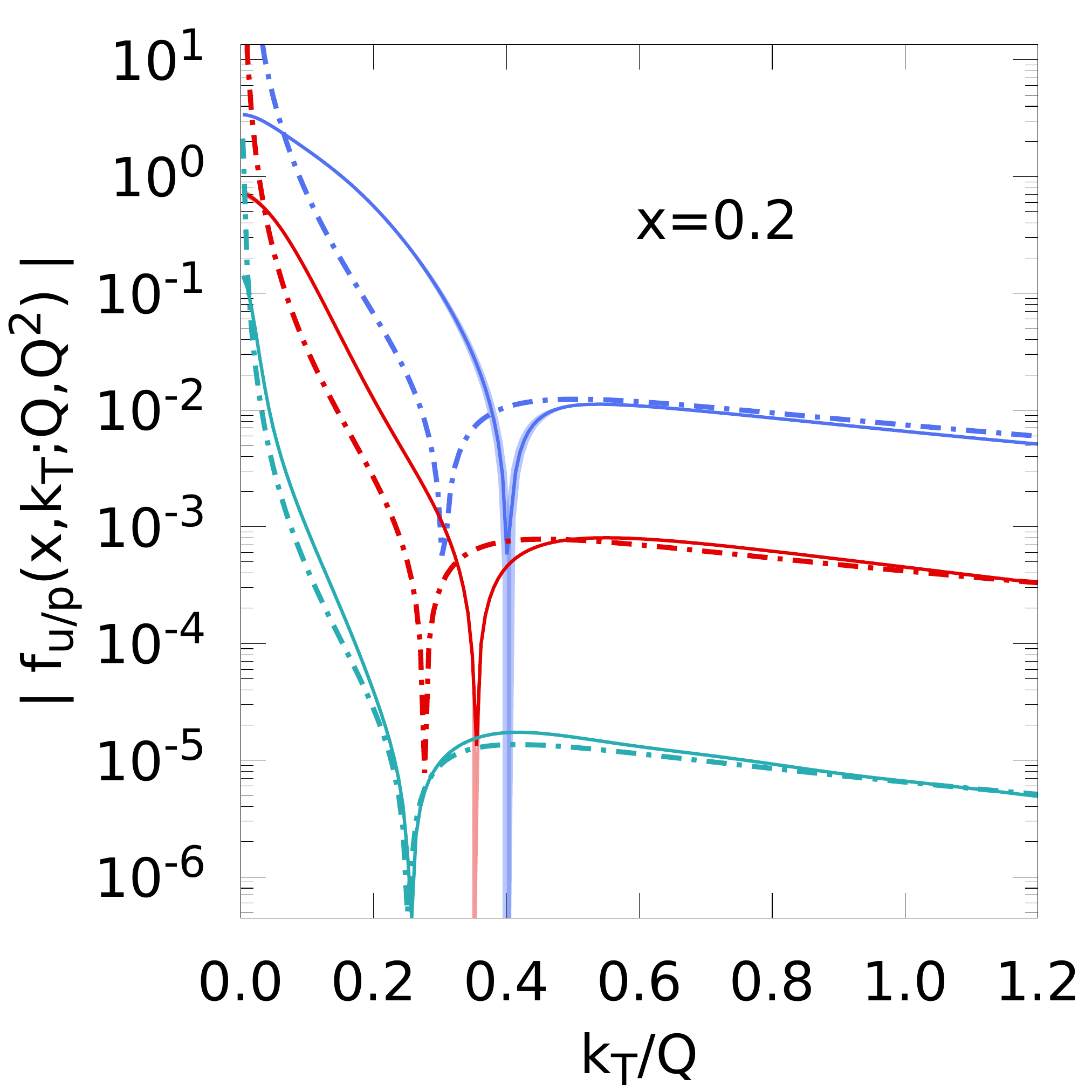}
    \includegraphics[scale=0.1667]{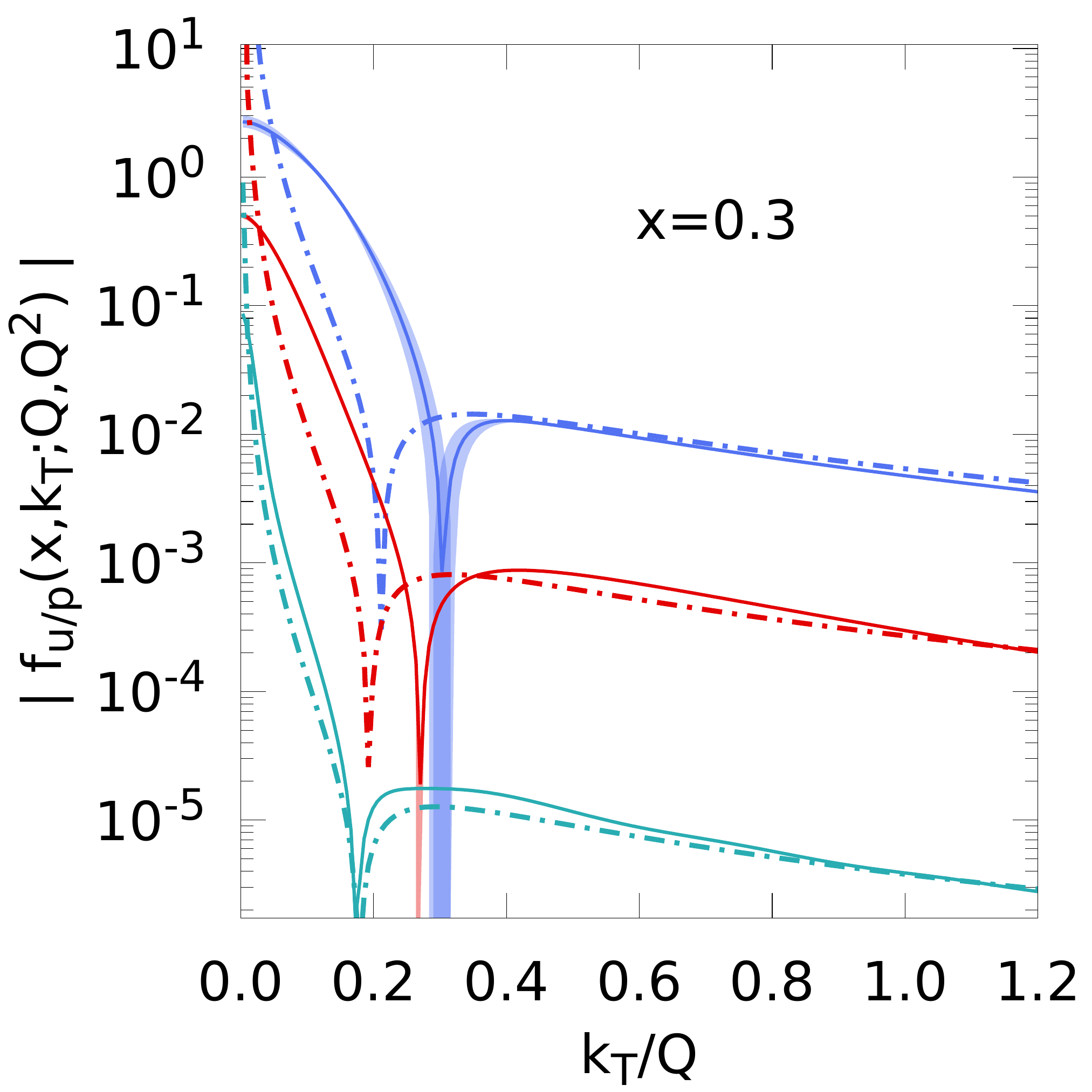}
    \includegraphics[scale=0.1667]{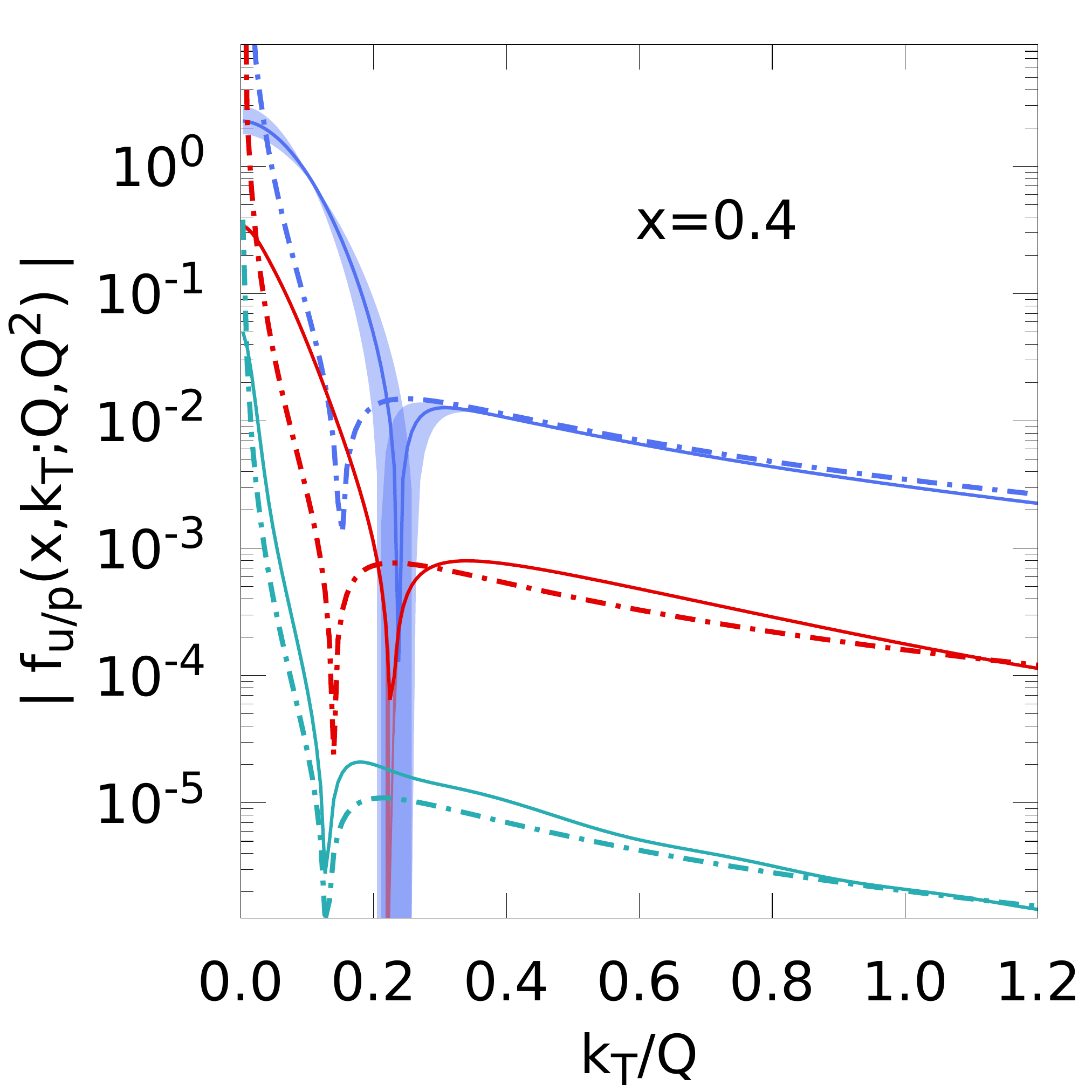}
    \includegraphics[scale=0.1667]{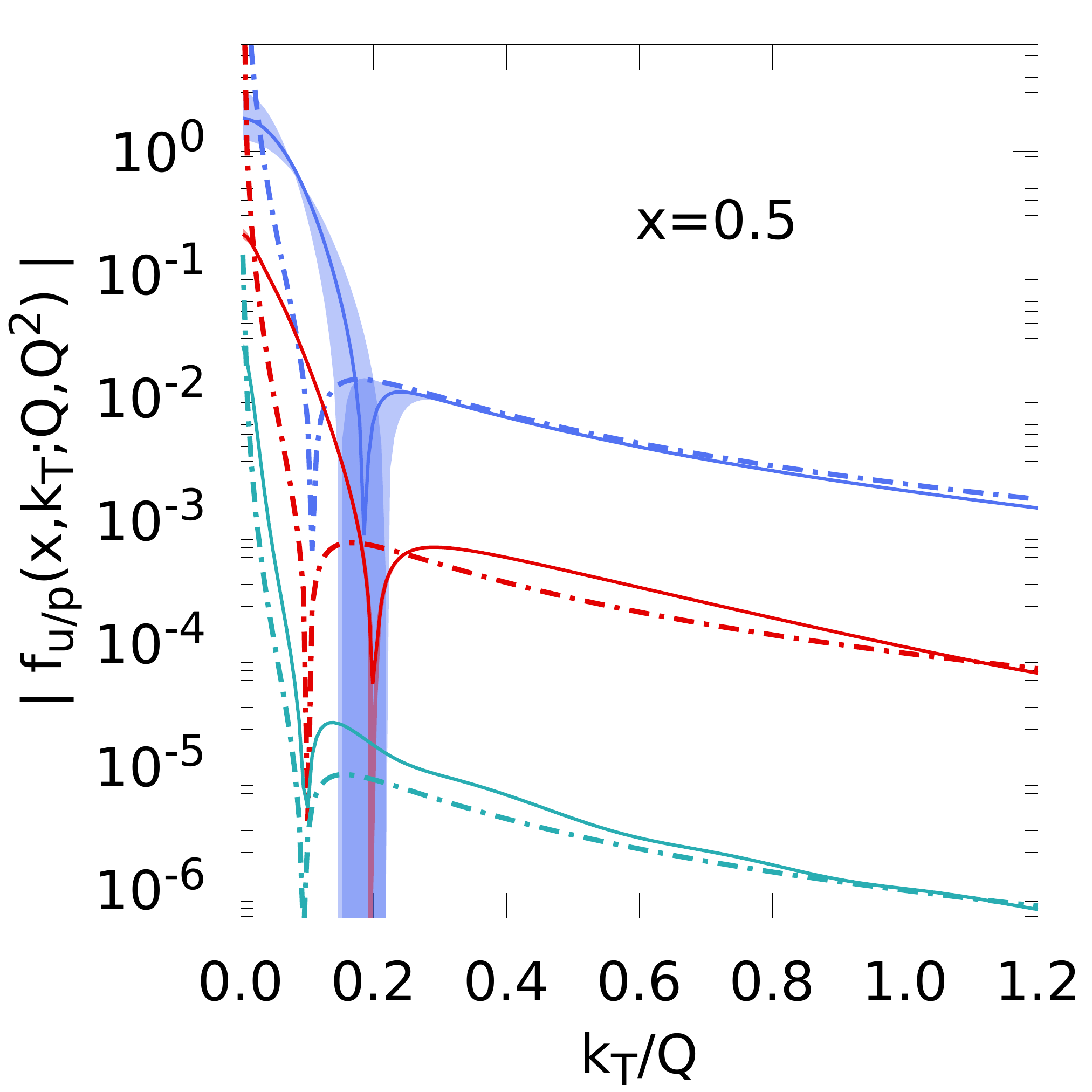}
    \includegraphics[scale=0.1667]{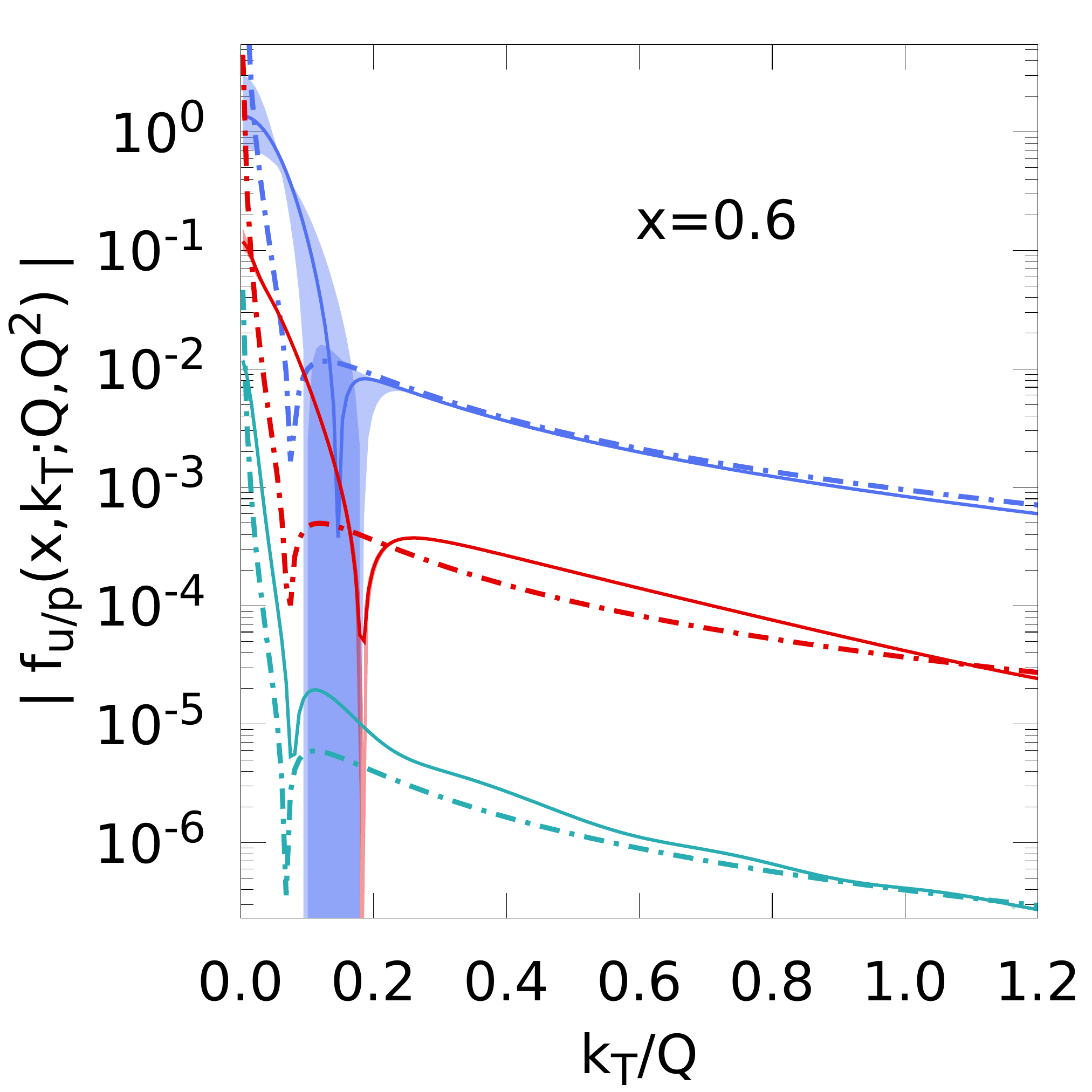}
    \caption{TMD pdfs obtained from fitting E288 data, with the Gaussian model of \eref{npmodels}, including RG improvements of \eref{evolvedd3p}. 
    The panels show values of $\xbj$ in the region of the data sets. We show central values (solid lines) and Hessian uncertainty bands, as described in~\sref{fittingmodQ}, for three different values of the hard scale, $Q=Q_0=4\,\text{GeV}$ (blue), $Q=14\,\text{GeV}$ (red), and $Q=91\,\text{GeV}$ (green). The TMD pdfs are compared to the perturbative tail of~\eref{pertf} (dot-dash lines).}
    \label{f.tmdbands}
\end{figure}
%%%%%%%%%%%%%%%%%%%%%%%%%%%

%%%%%%%%%%%%%%%%%%%%%%%%%%%%%%%%%%%%%%%%%%%
\section{Testing predictions at larger Q}
\label{s.test-large-Q}

Finally, we test the predictive power of the fits in the previous section. 
The steps are i) to extract the nonperturbative behavior of TMD pdfs and the CS kernel at the moderate energies above where sensitivity to nonperturbative effects is large (this step was completed in the previous section) and then ii) to evolve these extracted TMDs to higher energies and compare (postdict) 
higher energy data from the D0 and CDF experiments. This will test whether our
assumptions about the initial input parametrizations 
reasonably postdict 
experimental observations without any prior consideration of the final predicted data. 
In addition, by considering two different sources of moderate $Q$ data, one may examine how accurately and precisely the input assumptions postdict higher energy data \emph{independently} of the initially fitted data.
Our goal with this section is to illustrate how the predictive power of TMD factorization is brought to the surface within the HSO approach to phenomenological implementations. 
It is a somewhat different philosophy from many traditional global fitting frameworks, where there is generally no analogous postdiction stage.  
Note that statistical techniques such as cross-validation usually treat all data in the same footing, 
while the above emphasizes the special role of observables at low-to-moderate scales. 

Specifically, we compute theory curves for $Z^0 \to e^+e^-$ observables measured by the CDF I~\cite{CDF:1999bpw} and D0 I~\cite{D0:1999jba} collaborations. In order to evolve to larger $Q$, close to the $Z^0$ boson mass, we implement  RG improvements as discussed in~\sref{fullinptpara}, following the recipe from Sect.~VI of 
\cite{Gonzalez-Hernandez:2022ifv}. For the scale transformation of \eref{qbar_param_a}, we set a value of $a=Q_0=4\,\text{GeV}$. Both the CDF and D0 data sets are singly differential in $\Tsc{q}{}$, so one  must integrate over the kinematically allowed range of $\yh$. This region maps values of $\xbj$ as small as $10^{-3}$. Since the fixed-target data  used in our fits only cover the region  down to  $\xbj\approx 0.1$, in computing high energy observables we must 
extrapolate our TMD pdf model into unconstrained $\xbj$-kinematics. In the case of the Gaussian fits, we note that for $\xbj\approx 10^{-3}$, best fit values for $M_0$ and $M_1$ imply values for the Gaussian mass $M_F \approx 3\,\text{GeV}$. This would result in a bad agreement between  the  TMD pdf and its perturbative tail, close to the input scale $Q_0=4\,\text{GeV}$. To prevent this, we require that $M_F$ remains smaller than a typical nonperturbative mass, which we take to be the mass of the proton. Thus we require that $M_F<M_p$ in the $\xbj$-extrapolation region. This theory-motivated choice is  one of the assumptions  to be tested in our comparisons to CDF and D0 data. For the case of the spectator model, we keep the mass parameters constant in $\xbj$. 

In \fref{CDF_D0_comparison_1}, we compare the calculation obtained from the Gaussian E288 (left panels) and E605 (right panels) fits to the CDF~I (blue points) and D0~I (red points) sets. 
% The theory calculation only includes the contribution from the $Z^0\to e^+e^-$ channel to the differential cross section. 
Despite the overlapping kinematics, \fref{CDF_D0_comparison_1} suggests
that calculations using extractions from E288 better predict the $Z^0$-production data than extractions from E605, although in both cases the descriptions are qualitatively reasonable in the region $3\,\text{GeV}<\Tsc{q}{}<18\,\text{GeV}$.
Note that 
% while the theory bands obtained from the E605 fit (right panels) overlap with only a few points at small transverse momentum, 
the overall trend of the data-theory ratios is consistent over a wider range of $\Tsc{q}{}$ for the E288 fit (left panels), within an overall normalization error of 
% $\sim25\%$ 
$\sim10\%$
for CDF I data. We achieve a specially good postdiction of the D0~I data with the E288 fit.
% and $\sim15\%$ for D0 I data.  
In all cases, agreement between theory and data sharply deteriorates around $\Tsc{q}{} \approx 18\,\text{GeV}$, consistently with the $\Tsc{q}{}$ cut in~\eref{kincuts}.
%
%%%%%%%%%%%%%%%%%%%%%%%%%%%%%%%%%%%%%%%%%%%
\begin{figure}[t]
    \centering
    \includegraphics[scale=0.33]{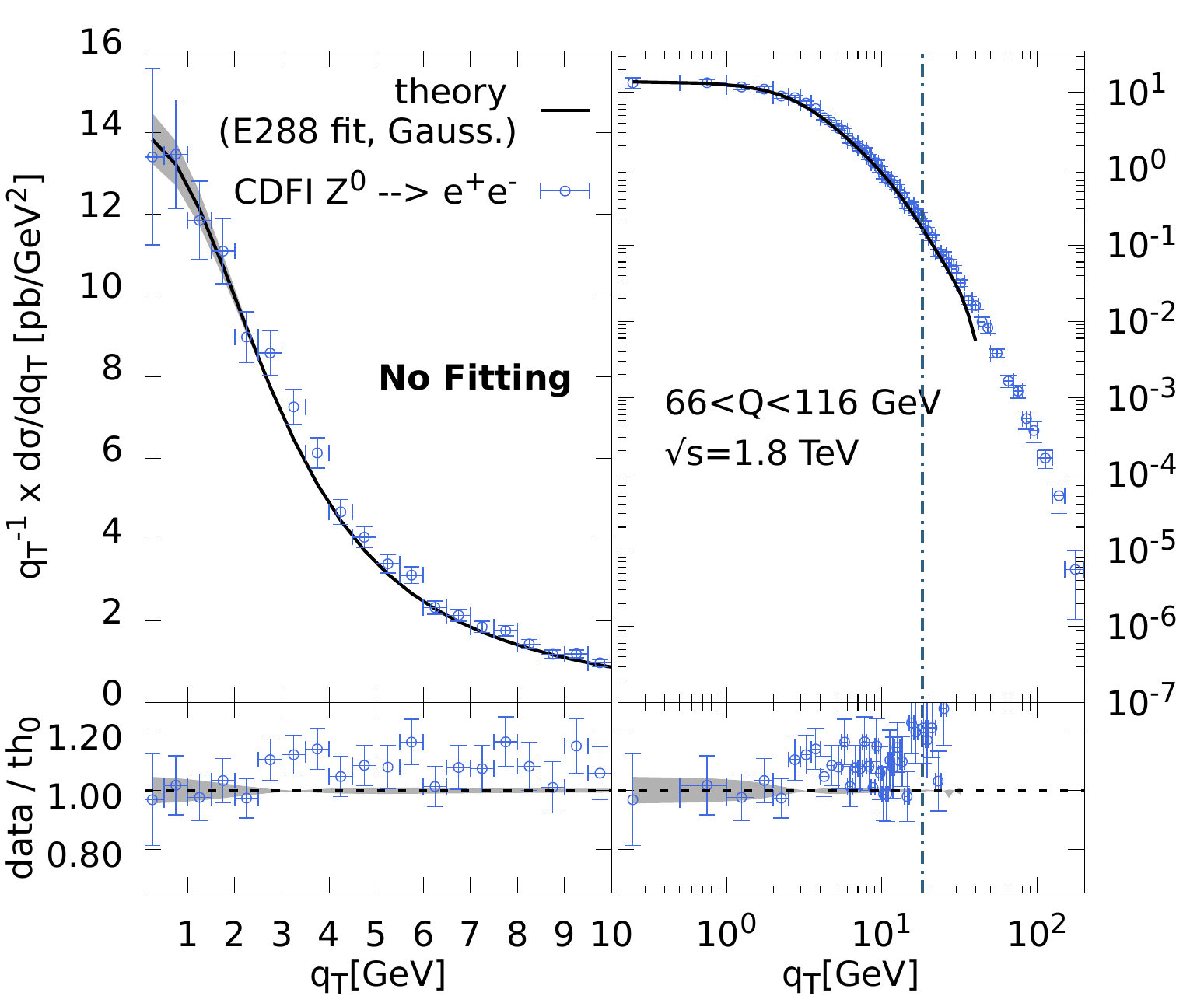}
    \includegraphics[scale=0.33]{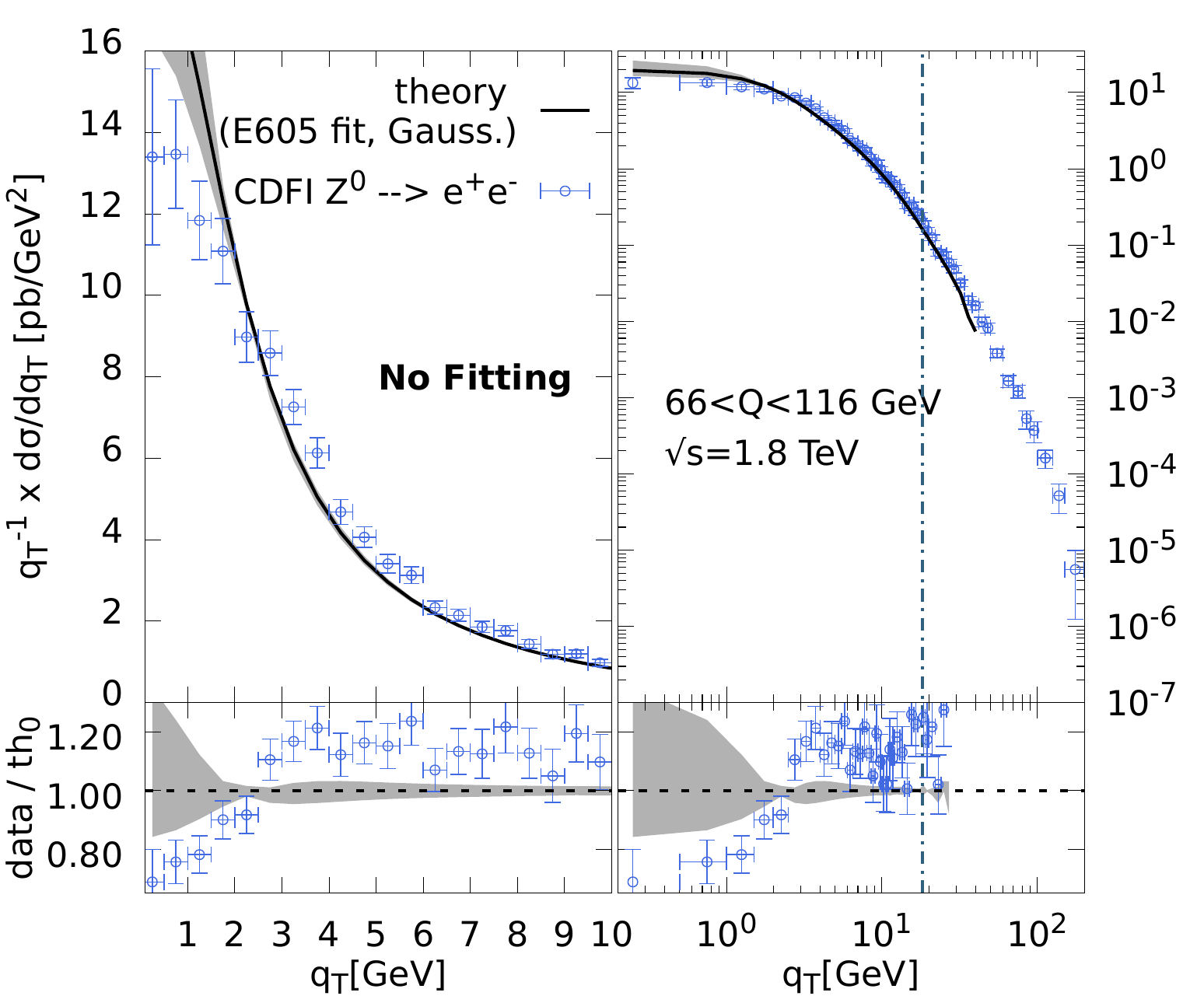}
    \includegraphics[scale=0.33]{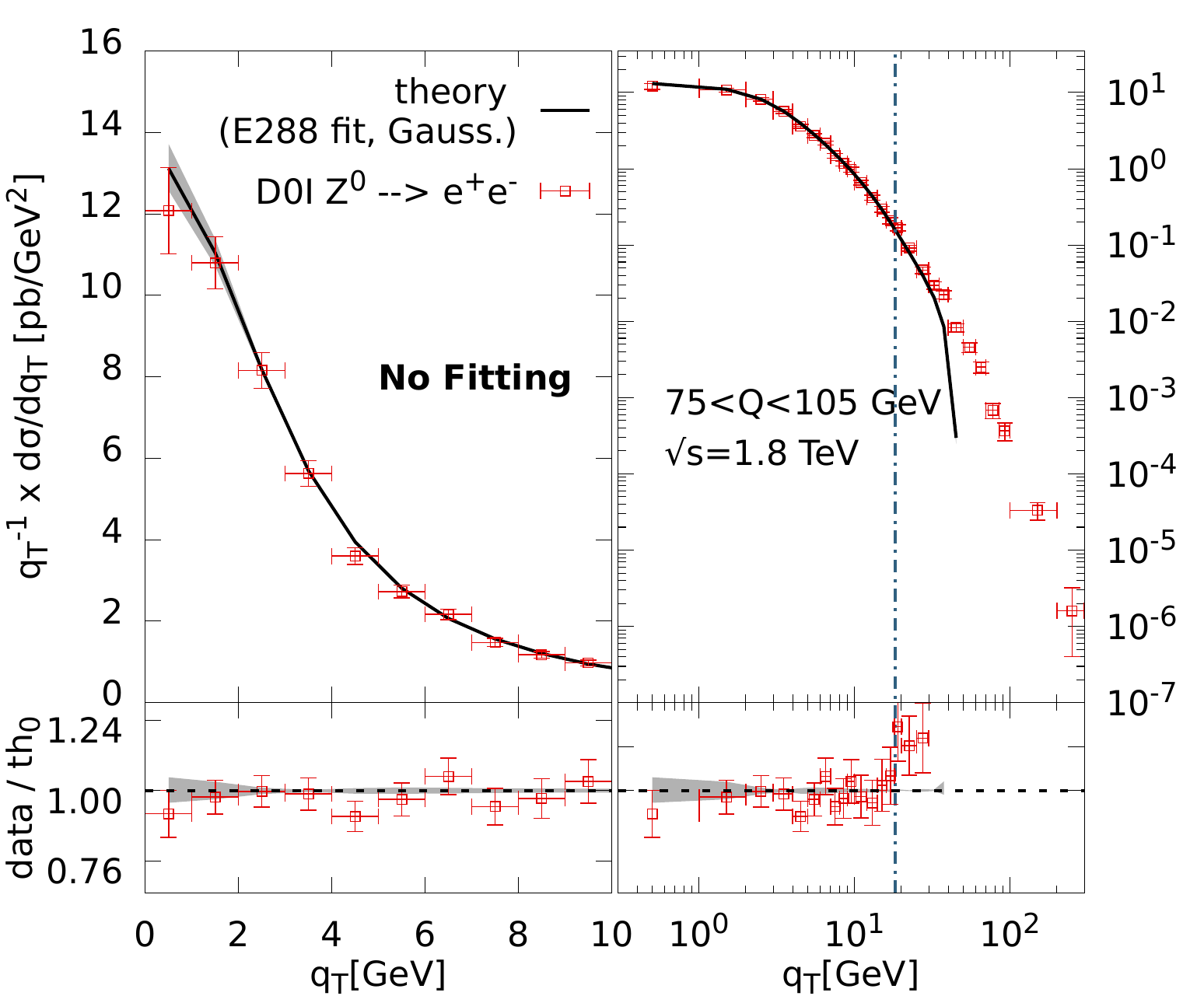}
    \includegraphics[scale=0.33]{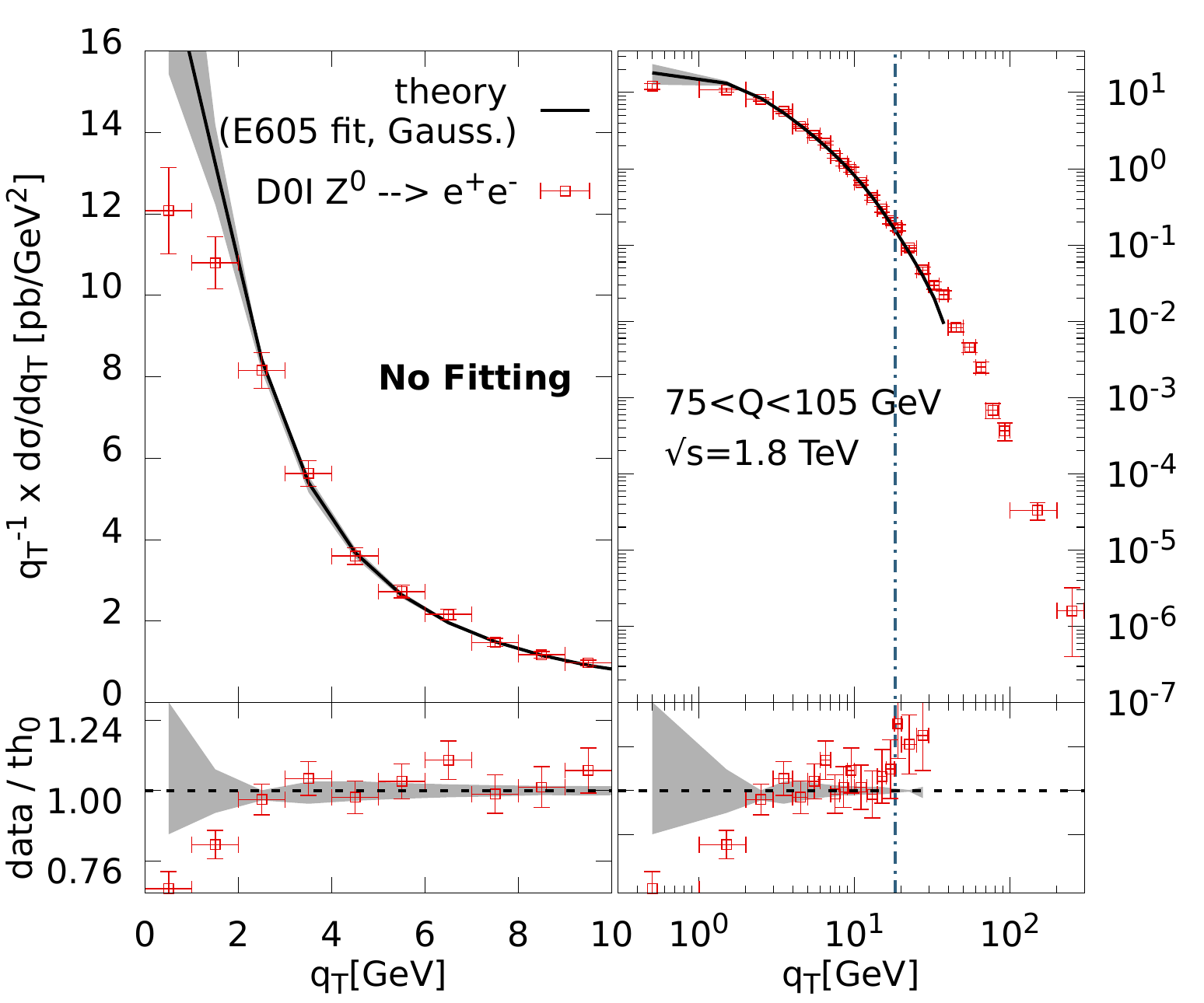}
    \caption{
    Testing the fits to E288 (left panels) and E605 (right panels) data. The  models of nonperturbative transverse momentum dependence, at an input scale $Q_0=4\,\text{GeV}$, are obtained by replacing the expressions in \eref{npmodels} and \eref{CSmodel} into \eref{candidateqpdf} and \eref{final_mom_K}, respectively. After RG improvements, the cross section is calculated at scales around the $Z_0$-boson mass and compared to CDF~I (top panels) and D0~I(bottom panels). Data errors do not include overall normalization uncertainty. 
    In each panel, the bottom of the plot shows the data and theory bands scaled by the central theory line. The vertical dot-dashed line indicates the value $\Tsc{q}{}=0.2 M_Z$.}
    \label{f.CDF_D0_comparison_1}
\end{figure}
%%%%%%%%%%%%%%%%%%%%%%%%%%%%%%%%%%%%%%%%%%%
%
%%%%%%%%%%%%comparison to spectator model fits
\begin{figure}[!h]
    \centering
    \includegraphics[scale=0.33]{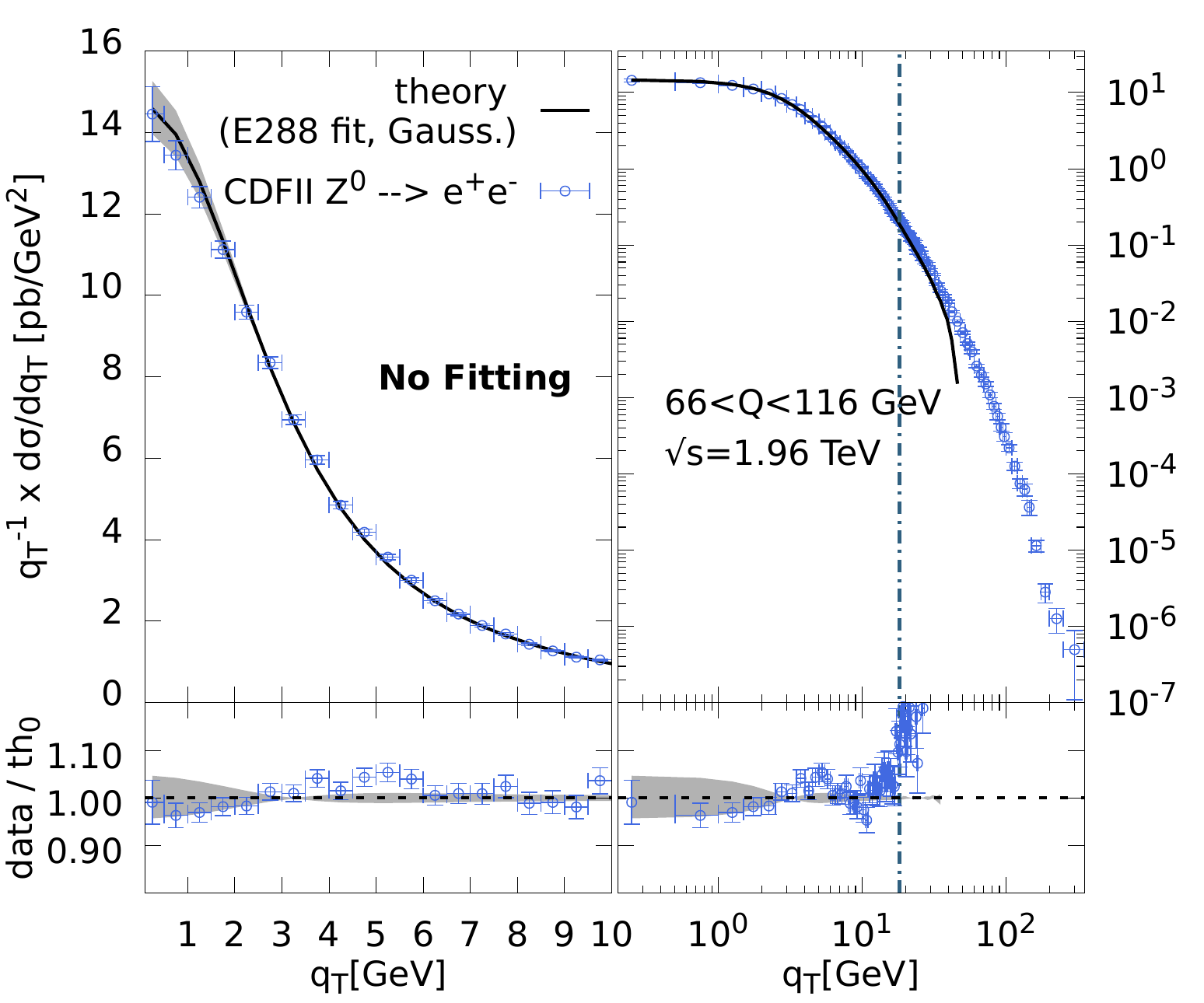}
    \includegraphics[scale=0.33]{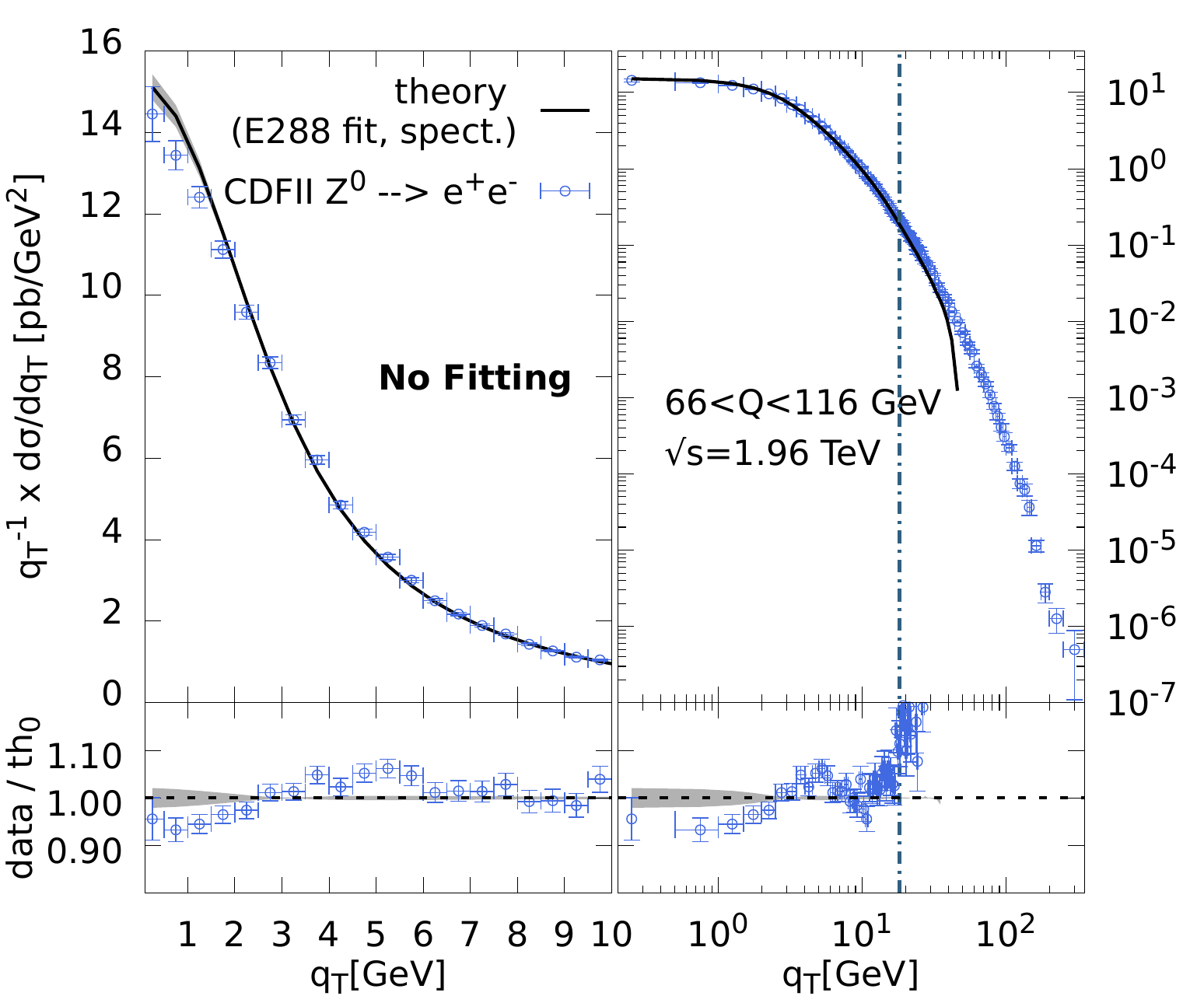}
    \caption{
    Comparison between fits using different models for the core function in \eref{candidateqpdf}. In each case, we fit only the E288 set. Settings for the evolution of the TMD pdfs are the same as in \fref{CDF_D0_comparison_1}. Theory calculations are compared now to the CDF~II data set. Systematic errors of the data are not shown. The vertical dot-dashed line indicates the value $\Tsc{q}{}=0.2 M_Z$.
    Left: Gaussian core function of \eref{npmodels}. Right: Spectator model of~\eref{spectatorpdf}.}
    \label{f.CDF_D0_comparison_2}
\end{figure}
%%%%%%%%%%%%%%%%%%%%%%%%%
%
%
%
%
It is perhaps more interesting to compare how different models perform at predicting $Z^0$-production data by comparing how the two different types of core parametrizations, Gaussian and spectator, considered in \sref{nonpertcore} perform. We use the E288 data to constrain the model parameters and compare with 
 the CDF II measurements
 % for $Z^0\to e^+e^-$ of
~\cite{CDF:2012brb}.
In~\fref{CDF_D0_comparison_2}, we compare how both models perform after evolution to scales $Q\approx M_Z$. 
Despite  significant qualitative differences in the functional forms of the core functions, both the Gaussian (left panel) and spectator (right panel) models seem to be 
% equally 
consistent with the CDF II data to a similar extent, as seen in the top panels.
After evolution to high $Q$, only the rough shape and order of magnitude, and not the details, of the nonperturbative modeling remain distinguishable.
Thus, any reasonable description of the low-to-moderate scale  data gives results similar those in~\fref{CDF_D0_comparison_2}, provided that fitted data constrains both the TMD pdf and the CS kernel sufficiently well. 

One may indeed attempt to discriminate  between models by quantifying the agreement of their postdiction, but this involves careful treatment of \emph{all} uncertainties, which is beyond the scope of this work. It is, however, instructive to consider a situation when this is done. Suppose we take as an indication the data/theory ratios in the bottom panels of \fref{CDF_D0_comparison_2}. They suggest that the Gaussian form performs better than the spectator model, so on the basis of the postdiction we would have been able to decide which extraction is more acceptable. Recall the values of the minimal $\chi^2$ for each fit were the same, and then it was not clear from the fit alone which version should be preferred. 

Contrast this with a more typical way of proceeding. Consider a case in which both the E288 and CDF~II data are fitted simultaneously, using the spectator model. In order to improve the description of the CDF II data, model parameters would be adjusted. But variations that are significant at large scales are in fact much more prominent at  lower scales. Then, a better fit to CDF~II would result in a poorer description of E288.  

In a typical global fitting strategy, 
one would increase the number of parameters in models of nonperturbative behavior until the desired agreement is achieved. But this may result in overfitting in the moderate-scale data, E288 in our example. 
While such a strategy might engineer better agreement with the large-scale data, it discards the constraints from moderate-scale fits where the main sensitivity to nonperturbative transverse momentum resides.
The cost is that the predictive power of nonpeturbative transverse momentum is largely lost. This type of problem worsens as one considers more data sets, especially when the only focus is to reduce the minimal global $\chi^2$. Note that in the two models that we tested the number of parameters is always $p=3$, both with a minimal value of $\chi^2_{\text{dof}}=1.04$, and this implies that no additional flexibility is necessary for describing the E288 measurement.

% %%%%%%%%%%%%%%%%%%%%%%%%%

\newpage

%%%%%%%%%%%%%%%%%%%%%%%%%%%%%%%%%%%%%%%%%%%%%%%%%%%%%
\section{The $g_K$ function and the nonperturbative CS kernel}
\label{s.gKfunction}

In this section we perform a further consistency check by translating our parametrization of the CS kernel into the conventional one that uses a $g_K$ function and a $\bmax$ (see \aref{translation}).
To order $\order{\alpha_s}$ the CS kernel parametrization is 
%%%%%%%%%%%%%%%%%
\begin{equation}
\tilde{K}\parz{\Tsc{b}{};\mu_{Q_0}} = \frac{2 \alpha_s(\mu_{\overline{Q}_0}) C_F}{\pi} \left[ K_0\parz{\Tsc{b}{} m_K} + \ln \parz{\frac{m_K}{\mu_{\overline{Q}_0}}} \right] - \int_{\mu_{\overline{Q}_0}}^{\mu_{Q_0}} \frac{\diff{\mu'}{}}{\mu'} \frac{2 C_F \alpha_s(\mu')}{\pi} - b_k \parz{1 - e^{-m_k^2 \Tscsq{b}{}}} \, . \label{e.Kforplots}
\end{equation}
%%%%%%%%%%%%%%%%%
(Although it is straightforward to extend \eref{Kforplots} to $\order{\alpha_s^2}$ using the expressions in \sref{CSinput} and \eref{evol_paramb}, we leave an implementation of this  
to future work.) 
The ``nonperturbative'' contribution\footnote{Note the scare quotes on ``nonperturbative.'' This is because $g_K(\Tsc{b}{};\bmax)$ has perturbatively calculable contributions, particularly at small $\Tsc{b}{}$.} to the CS kernel that is traditionally expressed as a ``$g_K$'' function is defined (see \aref{translation}) as
%%%%%%%%%%%%%%%%%
\begin{equation}
g_K(\Tsc{b}{};\bmax) = \tilde{K}\parz{\bstarsc(\Tsc{b}{});\mu_{Q_0}} - \tilde{K}\parz{\Tsc{b}{};\mu_{Q_0}} \, .
\label{e.gKforplots}
\end{equation}
%%%%%%%%%%%%%%%%%
While making a transformation to a $g_K$ function is unnecessary in the HSO approach, doing so helps for comparing with earlier treatments that are organized around $g$-functions. For a given parametrization of $\tilde{K}\parz{\Tsc{b}{};\mu_{Q_0}}$ and choice of $\bmax$, substituting \eref{Kforplots} into \eref{gKforplots} uniquely determines the $g_K$ function, which may be compared with the $g_K$ parametrizations obtained from other treatments.

\begin{figure}[t]
    \centering
        \includegraphics[scale = 0.35]{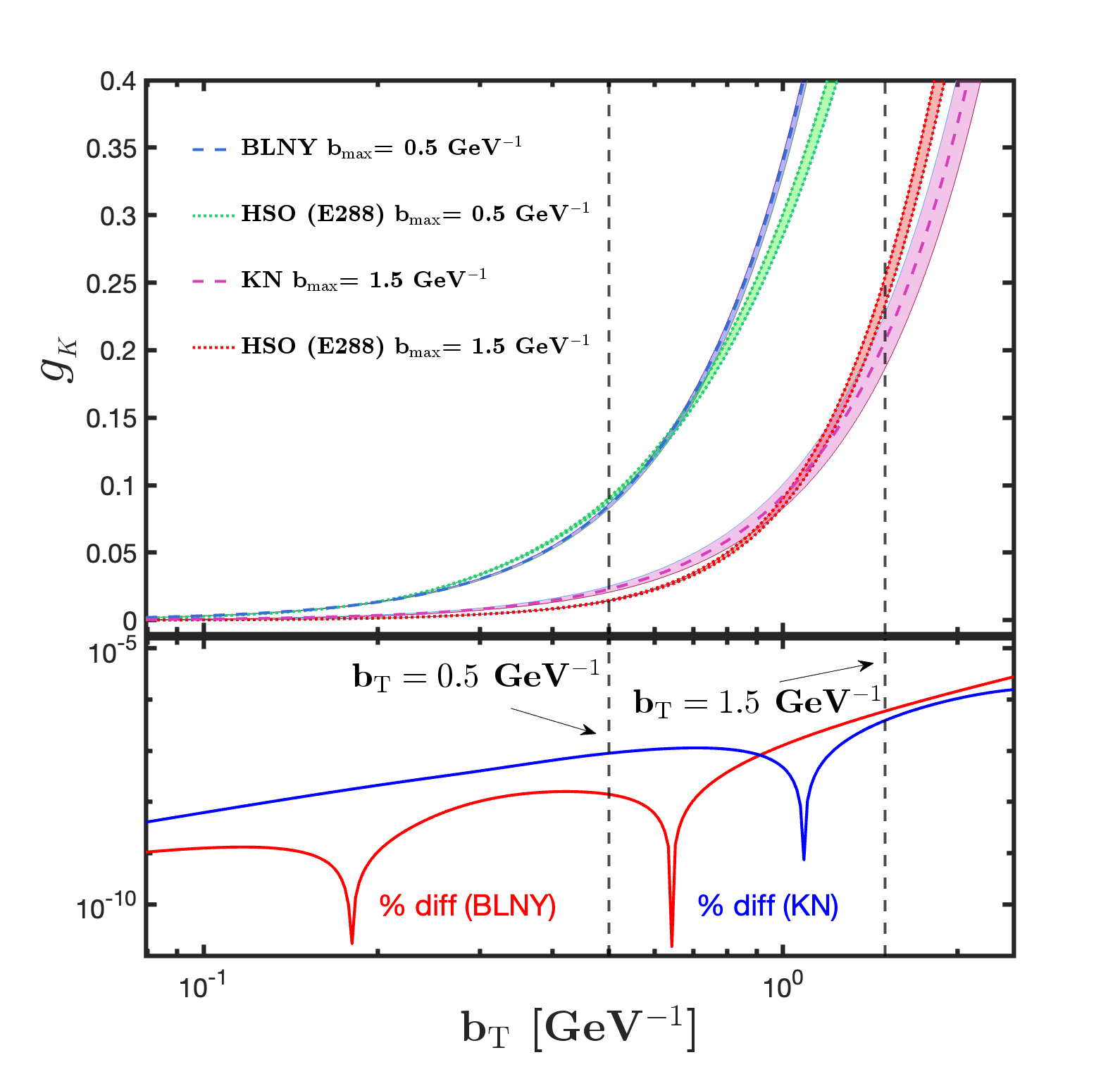}
    \caption{ Comparison between the $g_K$ functions obtained by BLNY \cite{Landry:2002ix} (dashed blue) and KN \cite{Konychev_2006} (dashed magenta) with the HSO version at $b_{\rm max}=0.5$ GeV$^{-1}$ (dotted green) and at $b_{\rm max}=1.5$ GeV$^{-1}$ (dotted red), as defined in~\erefs{Kforplots}{gKforplots}. For BLNY and KN, the shaded areas correspond to the envelope obtained from the parameter errors reported in \cite{Landry:2002ix,Konychev_2006}. The bands for the HSO functions are calculated as described in~\sref{fittingmodQ}. The two vertical dashed lines indicate the values of $\bmax$ chosen by  BLNY and KN. Both parametrizations eventually depart from  $g_K^{\rm HSO}$  since the HSO parametrization levels off to a constant at very large $\Tsc{b}{}$. This is shown in the increasing percentage difference defined as $2\,|g_K^{\rm HSO}-g_K^{\rm BLNY/KN}|/(g_K^{\rm HSO}+g_K^{\rm BLNY/KN})$ (solid red/solid blue) curves in the lower subplot. All lines in the lower panels are obtained using central values.}
    \label{fig:gK_HSO_vs_BLNY_and_KN}
\end{figure}

In the context of this paper, the $g_K$ parametrizations extracted in \cite{Landry:2002ix,Konychev_2006} are ideal because they use predominantly the same data sets that we have used in our analysis (although Refs.~\cite{Landry:2002ix,Konychev_2006} are of course much more thorough global analyses). Figure~\ref{fig:gK_HSO_vs_BLNY_and_KN} shows a comparison between the $g_K$ parametrization obtained from our fits of \eref{Kforplots} and those obtained in \cite{Landry:2002ix,Konychev_2006}, which we refer to as the BLNY and KN fits respectively. They are
%%%%%%%%%%%%%%%%%
\begin{equation}
\begin{split}
    g_K^{\rm BLNY} &= \frac{g_2}{2}b_{\rm T}^2,\quad g_2 = 0.68^{+0.01}_{-0.02}\,\text{GeV}^{2},\quad (b_{\rm max}=0.5~\text{GeV}^{-1}),\\
    g_K^{\rm KN} &= \frac{a_2}{2}b_{\rm T}^2,\quad a_2 = \left(0.184\pm0.018\right)\,\text{GeV}^{2},\quad (b_{\rm max}=1.5~\text{GeV}^{-1}).\\
\end{split}
\label{e.gK_BLNY_KN}
\end{equation} 
%%%%%%%%%%%%%%%%%

%%%%%%%%%%%%%%%%%%%%%%%%%%%%%%%%%%

%%%%%%%%%%%%%%%%%%%%%%%%%%%%%%%%%%
% %%%%%%%%%%%%%%%%%
The graphs show reassuring qualitative agreement between the three extractions, at least for $\Tsc{b}{} \lesssim \bmax$. There are significant differences only in the very large $\Tsc{b}{}$ regions, mostly beyond the primary range of sensitivity for the regions of $Q$ that we consider here.

To make the comparison slightly more quantitative, we consider that the important region of $\Tsc{b}{}$ for fitting at moderate-to-large $Q$ is around $\Tsc{b}{} \lesssim \bmax$. There, powers of $\Tsc{b}{}$ are not so small as to be entirely negligible, but not so large that the contributions from $g_K$ are completely suppressed by evolution. For that limited range of $\Tsc{b}{}$, we should therefore expect a power law expansion in $\Tscsq{b}{}/\bmax^2$ to provide a reasonable approximation of the effect of the full $g_K$ parametrization. Thus, we take  
%%%%%%%%%%%%%%%%%%%%%%%%%%%%%%%%%%
\begin{align}
g_K(\Tsc{b}{};\bmax) &{}\approx \frac{1}{2} c(\bmax) \; \Tscsq{b}{} \, , %\\
\label{e.gK_conventional}
\end{align}
%%%%%%%%%%%%%%%%%%%%%%%%%%%%%%%%%%
with $c(\bmax)$ to be a parameter that we may estimate from our full $g_K$ parametrization.
For a very crude estimate of $c$, we take few points around $\Tsc{b}{}= b_{\rm max}$ within a radius of $\Delta \Tsc{b}{}=0.4$ GeV$^{-1}$ and fit with a parabola. We find 
%%%%%%%%%%%%%%%%%%%%%%%%%%%%%%%%%%
\begin{equation}
    \begin{split}
        c(b_{\rm max}=0.5\,\text{GeV}^{-1})\approx 0.68~\text{GeV}^2\quad \text{and}\quad c(b_{\rm max}=1.5 \,\text{GeV}^{-1})\approx 0.22~\text{GeV}^2 \, , 
    \end{split}
\end{equation}
%%%%%%%%%%%%%%%%%%%%%%%%%%%%%%%%%%
which compare well with the values in \eref{gK_BLNY_KN} and confirms the qualitative agreement visually observable from \ref{fig:gK_HSO_vs_BLNY_and_KN}. The recovery of general features of early applications of the CSS formalism serves as an overall sanity check for our present approach.

Going beyond this cursory check will require a more thorough analysis than we are able to accommodate in this paper, and will involve more recent and complex model parametrizations of $g_K$. For now, we remark that preliminary attempts to 
compare with the extraction by the  MAP22~\cite{Bacchetta:2022awv}collaboration shows only a rather weak agreement with their parametrization at very small $b_{\rm T}\lesssim 0.15~$GeV$^{-1}$. Even worse agreement is found with that of the MSVZ23 collaboration~\cite{Moos:2023yfa}. We leave a full exploration of this to future work.

%%%%%%%%%%%%%%%%%%%%%%%%%%%%%%%%%%%%%%%%%%%%%%%%%%%
\section{Comparison with other recent treatments}
\label{s.comparison}

Next we compare the phenomenological approach above with other recent work that purports to isolate nonperturbative transverse momentum dependence in TMD factorization. 

For example, the main theoretical assertions put forward in work by the JAM collaboration in Ref.~\cite{Barry:2023qqh} rests upon the claim that the extracted TMD pdfs are mostly governed by the nonperturbative or intrinsic transverse momentum contributions and, conversely, that there is negligible sensitivity to the behavior of collinear pdfs and ffs. 
The role this plays in their physical interpretation is made explicit on page 4 of the article: ``Importantly, we have checked that the differences between the proton and pion $\langle \Tsc{b}{} | x \rangle $ are completely due to the nonperturbative TMD structure, independent of the collinear PDFs.'' Equation~(3) of Ref.~\cite{Barry:2023qqh} is meant to be the TMD pdf parametrization after evolution is applied, after application of an OPE in the small-$\Tsc{b}{}$ region is performed, and after the neglect of $\order{\bmax \Lambda_\text{QCD}}$ errors. In other words, it is a version of \eref{hadrotensold2} from \aref{translation}.  
Now recall that at large transverse momentum, $\Tsc{k}{} \approx Q$, the TMD pdf involves one and only one hard scale, and collinear factorization gives schematically (in JAM notation)
%%%%%%%%%%%%%%%%%
\begin{equation}
\label{e.btapprox}
\tilde{f}_{q/N(A)}(x,\Tsc{b}{}\approx 1/Q;\mu_Q,Q^2) \approx f_{q/N(A)}(x;Q) \, , 
\end{equation}
%%%%%%%%%%%%%%%%%
with errors that are subleading in $\alpha_s$ and powers of $1/Q$. Or, in transverse momentum space
%%%%%%%%%%%%%%%%%
\begin{equation}
\label{e.ktapprox}
f_{q/N(A)}(x,\Tsc{k}{}\approx Q;\mu_Q,Q^2) \approx \frac{1}{\Tscsq{k}{}} \sum \mathcal{C}(\ln (Q/\Tsc{k}{})) \otimes f_{q/N(A)}(x;Q) \, , 
\end{equation}
%%%%%%%%%%%%%%%%%
where $\mathcal{C}(\ln (Q/\Tsc{k}{}))$ is a hard coefficient that can only depend logarithmically on $Q/\Tsc{k}{}$.  
Likewise, the nonperturbative TMD parametrization must satisfy \erefs{fc_deff}{cutpdf}
%%%%%%%%%%%%%%%%%
\begin{equation}
\label{e.intcheck}
\pi \int_0^{Q^2} \diff{\Tscsq{k}{}} f_{q/N(A)}(x,\Tsc{k}{};\mu_Q,Q^2) \approx f_{q/N(A)}(x;\mu) \, .
\end{equation}
%%%%%%%%%%%%%%%%%
Reference \cite{Barry:2023qqh} does not enforce the consistency constraints of \erefs{btapprox}{intcheck}, and as a result the fitting at moderate scales is mostly controlled by the nonperturbative parts of \cite[Eq.(3)]{Barry:2023qqh}, even in regions of transverse momentum where that is no longer reasonable. %The lack of sensitivity to the collinear pdfs here is by construction, since the collinear factorization is not required to describe the collinear region, so the above quote is misleading. This is one example of how overfitting obscures a reasonable interpretation of fits.  
Thus, the potential sensitivity to collinear factorization is discarded by construction in that treatment, though that is obscured by the surface appearance of collinear pdfs in the fitting formula. The apparent lack of sensitivity to collinear factorization in the fits does not necessarily imply that collinear pdfs do not contribute. Rather, it likely means that the fitting of parameters for intrinsic nonperturbative transverse momentum dependence has been extended into regions where it is no longer reasonable. In other words, with the methods of Ref.~\cite{Barry:2023qqh} it is not possible to assess whether the collinear pdfs and collinear factorization are or are not relevant with a reasonable degree of accuracy. Therefore, the claimed physical interpretation put forward in~\cite{Barry:2023qqh} is misleading. In \cite{Barry:2023qqh}, a lack of sensitivity to collinear pdfs is interpreted to indicate dominance by nonperturbative transverse momentum structures. However, sensitivity to collinear pdfs should be large when nonperturbative transverse momentum structures dominate. It is simplest to see this in the case of a superrenormalizable theory rather than QCD. Then the equation 
\begin{equation}
\label{e.int}
f(x) = \pi \int_0^\infty \diff{\Tscsq{k}{}} f(x,\Tsc{k}{}) 
\end{equation}
would be completely exact. There would be no ultraviolet divergent contribution and all transverse momentum would be intrinsic. Then it would be a paradox to change $f(x)$ and find no change at all in $f(x,\Tsc{k}{})$ because one cannot change the outcome of a definite integral without making some change to the integrand. To find such behavior in a parametrization would indicate the presence of an inconsistency in the implementation, not confirmation that everything is intrinsic. The same basic issue applies in QCD, but it is less obvious because of the presence of divergences and the need for cutoffs.

Notice that, because of \eref{intcheck}, the specific type of collinear pdf parametrizations used in fits affect even (or especially) the $\Tsc{k}{} \approx 0$ region, not simply the tail at large transverse momentum. That connection must be preserved explicitly in the parametrization when the goal is to identify the separation between perturbative and nonperturbative transverse momentum dependence.
One way to demonstrate this is to fix nonperturbative TMD parameters and observe the effect of changing only the collinear pdfs.
As an example, 
we consider two cases in which the collinear functions are different enough so that they visibly modify the large-$\Tsc{k}{}$ behaviour of the TMDs and show how this also affects their behavior in the small-$\Tsc{k}{}$ region. To make the plots, we use the LHAPDF members $m=41$ and $m=42$ of the MMHT2014 set as a proxy. In each case, we compute the TMD pdf of~\eref{candidateqpdf} for strange quarks in a proton, using the results of our Gaussian fit to E288 data. In~\fref{tmdcollsensitivity} we show the two resulting TMD pdfs at the input scale $Q_0=4\,\text{GeV}$ and $x=0.1$.
Because of \eref{intcheck}, modifying the strange quark collinear pdf affects the strange quark TMD pdf by a significant amount, even at small $\Tsc{k}{}$. If one tunes the parameters of the nonperturbative transverse momentum dependence to bring the pink and blue curves into agreement at $\Tsc{k}{} \approx 0$, the effect unavoidably propagates to the large transverse momentum tails. Therefore, when a particular fit achieves agreement at large transverse momentum but without imposing \eref{intcheck}, it is as likely that it is \emph{because of} the fitting at small $\Tsc{k}{}$ rather than \emph{independent of} the fitting at small $\Tsc{k}{}$.

%%%%%%%%%%%%comparison to spectator model fits
\begin{figure}
    \centering
    \includegraphics[scale=0.5]{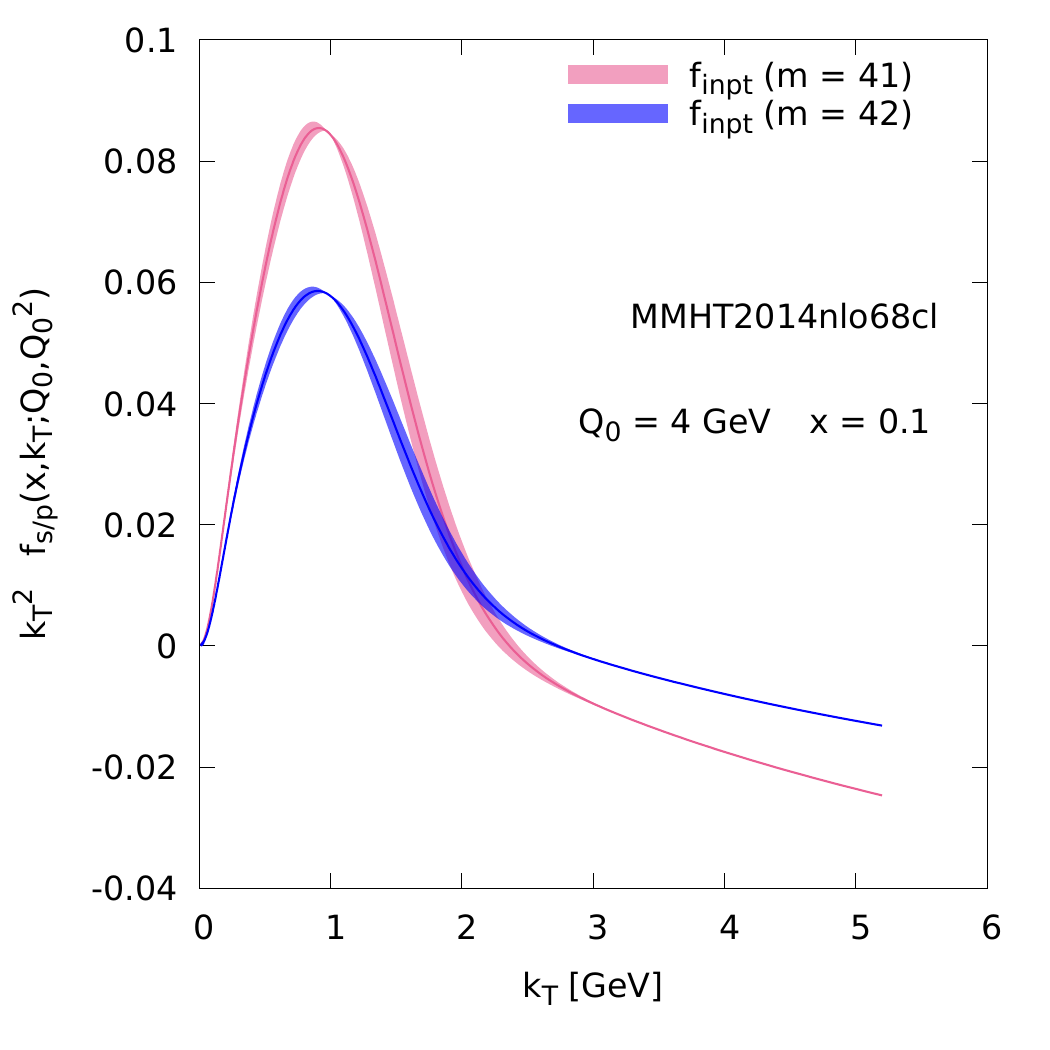}
    \caption{The TMD pdf for the strange quark in a proton at the input scale $Q_0=4\,\text{GeV}$ and $x=0.1$, computed according to \eref{candidateqpdf} using the results of our Gaussian fit to E288 data and the MMHT2014 set of collinear pdfs \cite{Harland-Lang:2014zoa}. The pink and blue lines correspond to {\tt LHAPDF} members $41$ and $42$ respectively. This plot clearly shows that changing the collinear pdf significantly affects the corresponding TMD even at very small $\Tsc{k}{}$.
    }
    \label{f.tmdcollsensitivity}
\end{figure}
%%%%%%%%%%%%%%%%%%%%%%%%%%%%%%%%%%%%%%%%%%%%%%%%%%%

\begin{figure}
    \centering
    \includegraphics[scale=0.166667]{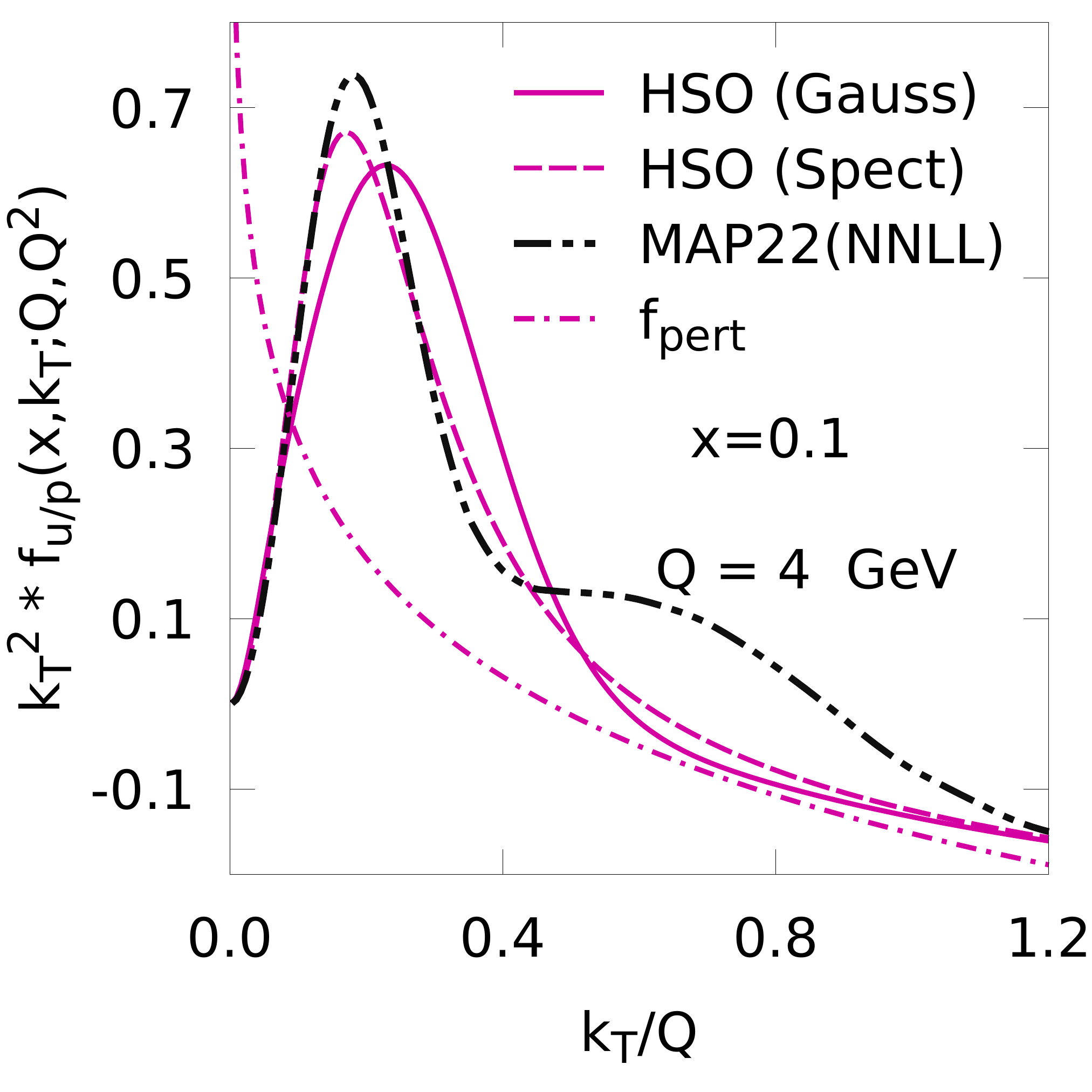}
    \includegraphics[scale=0.166667]{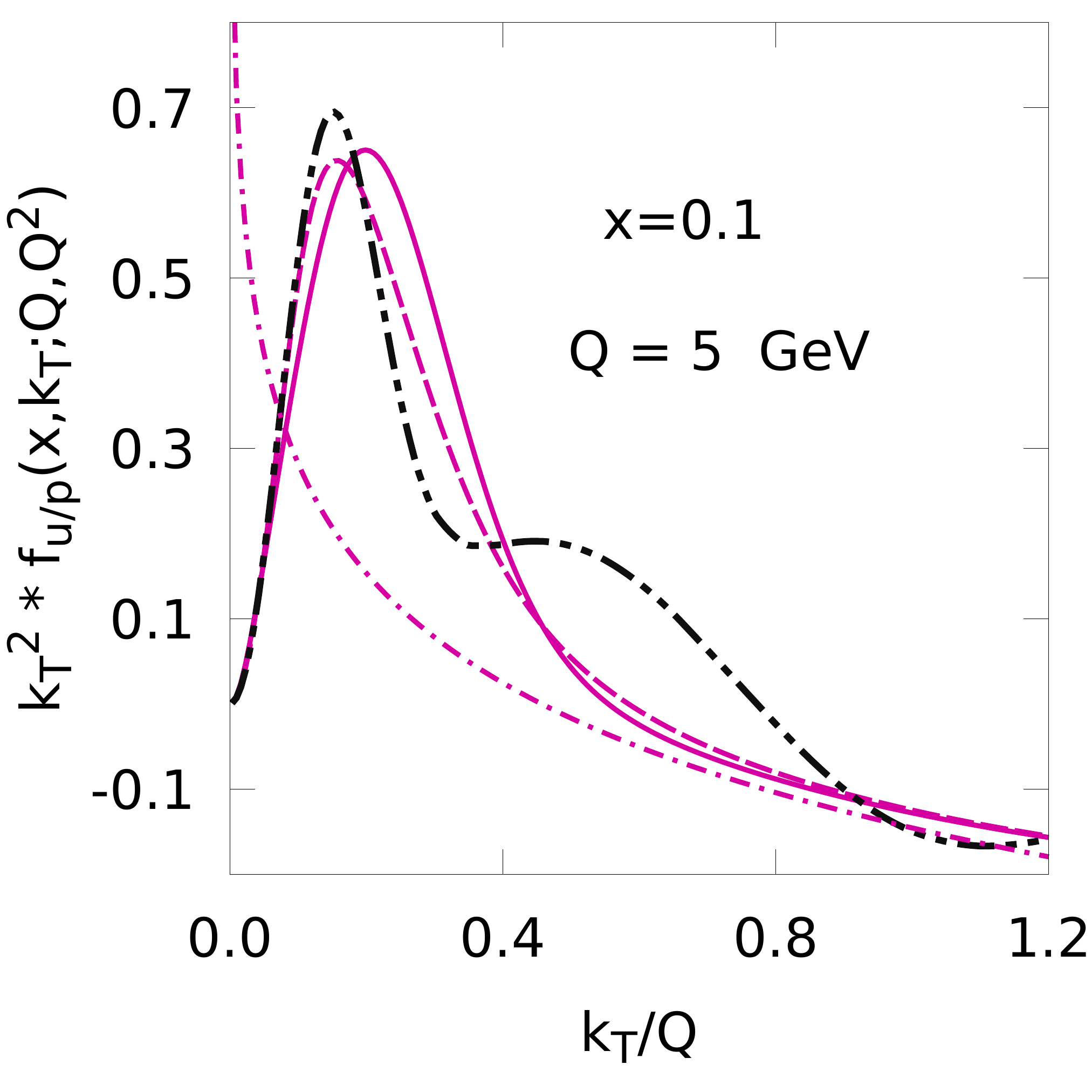}
    \includegraphics[scale=0.166667]{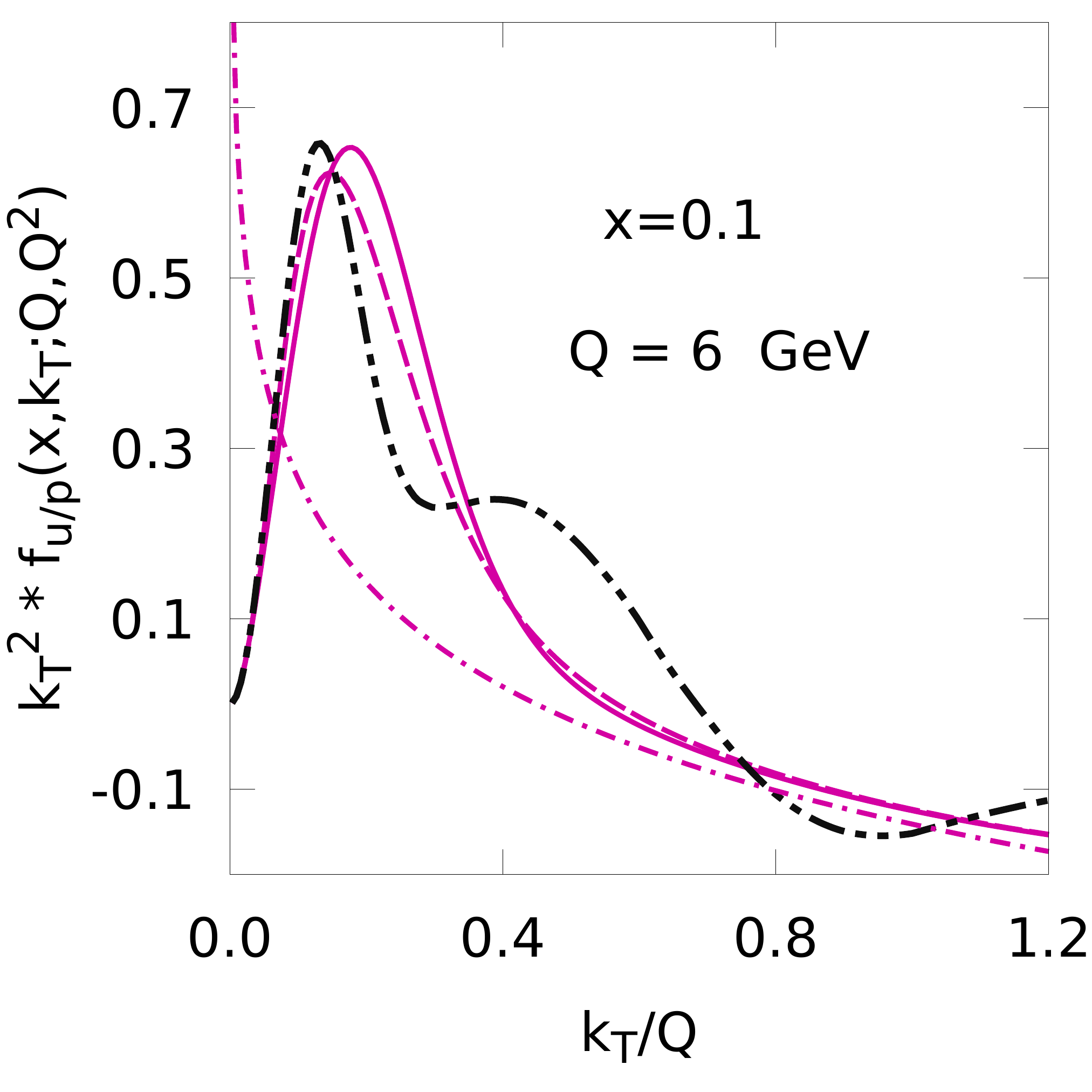}
    \includegraphics[scale=0.166667]{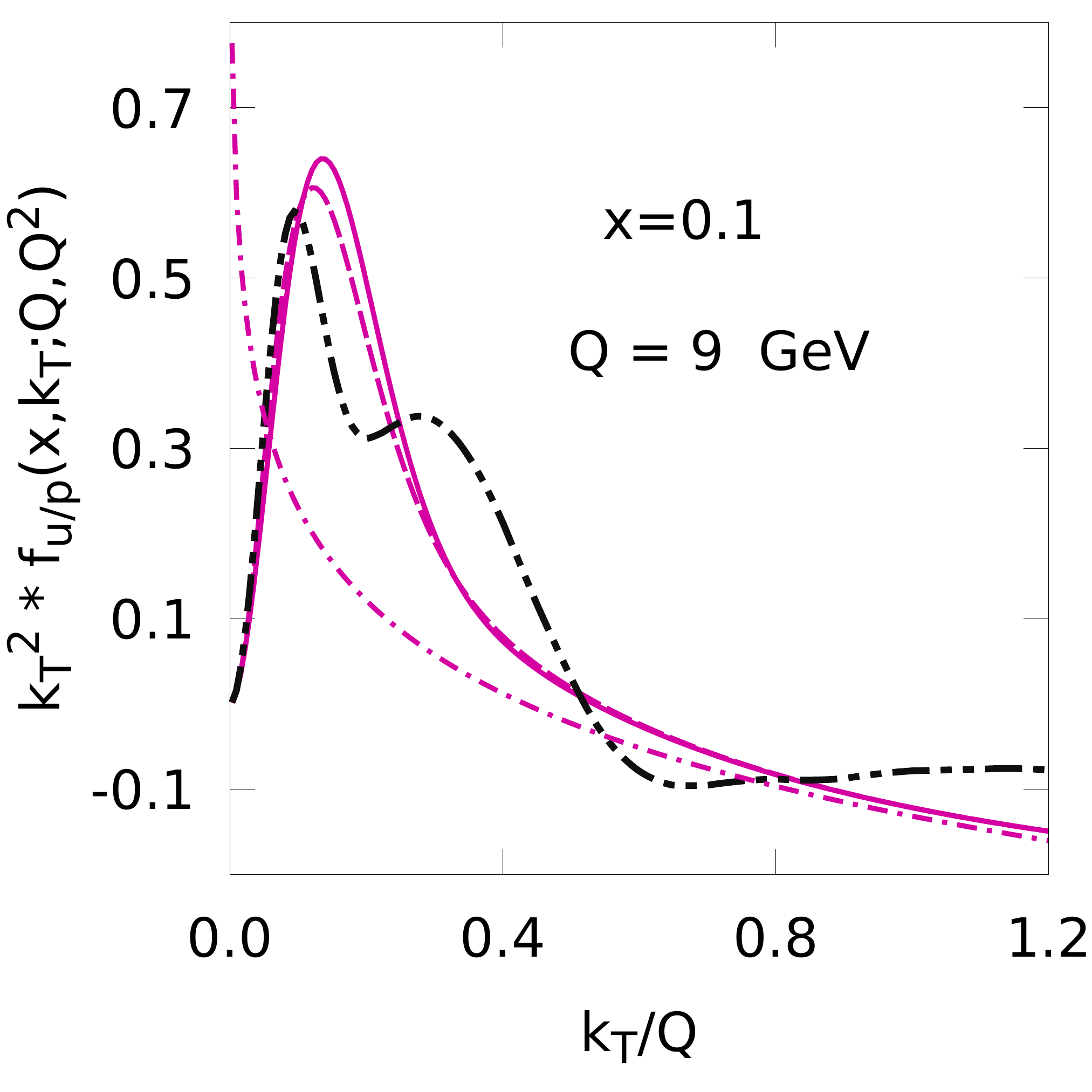}
    \includegraphics[scale=0.166667]{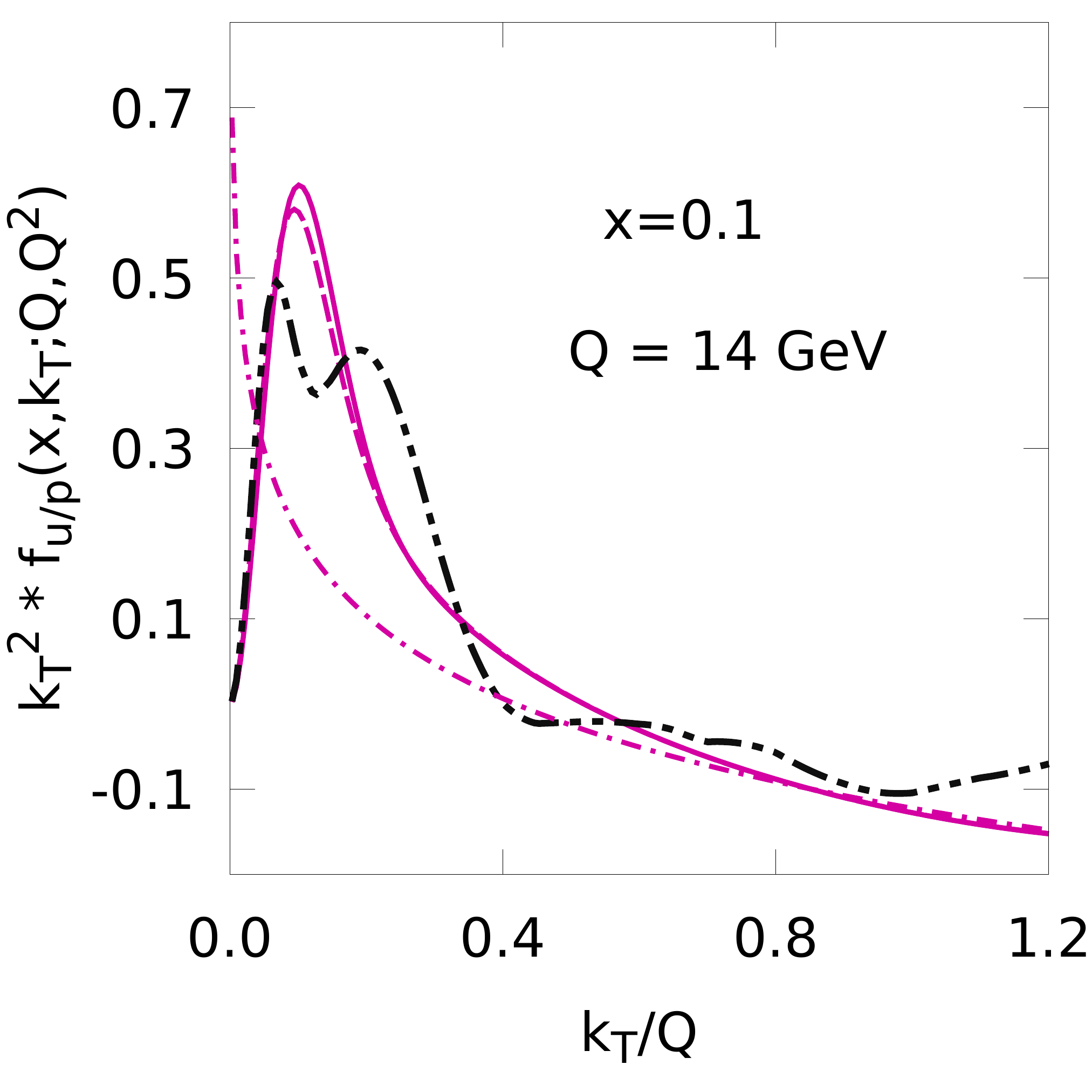}
    \includegraphics[scale=0.166667]{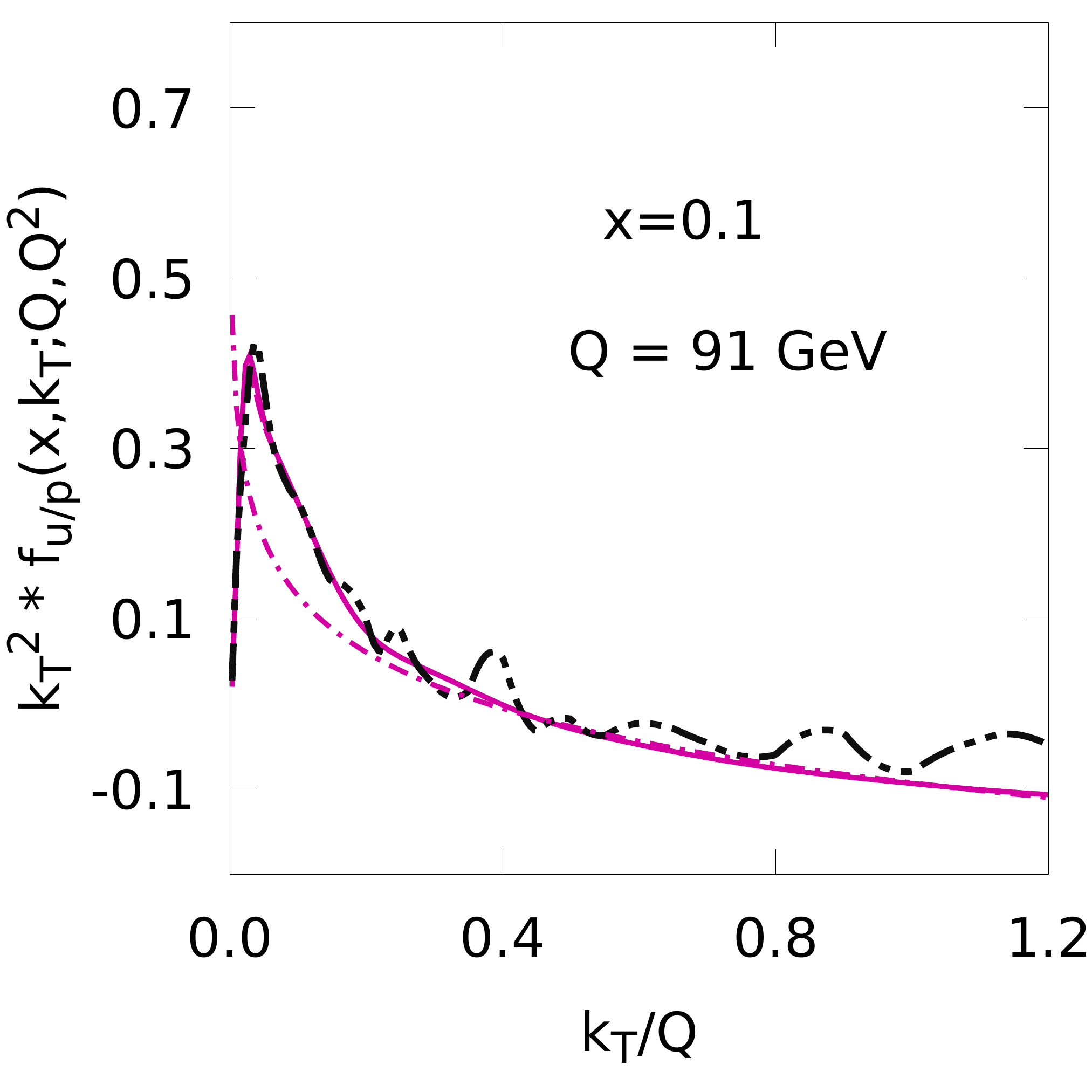}
    \caption{The up-flavor TMD pdfs obtained within the HSO approach from this paper compared those of the MAP22 collaboration, at $x=0.1$ and six different values of $Q$, as indicated in the legend inside each panel. The solid and dashed purple lines are the HSO parametrization of $\Tscsq{k}{}  f_{u/p}(x,\Tsc{k}{};Q,Q^2)$, as obtained by using the Gaussian and the spectator nonperturbative models, while the dashed-dotted purple lines represent the corresponding perturbative (large $\Tsc{k}{}$) behaviour. Black lines represent the NNLL MAP22~\cite{Bacchetta:2022awv} results  and were produced with \tt{NangaParbat}.}
    \label{f.mapcomparison}
\end{figure}

We next compare with another recent extraction of TMD pdfs in global fitting, the MAP22 NNLL analysis in Ref.~\cite{Bacchetta:2022awv}. Figure \ref{f.mapcomparison} shows the TMD pdfs obtained using the Gaussian and the spectator nonperturbative models within the HSO approach of this paper (solid and dashed purple lines), the perturbative tail (dot-dashed purple line), and the MAP22 result (dot-dot-dashed black line)\footnote{MAP22 result are obtained with \tt{NangaParbat}: https://github.com/MapCollaboration/NangaParbat}. Lines are for $x=0.1$ and six different scales from $4$~GeV to $91$~GeV.
Since one of our purposes is to compare differences in the large transverse momentum $\Tsc{k}{}$-tails, 
we have multiplied the TMD pdfs by $\Tscsq{k}{}$ to amplify the large-$\Tsc{k}{}$ region.\footnote{Using a logarithmic axis would more effectively magnify the tail, but we also wish to keep differences at moderate $\Tsc{k}{}/Q$ visible. Weighting by $\Tscsq{k}{}$ is sufficient to do both.}
At an input scale of $Q=Q_0=4$~GeV, the upper left plot in \fref{mapcomparison} shows that both the HSO Gaussian and spectator core models merge relatively quickly with the perturbative tail at moderate values of $\Tsc{k}{}/Q$. By contrast, the MAP22 line exhibits a significant enhancement in this region, suggesting that the nonperturbative model in Ref.~\cite{Bacchetta:2022awv} has a large effect in regions where the expectation is that transverse momentum should have a largely perturbative nature. This aspect of the MAP22 analysis also affects the shape of the evolved lines. At $Q=14\,\text{GeV}$, the MAP22 extraction retains its bimodal shape at small transverse momentum, so that its treatment of nonperturbative behavior (choices of model, $\bmin$, $\bmax$, etc.) seems to strongly determine the TMD pdf profile. By contrast, the central lines of the HSO (Gaussian and spectator core models) calculations at this same scale merge and closely agree at all ranges of $\Tsc{k}{}/Q$. Although the precise value of $Q$ where this should happen is not obvious, the expectation is that perturbative effects slowly dominate as $Q$ is evolved upwards until sensitivity to nonperturbative input parameters becomes very weak.
Thus, the MAP22 curve in the bottom-central panel of~\fref{mapcomparison} is another symptom that nonperturbative effects are leaking into the perturbative region.

Although the overall size of our up quark TMDs is, on average, similar to that obtained by MAP22, 
it is evident that the latter, which exploits a parametrization with a very large number of free parameters, has a very different shape: it results in a multimodal distribution, with a very slow convergence to the perturbative tail at large $\Tsc{k}{}$, even accounting for oscillations at large $\Tsc{k}{}/Q$, due to artifacts of the numerical integration routines\footnote{As explained in private communication with members of the MAP collaboration.}. As mentioned in \sref{intro}, the analysis of \cite{Bacchetta:2022awv} aims at
fitting the largest possible amount of data, thus requiring the use of a very flexible TMD parametrization. This excessive plasticity can cause a significant intrusion of the nonperturbative model in the kinematical region where the behaviour of the TMD should be dominated by perturbative physics, and lead to an inherent difficulty in the interpretation of the final results of the phenomenological analysis.

%%%%%%%%%%%%comparison to spectator model fits
%%%%%%%%%%%%%%%%%%%%%%%%%%%%%%%%%%%%%%%%%%%%%%%%%%%

The examples extend to treatments of spin-dependent observables. For instance, the phenomenological analysis of transverse single spin asymmetries from the JAM collaboration in Ref.~\cite{Cammarota:2020qcw} is accompanied by the strong interpretation that it ``indicates single transverse-spin asymmetries in high-energy collisions have a common origin."  
In support of this claim, Ref.~\cite{Cammarota:2020qcw} uses a hybrid of rather different theoretical formalisms, approximations, and simplifying assumptions in the calculations they use for their phenomenological analysis, with TMD factorization appearing as only one component. 
The range of $Q$ is large, extending as low as $Q \approx 1.4$~GeV, near the boundary where factorization starts to be questionable, and as high as $80$~GeV, so that at least some nontrivial effects from evolution might be expected. The fits appear reasonable by the standards of $\chi^2$ minimization, but it is unclear how sensitive this outcome is to the underlying theoretical hypotheses and assumptions that they purport to test. Therefore, the claims of predictive power given there are difficult to assess. Notably, however, the fits in Ref.~\cite{Cammarota:2020qcw} fail significantly at predicting much of both the isolated and full pion data in subsequent measurements from the STAR collaboration~\cite{STAR:2020nnl}. 

The approach we have discussed in this paper extends naturally to spin dependent observables like those considered in Ref.~\cite{Cammarota:2020qcw} while providing a much more systematic way to frame and assess claims of predictive power.

%%%%%%%%%%%%%%%%%%%%%%%%%%%%%%%%%%
\section{Conclusions}
\label{s.conclusion}

We have presented a first practical implementation of the steps for TMD phenomenology that were formulated and discussed more abstractly in Refs.~\cite{Gonzalez-Hernandez:2022ifv,Gonzalez-Hernandez:2023iso}. The primary focus of this ``HSO'' approach, as reviewed in the introduction, is to isolate and identify the non-perturbative input parametrizations needed to fully characterize the fundamental operator matrix elements in a TMD factorization framework, and in such a way as to allow for testable or falsifiable predictions for future measurements. 
We exhibited predictive power by first constraining  model parameters in a class of Drell-Yan measurements at moderate $Q$, and then evolving upwards in $Q$ to postdict reasonable agreement with measurements of $Z^0$ boson production. While doing this, we emphasized the modular nature of the nonperturbative input in the HSO approach, by swapping Gaussian and spectator model descriptions of the very small transverse momentum behavior. Our hope is that this feature of the HSO approach will open the way for a more direct 
incorporation of specific theoretical treatments of nonperturbative transverse momentum dependent structures in the future. Our examples illustrate how the sensitivity to some features of nonperturbative structure varies with $Q$ and $x$. 

An extraction of the nonperturbative function that is normally called $g_K$ in earlier literature is shown in \erefs{Kforplots}{gKforplots}. The $g_K$ function is very strongly universal, and thus its measurement leads to important predictions. The same function arises in a wide range of rather different types of observables, including SIDIS~\cite{Bacchetta:2022awv,Scimemi:2019cmh,Vladimirov:2020umg} and inclusive $e^+ e^-$ annihilation into back-to-back hadrons~\cite{Moffat:2019pci} or into one single hadron~\cite{Boglione:2017jlh,Boglione:2020cwn,Boglione:2020auc,Boglione:2023duo}, therefore finding agreement across all such processes constitutes an exciting and highly nontrivial test of the TMD factorization's potential for describing properties of the QCD vacuum and relating it to concrete physical observables. Implementations over a wider range of processes and with the full $\order{\alpha_s^2}$ expression are still needed, but we are reassured that the treatment in this paper finds good qualitative agreement with earlier pioneering extractions of $g_K$. 

We emphasize again that there is no difference between the theoretical underpinnings of the standard TMD factorization based on the CSS formalism and its extensions and the HSO style of working that we advocate in this paper. The differences lie only in the steps for implementing the formalism phenomenologically. As such, it is straightforward to translate between our steps and those of more typical TMD/CSS implementations, and some of the details for doing this are reviewed in \aref{translation}. However, there are major differences in philosophy between our approach and others, and we have argued that these can have significant practical consequences. For further commentary on the differences, we refer the reader to \cite{Gonzalez-Hernandez:2022ifv} and to a discussion of logarithmic power counting in \aref{logcounting}.

Observing more detailed features of nonperturbative structures will require more comprehensive fitting than what we have done in this paper. Especially important are SIDIS measurements like those at Jefferson Lab, since they are expected to have greater sensitivity to nonperturbative behavior (see for example Ref.~\cite{Accardi:2023chb}, Sect. 5.1, and references therein). However, we have also pointed out the subtle nuances involved in the interpretation of the results produced by direct fitting, and we have emphasized the dangers of overfitting. We have argued that the most effective way to test the theoretical understanding of nonperturbative structures lies with their predictive power, and for this a careful and discriminating selection of measurements is sometimes more important than the total quantity of data included in a global fit. 

In our plans, the next step is the extension of our HSO analysis to data covering larger values of $Q$. Evolving to even higher scales would entail considering kinematic regions involving much smaller values of $x$, for which no experimental data at low-to-moderate values of $Q$ are available, leaving a lack of information on the non-perturbative behavior of the TMD at low $\Tsc{k}{}$. A phenomenological analysis of LHC data, for example, would possibly require some non-trivial modifications of the parametric form of the core function to correctly reproduce the small-$x$ region, for which no Drell-Yan measurements are available at low or moderate $Q$.

Over a longer time scale, we will broaden the range of processes included
at the fitting stage, and to employ a more diverse and sophisticated set of nonperturbative calculations and model assumptions in the treatment of nonperturbative structures. Beyond including unpolarized $e^+ e^-$-annihilation processes and SIDIS, it will be straightforward to apply the HSO methodology to spin dependent observables. A major aim is to produce a set of predictions for the EIC whose success or failure can readily be assessed and interpreted in ways that concretely refine the understanding of deeper theoretical underpinnings of QCD processes. 

Another issue still to be addressed in future work is the continuing difficulty, at moderate values of $Q$, in describing data for large $\Tsc{q}{}$ tails without allowing nonperturbative transverse momentum effects to migrate into unsettlingly large regions of $\Tsc{q}{}$. We end this paper with some speculations about how this might be confronted in future work that takes the HSO approach. Traditionally, extractions of collinear pdfs have been the implicit anchor around which phenomenological extractions of other nonperturbative objects (like nonperturbative TMD effects) were organized. From that perspective, a consistent TMD extraction is one that reproduces known collinear extractions at large $\Tsc{q}{}$. We suggest that inverting this view may eventually be necessary for making progress going forward. That is, extractions and parametrizations of the TMD pdfs should be taken as the primary objects, with collinear pdf extractions adjusted to conform to TMD expectations rather than vice-versa. Indeed, the latter perspective is more faithful to the underlying theoretical setup, where TMD correlation functions are the more fundamental objects, while collinear correlation functions emerge from them as a consequence of transverse momentum integrals. We speculate that only rather minor modifications to existing collinear parametrizations may be necessary to bring the large-$\Tsc{q}{}$ calculations into greater agreement with both phenomenological data and with expectations from collinear factorization. An improvement of this type would also lead to greater consistency in the transition to a fixed order $Y$-term at $\Tsc{q}{} \approx Q$. We leave an exploration of this and other extensions of HSO approach to future work.

%%%%%%%%%%%%%%%%%%%%%
\vskip 0.3in
%%%%%%%%%%%%%%%%%%%%%
\acknowledgments
We thank Pavel Nadolsky for useful discussions. We also thank Andrea Signori and Matteo Cerutti for clarifying aspects of the MAP22 analysis. 
M. Boglione and  J.O.~Gonzalez-Hernandez acknowledge funding from the European
Union’s Horizon 2020 research and innovation programme under grant agreement No 824093.
T.~Rogers, T.~Rainaldi and A. Simonelli were partially supported by the U.S. Department of Energy, Office of Science, Office of Nuclear Physics, under Award Number DE-SC0018106 and partially supported by the U.S. Department of Energy, Office of Science, Office of Nuclear Physics, under Award Number DE-SC0024715. F. Aslan was supported by the NSF Award No. 2111490, the DOE Quark-Gluon Tomography Topical Collaboration with Award No. DE-SC0023646  and the Center for Nuclear Femtography, Southeastern
Universities Research Association, Washington D.C.. 
This work 
was also supported by the DOE Contract No. DE- AC05-06OR23177, under which 
Jefferson Science Associates, LLC operates Jefferson Lab. 
%%%%%%%%%%%%%%%%%%%%%

%==================================================

\begin{appendix}

%%%%%%%%%%%%%%%%%%%%%
\section{Expressions extended to $\order{\alpha_s^2}$}
\label{a.higherorder}

In this appendix we write out certain quantities from the main text to the next order in $\alpha_s$. Original results for some higher orders can be found in Refs.~\cite{Davies:1984hs,Moch:2005id,Li:2016ctv}.
The next order in the hard part (\eref{Hardfactor}) is~\cite{Moch:2005id} 
%%%%%%%%%%%%%
\begin{align}
  H_{j \bar{\jmath}}^{(2)}
  ={}&
        \delta_{j \bar{\jmath}} C_F^2 \left[
                   \frac{511}{4}
                   - \frac{83\pi^2}{3}
                   + \frac{67\pi^4}{30}
                   - 60 \zeta_3
        \right]
         + \delta_{j \bar{\jmath}} C_F C_A  \left[
               - \frac{51157}{324}
               + \frac{1061 \pi^2}{54}
               - \frac{8 \pi^4}{45}
               + \frac{626}{9} \zeta_3 
        \right]
\nonumber\\&
         +  \delta_{j \bar{\jmath}} n_f C_F \left[
                    \frac{4085}{162}
                    - \frac{91 \pi^2}{27}
                    + \frac{4}{9} \zeta_3 
        \right] \, .
\label{eq:H.hat.2}
\end{align}
%%%%%%%%%%%%%
The expression for $\tilde{K}(\Tsc{b}{}; \mu)$ (see \eref{pert_K}) to $\order{\alpha_s^2}$ is
%%%%%%%%%%%%%
\begin{align}
\label{e.pert_Ka}
\tilde{K}(\Tsc{b}{}; \mu)   
={}& -\frac{2 C_F \alpha_s(\mu)}{\pi} \ln \left( \frac{\Tsc{b}{} \mu}{2 e^{-\gamma_E}} \right)  \nonumber \\
  & +  C_F \frac{\alpha_s(\mu)^2}{2 \pi^2} \left[ \left( \frac{2}{3} n_f  - \frac{11}{3} C_A \right) \ln^2 \left( \frac{\Tsc{b}{} \mu}{2 e^{-\gamma_E}} \right) 
                        \right. 
\nonumber \\ 
 & \qquad \qquad \qquad
+ \left. 
             \left( -\frac{67}{9}C_A + \frac{\pi^2}{3} C_A +  \frac{10}{9} n_f \right) \ln \left( \frac{\Tsc{b}{} \mu}{2 e^{-\gamma_E}} \right) 
               + \left( \frac{7}{2} \zeta_3 - \frac{101}{27} \right)
               C_A 
               + \frac{14}{27}  n_f
      \right]  \, \nonumber \\
    &  + O(\alpha_s^3) \, ,
\end{align}
%%%%%%%%%%%%%
and its anomalous dimension is
%%%%%%%%%%%%%
\begin{align}
  \gamma_K ={}&   
    \frac{2 C_F \alpha_s}{\pi}
    + \frac{\alpha_s^2}{16 \pi^2}
      \left[
             C_A C_F \left( \frac{536}{9}- \frac{8\pi^2}{3} \right)
             - \frac{80}{9} C_F n_f
      \right] + \order{\alpha_s^3} \, .
\end{align}
%%%%%%%%%%%%%

Next, we repeat exactly the steps of \sref{CSinput}, but now with the $\tilde{K}(\Tsc{b}{}; \mu)$ and $\gamma_K$ kept through $\order{\alpha_s^2}$. For the complete range of $0 < \Tsc{b}{}< \infty$, the HSO approach requires that the renormalization group equation be satisfied through order $\alpha_s(\mu)^2$, 
%%%%%%%%%%%%%
\begin{align}
&{}\frac{\diff{}{}}{\diff{\ln \mu}{}} \tilde{K}(\Tsc{b}{};\mu) = - \gamma_K(\alpha_s(\mu)) \no
&{}= -\frac{2 C_F \alpha_s(\mu)}{\pi} - \frac{\alpha(\mu)^2}{16 \pi^2} \left[ C_A C_F \parz{\frac{536}{9} - \frac{8 \pi^2}{3}} - \frac{80}{9} C_F n_f \right] + \order{\alpha_s^3} \label{e.K_RGa}\, .
\end{align}
%%%%%%%%%%%%%
We write the input transverse momentum space CS kernel as 
%%%%%%%%%%%%%
\begin{align}
&{}\inpt{K}(\Tsc{k}{};\mu_{Q_0}) = & \no
&{} \left[A_K^{(1)}(\mu_{Q_0}) + A_K^{(2)}(\mu_{Q_0}) \right] \frac{1}{\Tscsq{k}{} + m_K^2} 
+  B_K^{(2)}(\mu_{Q_0}) \frac{1}{\Tscsq{k}{} + m_K^2} \ln \parz{\frac{\mu_{Q_0}^2}{\Tscsq{k}{} + m_K^2}} + K_\text{core}(\Tsc{k}{}) + D_K(\mu_{Q_0}) \delta^{(2)}\parz{\T{k}{}} \, , \label{e.final_mom_Ka}
\end{align}
%%%%%%%%%%%%%
where 
%%%%%%%%%%%%%
\begin{align}
A_K^{(1)}(\mu_{Q_0}) &= \frac{\alpha_s(\mu_{Q_0}) C_F}{\pi^2} \, , \\
A_K^{(2)}(\mu_{Q_0}) &= -\frac{\alpha_s(\mu_{Q_0})^2 C_F}{4 \pi^3} \parz{-\frac{67}{9} C_A + \frac{\pi^2}{3} C_A + \frac{10}{9} n_f } \, , \\
B_K^{(2)}(\mu_{Q_0}) &= -\frac{\alpha_s(\mu_{Q_0})^2 C_F}{4 \pi^3} \parz{\frac{2}{3} n_f - \frac{11}{3} C_A} \, .
\end{align}
%%%%%%%%%%%%%
The function $K_\text{core}(\Tsc{k}{})$ is analogous to 
$\np{f}{i/p}(\xbj,\Tsc{k}{};Q_0^2)$. It is used to tune the very large $\Tsc{b}{}$ behavior and it vanishes like a power at small $\Tsc{b}{}$. It will introduce at least one extra parameter beyond $m_K$. In coordinate space, we will demand that $\tilde{K}(\Tsc{b}{};\mu_{Q_0})$ approach a negative constant $b_K$ as $\Tsc{b}{} \to \infty$~\cite{Collins:2014jpa}. We will use a Gaussian, 
%%%%%%%%%%%%%
\begin{equation}
\label{e.CSmodela}
K_\text{core}(\Tsc{k}{}) = \frac{b_K}{4 \pi m_K^2} e^{-\frac{\Tscsq{k}{}}{4 m_K^2}} \, .
\end{equation}
%%%%%%%%%%%%%
The last term in \eref{final_mom_K} has a $D_K(\mu_{Q_0})$ which is
%%%%%%%%%%%%%
\begin{align}
D_K(\mu_{Q_0}) &{}= - b_K \no
&{}+ \frac{2 \alpha_s(\mu_{Q_0}) C_F}{\pi} \ln \parz{\frac{m_K}{\mu_{Q_0}}} \no
&{}+ \frac{C_F \alpha_s(\mu_{Q_0})^2}{2 \pi^2}\left[ C_A \parz{\frac{7}{2}\zeta_3 - \frac{101}{27}} + \frac{14}{27} n_f \right] \no
&{}- \frac{\alpha_s(\mu_{Q_0})^2 C_F}{18 \pi^2} \ln \parz{\frac{m_K}{\mu_{Q_0}}} \left[ \parz{33 C_A - 6 n_f} \ln \parz{\frac{m_K}{\mu_{Q_0}}} + \parz{3 \pi^2 - 67} C_A + 10 n_f \right] \, .
\end{align}
%%%%%%%%%%%%%
Transforming \eref{final_mom_Ka} into coordinate space gives
%%%%%%%%%%%%%
\begin{align}
\label{e.final_coord_K2a}
&{} \inpt{\tilde{K}}(\Tsc{b}{};\mu_{Q_0}) = \no
&{} \, 2 \pi \left[A_K^{(1)}(\mu_{Q_0}) + A_K^{(2)}(\mu_{Q_0}) \right] K_0\parz{m_K \Tsc{b}{}} + 2 \pi B_K^{(2)}(\mu_{Q_0}) K_0\parz{m_K \Tsc{b}{}} \ln \parz{\frac{\mu_{Q_0}^2 \Tsc{b}{}}{2 m_K e^{-\gamma_E}}} + b_K e^{-m_K^2 \Tscsq{b}{}} + D_K(\mu_{Q_0}) \, .
\end{align}
%%%%%%%%%%%%%
It is straightforward to verify that \eref{final_coord_K2a} 
equals \eref{pert_K} when $m_k \Tsc{b}{} \to 0$. Using
%%%%%%%%%%%%%
\begin{equation}
\label{e.A_K_RGa}
\frac{\diff{A_K^{(1)}(\mu_{Q_0})}{}}{\diff{\ln \mu_{Q_0}}{}} = -2 B^{(2)}_K(\mu_{Q_0}) + \order{\alpha_s(\mu_{Q_0})^3} \, 
\end{equation}
%%%%%%%%%%%%%
also shows that it satisfies \eref{K_RG} for all $\Tsc{b}{}$. The large-$\Tsc{b}{}$ limit of the CS kernel in \eref{final_coord_K2} is 
%%%%%%%%%%%%%
\begin{equation}
\label{e.CS_limita}
\lim_{\Tsc{b}{} \to \infty} \inpt{\tilde{K}}(\Tsc{b}{};\mu_{Q_0}) = D_{K}(\mu_{Q_0}) = -b_K + \order{\alpha_s(\mu_{Q_0})} \, . 
\end{equation}
%%%%%%%%%%%%%

The remaining $\order{\alpha_s(\mu)^2}$ expression needed for evolution is the next order version of \eref{gammaexp}, 
%%%%%%%%%%%%%
\begin{align}
\gamma(\alpha_s(\mu);1)    
&{}= \frac{3 C_F \alpha_s(\mu)}{2 \pi}  \no
&{} + \frac{\alpha_s(\mu)^2}    {16 \pi^2}
      \left[
           C_F^2 \left( 3 - 4 \pi^2 + 48 \zeta_3 \right)
           + C_FC_A \left( \frac{961}{27} + \frac{11 \pi^2}{3} - 52 \zeta_3 \right)
           + C_Fn_f \left( - \frac{130}{27} - \frac{2\pi^2}{3} \right)
     \right] \no
&{}+ \order{\alpha_s(\mu)^3} \, . \label{e.gammaexpa}
\end{align}
%%%%%%%%%%%%%

For implementing evolution at $\order{\alpha_s(\mu)^2}$, the steps from the main body of the paper apply unmodified with the above expressions. Note, however, that a full $\order{\alpha_s(\mu)^2}$ treatment requires also the large $\Tsc{k}{}$ tail expressions to order $\alpha_s(\mu)^2$, which we have not included above.

%%%%%%%%%%%%%%%%%%%%%
\section{Translation to ``$g$-functions''}
\label{a.translation}

In classic phenomenological treatments of TMD factorization based upon the CSS formalism, the expression for the cross section takes a somewhat different form from \eref{Fuu}. The more common organization has a small $\Tsc{b}{}$ OPE already applied explicitly, and all the presumably nonperturbative transverse momentum behavior is isolated in exponentials of separate functions usually labeled by $g$. (For a typical example, see Eq.(22) of Ref.~\cite{Collins:2014jpa}.) In fact, the parametrization from this paper can be recast in exactly this form as well, and we show the steps here. The discussion below will be kept relatively brief, while a more detailed explanation is to be found in \cite[Sect.~IX]{Gonzalez-Hernandez:2022ifv}.

The sequestration of nonperturbative parts into $g$-functions begins~\cite{Collins:1981va}  by defining an arbitrary function, $\bstar(\Tsc{b}{})$, with the properties
%%%%%%%%%%%%%%%%%%%%%
\begin{equation}
\bstar = \bstar(\Tsc{b}{}) = 
\begin{dcases}
\T{b}{} & b_T \ll b_{\rm max} \\
\vect{b}_{\rm max} & b_T \gg b_{\rm max} \,  \label{e.bdefold}
\end{dcases}\, ,
\end{equation}
%%%%%%%%%%%%%%%%%%%%%
where $\bmax$ is an equally arbitrary transverse size dividing what are considered large and small $\Tsc{b}{}$. 
A very common choice~\cite{Collins:1984kg} is
%%%%%%%%%%%%%%%%%
\begin{align}
\label{e.bstar}
  \bstar = \frac{ \T{b}{} }{ \sqrt{ 1 + \Tscsq{b}{}/\bmax^2} } \, , 
\end{align}
%%%%%%%%%%%%%%%%%%%%
though any other smooth function that satisfies \eref{bdefold} is equally valid. 
In addition, one defines a renormalization scale 
%%%%%%%%%%%%%%%%%%%%
\begin{equation}
\label{e.mubst}
\mu_{\bstarsc} \equiv C_1/\bstarsc \, ,
\end{equation}
%%%%%%%%%%%%%%%%%%%%
that optimizes perturbation theory in the OPE for the small-$\Tsc{b}{}$ limit. Next, the evolution equation \eref{CSeq}
gives
%%%%%%%%%%%%%%%%%%%%
\begin{equation}
\tilde{f}_{j/p}(x,{\Tsc{b}{}}{};\mu,\zeta) = \tilde{f}_{j/p}(x,{\Tsc{b}{}}{};\mu,Q_0^2) \exp \left\{ \tilde{K}(\Tsc{b}{};\mu) \ln \parz{\frac{\sqrt{\zeta}}{Q_0}} \right\} \, , \label{e.CSevolve1}
\end{equation}
%%%%%%%%%%%%%%%%%%%%
and since this is independent of $\Tsc{b}{}$ one may also write
%%%%%%%%%%%%%%%%%%%%
\begin{equation}
\tilde{f}_{j/p}(x,\bstarsc;\mu,\zeta) = \tilde{f}_{j/p}(x,\bstarsc;\mu,Q_0^2) \exp \left\{ \tilde{K}(\bstarsc;\mu) \ln \parz{\frac{\sqrt{\zeta}}{Q_0}} \right\} \, . \label{e.CSevolve2}
\end{equation}
%%%%%%%%%%%%%%%%%%%%
Dividing \eref{CSevolve1} by \eref{CSevolve2} leads to some useful simplifications, 
%%%%%%%%%%%%%%%%%%%%
\begin{equation}
\frac{\tilde{f}_{j/p}(x,{\Tsc{b}{}}{};\mu,\zeta)}{\tilde{f}_{j/p}(x,\bstarsc;\mu,\zeta)} = \frac{\tilde{f}_{j/p}(x,{\Tsc{b}{}}{};\mu_{Q_0},Q_0^2)}{\tilde{f}_{j/p}(x,\bstarsc;\mu_{Q_0},Q_0^2)} \exp \left\{ - \left[ \tilde{K}\parz{\bstarsc;\mu_{Q_0}} - \tilde{K}\parz{\Tsc{b}{};\mu_{Q_0}} \right] \ln \parz{\frac{\sqrt{\zeta}}{Q_0}} \right\} \, . \label{e.firstratio}
\end{equation}
%%%%%%%%%%%%%%%%%%%%
From the evolution equations \erefs{CSrg}{TMDrg}, the $\mu$ dependence exactly cancels in the ratio of $\tilde{f}_{j/p}$'s and in the differences of $\tilde{K}$'s, 
%%%%%%%%%%%%%%%%%%%%
\begin{equation}
\frac{\tilde{f}_{j/p}(x,{\Tsc{b}{}}{};\mu,Q_0^2)}{\tilde{f}_{j/p}(x,\bstarsc;\mu,Q_0^2)} = \frac{\tilde{f}_{j/p}(x,{\Tsc{b}{}}{};\mu_{Q_0},Q_0^2)}{\tilde{f}_{j/p}(x,\bstarsc;\mu_{Q_0},Q_0^2)}\, , \qquad 
\tilde{K}\parz{\bstarsc;\mu} - \tilde{K}\parz{\Tsc{b}{};\mu} = \tilde{K}\parz{\bstarsc;\mu_{Q_0}} - \tilde{K}\parz{\Tsc{b}{};\mu_{Q_0}} \, ,
\end{equation}
%%%%%%%%%%%%%%%%%%%%
so for definiteness we have set $\mu = \mu_{Q_0}$ on the right side of \eref{firstratio}. Equation~\eqref{e.firstratio} may be re-expressed as 
%%%%%%%%%%%%%%%%%%%%
\begin{equation}
\tilde{f}_{j/p}(x,{\Tsc{b}{}}{};\mu,\zeta) = \tilde{f}_{j/p}(x,\bstarsc;\mu,\zeta) \exp \left\{ -g_{i/p}(x,\Tsc{b}{}) - g_K(\Tsc{b}{}) \ln \parz{\frac{\sqrt{\zeta}}{Q_0}} \right\} \, , \label{e.gintro}
\end{equation}
%%%%%%%%%%%%%%%%%%%%
where we have defined 
%%%%%%%%%%%%%%%%%%%%
\begin{equation}
g_{i/p}(x,\Tsc{b}{}) \equiv - \ln \parz{\frac{\tilde{f}_{j/p}(x,{\Tsc{b}{}}{};\mu_{Q_0},Q_0^2)}{\tilde{f}_{j/p}(x,\bstarsc;\mu_{Q_0},Q_0^2)}} \qquad \& \qquad g_K(\Tsc{b}{}) \equiv 
\tilde{K}\parz{\bstarsc;\mu_{Q_0}} - \tilde{K}\parz{\Tsc{b}{};\mu_{Q_0}} \, . \label{e.gdefs}
\end{equation}
%%%%%%%%%%%%%%%%%%%%
Note that neither $g_{i/p}(x,\Tsc{b}{})$ nor $g_K(\Tsc{b}{})$ depends on a renormalization scale $\mu$. (Although both will generally depend on $\bmax$ and $g_{i/p}(x,\Tsc{b}{})$ will depend on the input hard scale $Q_0$.)
By construction, these $g$ functions vanish like a power of $\Tsc{b}{}$ as $\Tsc{b}{} \to 0$, while at large $\Tsc{b}{}$ they are sensitive to nonperturbative effects. The TMD pdf factor  $\tilde{f}_{j/p}(x,\bstarsc;\mu,\zeta)$ out front on the right side of \eref{gintro} is restricted to be evaluated only at values of $\Tsc{b}{}$ that are smaller than $\approx \bmax$. We may evolve it to the scale $\mubstar$, 
%%%%%%%%%%%%%
\begin{align}
&\tilde{f}_{i/p}(\xbj,\bstarsc;\mu,\zeta) \no
&= \tilde{f}_{i/p}(\xbj,\bstarsc;\mubstar,\mubstar^2)
\exp \left\{
\int_{\mubstar}^{\mu} \frac{d \mu^\prime}{\mu^\prime} \left[\gamma(\alpha_s(\mu^\prime);1) 
- \ln \left(\frac{\sqrt{\zeta}}{\mu^\prime} \right)\gamma_K(\alpha_s(\mu^\prime))
  \right] +\ln \left(\frac{\sqrt{\zeta}}{\mubstar} \right)\tilde{K}(\Tsc{b}{};\mubstar) \right\} \, . \label{e.evolveddbstar}
\end{align} 
%%%%%%%%%%%%%
Evolving to $\mu = \sqrt{\zeta} = Q = \mu_Q$ then gives for \eref{gintro}
%%%%%%%%%%%%%
\begin{align}
& \tilde{f}_{j/p}(x,{\Tsc{b}{}}{};\mu_Q,Q^2) = \no
& \;\; \tilde{f}_{i/p}(\xbj,\bstarsc;\mubstar,\mubstar^2)
\exp \left\{
\int_{\mubstar}^{\mu_Q} \frac{d \mu^\prime}{\mu^\prime} \left[\gamma(\alpha_s(\mu^\prime);1) 
- \ln \left(\frac{Q}{\mu^\prime} \right)\gamma_K(\alpha_s(\mu^\prime))
  \right] +\ln \left(\frac{Q}{\mubstar} \right)\tilde{K}(\Tsc{b}{};\mubstar) \right\} \no
&  \;\; \times \exp \left\{ -g_{i/p}(x,\Tsc{b}{}) - g_K(\Tsc{b}{}) \ln \parz{\frac{Q}{Q_0}} \right\} \, . \label{e.CSSform1}
\end{align}
%%%%%%%%%%%%%
As final step, $\tilde{f}_{i/p}(\xbj,\bstarsc;\mubstar,\mubstar^2)$ is typically expanded in a small-$\Tsc{b}{}$ OPE, 
%%%%%%%%%%%%%
\begin{equation}
\tilde{f}_{i/p}(\xbj,\bstarsc;\mubstar,\mubstar^2) = 
\tilde{f}^\text{OPE}_{i/p}(\xbj,\bstarsc;\mubstar,\mubstar^2) + \order{\Lambda_\text{QCD}^2 \bmax^2 } \, ,
\end{equation}
%%%%%%%%%%%%%
Then \eref{CSSform1} becomes 
%%%%%%%%%%%%%
\begin{align}
& \tilde{f}_{j/p}(x,{\Tsc{b}{}}{};\mu_Q,Q^2) \approx \no
& \;\; \tilde{f}^\text{OPE}_{i/p}(\xbj,\bstarsc;\mubstar,\mubstar^2)
\exp \left\{
\int_{\mubstar}^{\mu_Q} \frac{d \mu^\prime}{\mu^\prime} \left[\gamma(\alpha_s(\mu^\prime);1) 
- \ln \left(\frac{Q}{\mu^\prime} \right)\gamma_K(\alpha_s(\mu^\prime))
  \right] +\ln \left(\frac{Q}{\mubstar}\right) \tilde{K}(\Tsc{b}{};\mubstar) \right\} \no
&  \;\; \times \exp \left\{ -g_{i/p}(x,\Tsc{b}{}) - g_K(\Tsc{b}{}) \ln \parz{\frac{Q}{Q_0}} \right\} \, ,\label{e.CSSform2}
\end{align}
%%%%%%%%%%%%%
with errors suppressed by powers of $\order{\Lambda_\text{QCD} \bmax }$. Substituting \eref{CSSform1} into the second line of \eref{hadrotens} gives the cross section in the form that is more familiar from earlier literature,
%%%%%%%%%%%%%
\begin{align}
&{} W^{\mu \nu}(x_a,x_b,Q,\T{q}{h}) \no
&{}=\sum_{j} H_{j\bar{\jmath}}^{\mu \nu} \int \frac{\diff[2]{\T{b}{}}}{(2 \pi)^2}
    ~ e^{i\T{q}{h}\cdot \T{b}{} } e^{-g_{i/h_a}(x_a,\Tsc{b}{})} \tilde{f}_{i/h_a}(x_a,\bstarsc;\mubstar,\mubstar^2) e^{-g_{i/h_b}(x_b,\Tsc{b}{})} \tilde{f}_{i/h_b}(x_b,\bstarsc;\mubstar,\mubstar^2) \no
&{} \; \; \times \exp \left\{
2 \int_{\mubstar}^{\mu_Q} \frac{d \mu^\prime}{\mu^\prime} \left[\gamma(\alpha_s(\mu^\prime);1) 
- \ln \left(\frac{Q}{\mu^\prime} \right)\gamma_K(\alpha_s(\mu^\prime))
  \right] +\ln \left(\frac{Q^2}{\mubstar^2}\right) \tilde{K}(\Tsc{b}{};\mubstar)  - g_K(\Tsc{b}{}) \ln \parz{\frac{Q^2}{Q_0^2}} \right\} \no
&{} \; \;  + \left(a\longleftrightarrow b\right) + \order{\Lambda_\text{QCD}^2/Q^2} \, .
\label{e.hadrotensold1}
\end{align}
%%%%%%%%%%%%%
Or, utilizing the typical OPE approximations for $\tilde{f}_{i/h_a}(x_a,\bstarsc;\mubstar,\mubstar^2)$ and 
$\tilde{f}_{i/h_b}(x_b,\bstarsc;\mubstar,\mubstar^2)$, 
%%%%%%%%%%%%%
\begin{align}
&{} W^{\mu \nu}(x_a,x_b,Q,\T{q}{h}) \no
&{}=\sum_{j} H_{j\bar{\jmath}}^{\mu \nu} \int \frac{\diff[2]{\T{b}{}}}{(2 \pi)^2}
    ~ e^{i\T{q}{h}\cdot \T{b}{} } e^{-g_{i/h_a}(x_a,\Tsc{b}{})} \tilde{f}^\text{OPE}_{i/h_a}(x_a,\bstarsc;\mubstar,\mubstar^2) e^{-g_{i/h_b}(x_b,\Tsc{b}{})} \tilde{f}^\text{OPE}_{i/h_b}(x_b,\bstarsc;\mubstar,\mubstar^2) \no
&{} \; \; \times \exp \left\{
2 \int_{\mubstar}^{\mu_Q} \frac{d \mu^\prime}{\mu^\prime} \left[\gamma(\alpha_s(\mu^\prime);1) 
- \ln \left(\frac{Q}{\mu^\prime} \right)\gamma_K(\alpha_s(\mu^\prime))
  \right] +\ln \left(\frac{Q^2}{\mubstar^2}\right) \tilde{K}(\Tsc{b}{};\mubstar)  - g_K(\Tsc{b}{}) \ln \parz{\frac{Q^2}{Q_0^2}} \right\} \no
&{} \; \;  + \left(a\longleftrightarrow b\right) + \order{\Lambda_\text{QCD}^2 \bmax^2} \, ,
\label{e.hadrotensold2}
\end{align}
%%%%%%%%%%%%%
which is how the Drell-Yan cross section has been typically expressed in the literature. 

The steps leading to \erefs{hadrotensold1}{hadrotensold2} apply to the precise operator definitions of the correlation functions, but they also remain exact for the expressions in the HSO approach at order $n$ so long as all quantities, including kernels and anomalous dimensions are exactly the order-$n$ expressions. As such, all the expressions from the main body of this paper can be used directly in \eref{hadrotensold1} or \eref{hadrotensold2}, and the  to the more familiar form is complete. For example, substituting our \eref{evolvedd3p} and \eref{evol_paramb} into \eref{gdefs} gives the $g$-functions in \eref{hadrotensold1}.

In the HSO approach, making the approximation leading from \eref{hadrotensold1} to \eref{hadrotensold2}, while allowed, is not necessary because the parametrizations of $\tilde{f}_{i/h}(x,\bstarsc;\mubstar,\mubstar^2)$ automatically interpolate to the OPE expressions in the small $\Tsc{b}{}$ limit. We will never make such an approximation in our applications. 

Notice that the TMD pdfs, and hence the cross section, are exactly independent of $\bmax$, 
%%%%%%%%%%%%%
\begin{equation}
\frac{\diff{}{}}{\diff{\bmax}{}} \tilde{f}_{j/p}(x,{\Tsc{b}{}}{};\mu_Q,Q^2) = 0 \, .
\end{equation}
%%%%%%%%%%%%%
Thus, despite how it is sometimes portrayed in the literature, $\bmax$ is \emph{not} a model parameter. Large sensitivity to $\bmax$ in an actual TMD parametrization is a symptom that the approximations leading from \eref{evolveddbstar} to \eref{CSSform2} are not under control. 

It is worthwhile to further discuss the various auxiliary parameters that enter both the more conventional approaches and the HSO approach and comparing their roles. This includes parameters like $\bmax$ and $\bmin$ from the conventional approaches and the parameter $a$ in \eref{qbar_param_a} from the HSO approach. All are auxiliary in the sense that the TMD pdfs are in principle exactly independent of them. Any dependence in practical parametrizations reflects sensitivity to approximations like truncating the order of perturbation theory. For example, there is exactly no $\bmax$ dependence in \eref{hadrotensold1}, either with exact expressions or with expressions in the HSO approach. However, some residual $\bmax$-dependence gets introduced when using 
\eref{hadrotensold2} because there is a perturbative OPE approximation applied to $\tilde{f}^\text{OPE}_{i/h}(x,\bstarsc;\mubstar,\mubstar^2)$ and $\order{\Lambda_\text{QCD} \bmax}$ terms are neglected. Potentially large errors are also introduced if an overly simplistic model is used for the $g$-functions. Since varying $\bmax$ amounts to shifting contributions between the ``perturbative'' parts of \eref{hadrotensold1} and the ``nonperturbative'' $g$-functions, then the ansatz for the $g$-functions cannot be guessed at independently of the perturbative contributions without introducing hard-to-control errors. This creates an awkward trade-off for the conventional approaches: On one hand, one may try to choose a very small $\bmax$ so as to ensure that dropping $\order{\Lambda_\text{QCD} \bmax}$ in \eref{hadrotensold2} is valid, but then the ostensibly nonperturbative $g$-functions will need to describe a wide region of $\Tsc{b}{}$ that is essentially perturbative, and simplistic $g$-ansatzes are bound to fail there. On the other hand, one may choose a large $\bmax$ to ensure that the $g$-functions are ``purely nonperturbative,'' but this then threatens to introduce large, uncontrolled errors in the transition from \eref{hadrotensold1} to \eref{hadrotensold2}. The HSO approach does not encounter this problem because it does not use the $\bstarsc$ separation. 

The $\overline{Q}_0$ that is used in the HSO approach is in some ways analogous to the $\mubstar$ in the conventional $\bstar$ approach in that both impose a renormalization scale transition to $\sim 1/\Tsc{b}{}$ at small $\Tsc{b}{}$. However, there are important differences. The $\bstar$ prescription does not merely provide an RG scale transformation at low $\Tsc{b}{}$, but rather it also fixes the transition between perturbative and nonperturbative descriptions. By contrast, the $\overline{Q}_0$ in the HSO approach is used \emph{only} to transform scales as sensitivity to UV transverse momentum grows. Notice that in the conventional approach there is not only a scheme change from $\mu$ to $\mubstar$ in \eref{hadrotensold2}, but also the $\Tsc{b}{}$ argument itself gets replaced by $\bstarsc$ in $\tilde{f}^\text{OPE}_{i/h}(x,\bstarsc;\mubstar,\mubstar^2)$.  There is no such analogous replacement in the HSO approach. Rather, there is an explicit description of the transition between the physics of the purely nonperturbative and safely perturbative regions. 

In other words, the use of the $\bstar$ prescription within the conventional approach couples the steps of 1.) transforming scales and 2.) of demarcating the perturbative and nonperturbative regions while the HSO approach maintains these as two completely different steps. One consequence is that sensitivity to $\overline{Q}_0$ is limited to renormalization scale sensitivity, and is therefore much weaker than the typical sensitivity to $\bmax$ in conventional steps. In principle, $\overline{Q}_0$ is eliminated completely by keeping higher orders and using sufficiently large $Q_0$. See, for example, Figures 11 and 12 of \cite{Gonzalez-Hernandez:2022ifv}. 

Another parameter that sometimes gets introduced in the 
more conventional organizations is a cutoff $\bmin$ on small transverse sizes~\cite{Collins:2016hqq}. At first sight, this appears to provide a convenient way to connect to collinear pdfs and to control UV contributions from $\Tsc{k}{} \gg Q$.  In a mirror image of the $\bmax$ approach, one replaces $\Tsc{b}{}$ by a function $b_c(\Tsc{}{})$ that is $\Tsc{b}{}$ at large $\Tsc{b}{}$ but approaches a $\bmin \sim 1/Q$ as $\Tsc{b}{} \to 0$. For example, 
%%%%%%%%%%%%%
\begin{equation}
b_c(\Tsc{b}{}) = \sqrt{\Tscsq{b}{} + C_1^2/Q^2} \, 
\end{equation}
%%%%%%%%%%%%%
so that $\bmin = C_1/Q$. In the Fourier transform that defines the momentum space TMD pdf, the replacement is  
$\Tsc{b}{} \to b_c(\Tsc{b}{})$ so that 
%%%%%%%%%%%%%
\begin{equation}
f_{j/h}(\xbj,\Tsc{k}{};\mu_Q,Q^2) \to \int \frac{\diff[2]{\T{b}{}}}{(2 \pi)^2} e^{i \T{k}{} \cdot \T{b}{}} \tilde{f}_{j/h}(x,b_c(\Tsc{b}{});\mu_Q, Q^2)  \, , 
\label{e.bcuttmd}
\end{equation}
%%%%%%%%%%%%%
and the resulting TMD pdf has the $\Tsc{b}{} \ll 1/Q$ contribution removed. Integrating this UV-regulated TMD pdf over all $\Tsc{k}{}$ gives a delta function that fixes $\Tsc{b}{} = 0$ and $b_c(\Tsc{b}{}) = \bmin$. From \eref{CSSform2} this gives, 
%%%%%%%%%%%%%
\begin{align}
&{} \tilde{f}^\text{OPE}_{j/h}(x,\bmin;\mu_Q, Q^2) 
\exp \left\{ -g_{i/p}(x,\bmin) - g_K(\bmin) \ln \parz{\frac{Q}{Q_0}} \right\} \no
&{} \qquad =  f_{j/h}(x,\mu_Q) + \order{\alpha_s(\mu_Q)}   + \order{\frac{\bmin^2}{\bmax^2}} + \order{\Lambda_{\text{QCD}}^2 \bmax^2} \, .
\end{align}
%%%%%%%%%%%%%
Thus, integrating \eref{bcuttmd} appears to give an approximation to the standard integral expectation relating TMD and collinear pdfs, 
%%%%%%%%%%%%%
\begin{equation}
\int \diff[2]{\T{k}{}} f_{j/h}(\xbj,\Tsc{k}{};\mu_Q,Q^2)
= f_{j/h}(x,\mu_Q) + \order{\alpha_s(\mu_Q)}  + 
\order{\frac{\bmin^2}{\bmax^2}} + \order{\Lambda_{\text{QCD}}^2 \bmax^2}
\label{e.intrelold}
\end{equation} 
%%%%%%%%%%%%%
if it is possible to justify an approximation wherein $\bmin \ll \bmax$. Recall, however, that $\bmax$ needs to be kept as small as possible to justify dropping  $\order{\Lambda_\text{QCD} \bmax}$ terms in \eref{CSSform2} while $\bmin$ must be $\order{1/Q}$ (above we have used $\bmin = C_1/Q$). Near the moderate scales where sensitivity to nonperturbative transverse momentum is most relevant, therefore, $\bmin \approx \bmax$ and the errors in \eref{intrelold} are not negligible. A $\bmin$ prescription is, therefore, not appropriate for treatments  
focused on extracting nonperturbative properties of TMD pdfs. It is only after having evolved to rather high $Q$ that one may justify a $\bmin \ll \bmax$ type of approximation. 

A $\bmin$ regulator is not necessary, and we do not use one in the HSO approach. 
%%%%%%%%%%%%%%%%%%%%%

%%%%%%%%%%%%%%%%%%%%%
\section{Comment on counting logarithms}
\label{a.logcounting}

High energy transverse momentum dependent cross sections are often discussed in the language of resummed logarithms of transverse momentum. These methodologies start from a purely collinear factorization framework, assuming $\Tsc{q}{}$ to be large enough to be perturbative, but still small enough that $\Tsc{q}{}/Q \ll 1$. This creates a situation in collinear factorization where it is ambiguous which of the two scales is an appropriate hard scale. Either case leads to logarithms of the form  
%%%%%%%%%%%%%
\begin{equation}
\alpha_s(Q) \ln^2 \parz{Q \Tsc{b}{}}
\end{equation}
%%%%%%%%%%%%%
in transverse coordinate space for the cross section with large contributions from the $\Tsc{b}{} \sim 1/\Tsc{q}{}$ region. A typical resummation approach organizes large logarithms of this type into series that are then summed and exponentiated to all orders.  Thus, one hears of leading and next-to-leading logarithm resummations, depending on the order of logarithms that are resummed. This is a reasonable approach provided the logarithms are never so large that the small coupling perturbation series expansion stops being valid. Its range of applicability is then limited to some region of $\Lambda_\text{QCD} \ll \Tsc{q}{} \ll Q$, a window that grows and shrinks depending on $Q$. As $\Tsc{q}{} \to 0$, collinear factorization breaks down entirely, while for $\Tsc{q}{} \to Q$ the logarithms of $\Tsc{q}{}/Q$ vanish and resummation becomes unnecessary. Therefore, ``counting logarithms'' alone can become a rather misleading characterization the accuracy or precision of a calculation. It is only when the TMDs are considered primarily as \emph{perturbative} objects in collinear factorization that the increasing size of the logarithms at large $\Tsc{b}{}$ implies a genuine problem with factorization. In treatments that are organized around a resummation methodology, one frequently starts with a descriptions of the $\Lambda_\text{QCD} \ll \Tsc{q}{} \ll Q$ region and incorporates nonperturbative fitting only as a correction to interpolate into the very small $\Tsc{q}{} \sim 0$ region.
Such a strategy \cite{Gonzalez-Hernandez:2022ifv} is reasonable at high energies, but it is arguably less suited for the treatment of more moderate energy data where detailed information about the structure of hadrons can become relevant.

The full TMD factorization and evolution treatment is more general and powerful than a basic resummation. There, the TMDs are taken from the outside as generally non-perturbative objects on the same footing as collinear pdfs, with specific operator definitions, evolution equations, etc. precision can be measured by the order of $\alpha_s$ in hard parts and evolution kernels rather than by the sizes of specific logarithms. The resummation style of treatments discussed above is recovered by TMD factorization, but as a special limiting case, emerging when the combination $a_s(Q) \, \log{(\Tsc{q}{}/Q)}$ 
is large
but fixed and $Q \to \infty$. There are no prior assumptions about the regions where resummation approximations work or how accurate they are.
For further discusions of these points, see for example Secs.~13.13.5 and 15.4 of Ref.\cite{Collins:2011qcdbook} and Appendix B of \cite{Gonzalez-Hernandez:2022ifv}.

The work in the main body of the paper is an example of this strategy, referred to as a ``bottom-up" approach in Ref\cite{Gonzalez-Hernandez:2022ifv}. We show how a successful description of experimental data can be obtained by 
expanding the perturbative part of the TMD in powers of the strong coupling, specifically up to $\mathcal{O}(\alpha_s)$. This does not prevent an improvement of the description of the data by extending calculations to next-to-leading-log accuracy, but it is because doing so introduces contributions of order $\mathcal{O}(\alpha_s^2)$.

\end{appendix}

%%%%%%%%%%%%%%%%%%%%%
\bibliography{bibliography}

%%%%%%%%%%%%%%%%%%%%%
\end{document}